\documentclass[longauth,traditabstract]{aa}

\usepackage[nonamebreak]{natbib}
\usepackage[stable]{footmisc}
\usepackage{amsmath}
\usepackage{txfonts}
\usepackage{placeins}
\usepackage{natbib}
\bibpunct{(}{)}{;}{a}{}{,}
\usepackage{graphicx}
\usepackage{epstopdf}
\usepackage{ifthen}
\usepackage{mathtools}
\usepackage[breaklinks, colorlinks, citecolor=blue]{hyperref}
\usepackage{fixltx2e}
\usepackage{url}
\usepackage[table,usenames,dvipsnames]{xcolor}
\usepackage{soul} 

\def\setsymbol#1#2{\expandafter\def\csname #1\endcsname{#2}}
\def\getsymbol#1{\csname #1\endcsname}

\def\Planck{\textit{Planck}}

\def\HeJT{$^4$He-JT}

\newbox\tablebox    \newdimen\tablewidth
\def\leaderfil{\leaders\hbox to 5pt{\hss.\hss}\hfil}
\def\endPlancktable{\tablewidth=\columnwidth 
    $$\hss\copy\tablebox\hss$$
    \vskip-\lastskip\vskip -2pt}

\def\tablenote#1 #2\par{\begingroup \parindent=0.8em
    \abovedisplayshortskip=0pt\belowdisplayshortskip=0pt
    \noindent
    $$\hss\vbox{\hsize\tablewidth \hangindent=\parindent \hangafter=1 \noindent
    \hbox to \parindent{$^#1$\hss}\strut#2\strut\par}\hss$$
    \endgroup}
\def\doubleline{\vskip 3pt\hrule \vskip 1.5pt \hrule \vskip 5pt}

\def\L2{\ifmmode L_2\else $L_2$\fi}

\def\DeltaT{\ifmmode \Delta T\else $\Delta T$\fi}
\def\deltat{\ifmmode \Delta t\else $\Delta t$\fi}
\def\fknee{\ifmmode f_{\rm knee}\else $f_{\rm knee}$\fi}
\def\Fmax{\ifmmode F_{\rm max}\else $F_{\rm max}$\fi}
\def\solar{\ifmmode{\rm M}_{\mathord\odot}\else${\rm M}_{\mathord\odot}$\fi}
\def\Msolar{\ifmmode{\rm M}_{\mathord\odot}\else${\rm M}_{\mathord\odot}$\fi}
\def\Lsolar{\ifmmode{\rm L}_{\mathord\odot}\else${\rm L}_{\mathord\odot}$\fi}
\def\inv{\ifmmode^{-1}\else$^{-1}$\fi}
\def\mo{\ifmmode^{-1}\else$^{-1}$\fi}
\def\sup#1{\ifmmode ^{\rm #1}\else $^{\rm #1}$\fi}
\def\expo#1{\ifmmode \times 10^{#1}\else $\times 10^{#1}$\fi}
\def\,{\thinspace}
\def\lsim{\mathrel{\raise .4ex\hbox{\rlap{$<$}\lower 1.2ex\hbox{$\sim$}}}}
\def\gsim{\mathrel{\raise .4ex\hbox{\rlap{$>$}\lower 1.2ex\hbox{$\sim$}}}}

\def\simprop{\mathrel{\raise .4ex\hbox{\rlap{$\propto$}\lower 1.2ex\hbox{$\sim$}}}}
\def\deg{\ifmmode^\circ\else$^\circ$\fi}
\def\pdeg{\ifmmode $\setbox0=\hbox{$^{\circ}$}\rlap{\hskip.11\wd0 .}$^{\circ}
          \else \setbox0=\hbox{$^{\circ}$}\rlap{\hskip.11\wd0 .}$^{\circ}$\fi}
\def\arcs{\ifmmode {^{\scriptstyle\prime\prime}}
          \else $^{\scriptstyle\prime\prime}$\fi}
\def\arcm{\ifmmode {^{\scriptstyle\prime}}
          \else $^{\scriptstyle\prime}$\fi}
\newdimen\sa  \newdimen\sb
\def\parcs{\sa=.07em \sb=.03em
     \ifmmode \hbox{\rlap{.}}^{\scriptstyle\prime\kern -\sb\prime}\hbox{\kern -\sa}
     \else \rlap{.}$^{\scriptstyle\prime\kern -\sb\prime}$\kern -\sa\fi}
\def\parcm{\sa=.08em \sb=.03em
     \ifmmode \hbox{\rlap{.}\kern\sa}^{\scriptstyle\prime}\hbox{\kern-\sb}
     \else \rlap{.}\kern\sa$^{\scriptstyle\prime}$\kern-\sb\fi}
\def\ra[#1 #2 #3.#4]{#1\sup{h}#2\sup{m}#3\sup{s}\llap.#4}
\def\dec[#1 #2 #3.#4]{#1\deg#2\arcm#3\arcs\llap.#4}
\def\deco[#1 #2 #3]{#1\deg#2\arcm#3\arcs}
\def\rra[#1 #2]{#1\sup{h}#2\sup{m}}

\def\dots{\relax\ifmmode \ldots\else $\ldots$\fi}
\def\WHzsr{\ifmmode $W\,Hz\mo\,sr\mo$\else W\,Hz\mo\,sr\mo\fi}
\def\mHz{\ifmmode $\,mHz$\else \,mHz\fi}
\def\GHz{\ifmmode $\,GHz$\else \,GHz\fi}
\def\mKs{\ifmmode $\,mK\,s$^{1/2}\else \,mK\,s$^{1/2}$\fi}
\def\muKs{\ifmmode \,\mu$K\,s$^{1/2}\else \,$\mu$K\,s$^{1/2}$\fi}
\def\muKRJs{\ifmmode \,\mu$K$_{\rm RJ}$\,s$^{1/2}\else \,$\mu$K$_{\rm RJ}$\,s$^{1/2}$\fi}
\def\muKHz{\ifmmode \,\mu$K\,Hz$^{-1/2}\else \,$\mu$K\,Hz$^{-1/2}$\fi}
\def\MJysr{\ifmmode \,$MJy\,sr\mo$\else \,MJy\,sr\mo\fi}
\def\MJysrmK{\ifmmode \,$MJy\,sr\mo$\,mK$_{\rm CMB}\mo\else \,MJy\,sr\mo\,mK$_{\rm CMB}\mo$\fi}
\def\microns{\ifmmode \,\mu$m$\else \,$\mu$m\fi}

\def\muK{\ifmmode \,\mu$K$\else \,$\mu$\hbox{K}\fi}
\def\microK{\ifmmode \,\mu$K$\else \,$\mu$\hbox{K}\fi}
\def\muW{\ifmmode \,\mu$W$\else \,$\mu$\hbox{W}\fi}
\def\kms{\ifmmode $\,km\,s$^{-1}\else \,km\,s$^{-1}$\fi}
\def\kmsMpc{\ifmmode $\,\kms\,Mpc\mo$\else \,\kms\,Mpc\mo\fi}
\providecommand{\sorthelp}[1]{}

\graphicspath{{./}} 

\newcommand{\citelowell}{\citetalias{planck2014-a10}}
\newcommand{\citedpc}{\citetalias{planck2016-l03}}

\defcitealias{planck2014-a10}{LowEll}
\defcitealias{planck2016-l03}{L03}

\newcommand{\CHANGE}[1]{{\color{black} #1}}
\newcommand{\NCHANGE}[1]{{\color{black} #1}}
\newcommand{\sroll}{{\tt SRoll}}
\newcommand{\srollone}{{\tt SRoll1}}
\newcommand{\srolltwo}{{\tt SRoll2}}
\newcommand{\sims}{{\tt SRoll2Sims}}
\newcommand{\stepone}{{\tt Step\#1}}
\newcommand{\steptwo}{{\tt Step\#2}}

\begin{document}

\title{\vglue -10mm \srolltwo: an improved mapmaking approach to reduce large-scale systematic effects in the \Planck\ High Frequency Instrument legacy maps}

\author{\small
J.-M.~Delouis\inst{1,2,3}~\thanks{Corresponding author: J.-M.~Delouis, jean.marc.delouis@ifremer.fr}
\and
L.~Pagano\inst{4,5,2,6}
\and
S.~Mottet\inst{2,3}
\and
J.-L.~Puget\inst{6,5,2}
\and
L.~Vibert\inst{5,2}
}

\institute{\small
Laboratoire d'Oc{\'e}anographie Physique et Spatiale (LOPS), Univ. Brest, CNRS, Ifremer, IRD, Brest, France\goodbreak
\and
Institut d'Astrophysique de Paris, CNRS (UMR7095), 98 bis Boulevard Arago, F-75014, Paris, France\goodbreak
\and
Sorbonne Universit\'{e}, UMR7095, 98 bis Boulevard Arago, F-75014, Paris, France\goodbreak
\and
Dipartimento di Fisica e Scienze della Terra, Universit\`a degli Studi di Ferrara and INFN -- Sezione di Ferrara, Via Saragat 1, I-44100 Ferrara, Italy \goodbreak
\and
Institut d'Astrophysique Spatiale, CNRS, Univ. Paris-Sud, Universit\'{e} Paris-Saclay, B\^{a}t. 121, 91405 Orsay cedex, France\goodbreak
\and
LERMA, Sorbonne Universit\'{e}, Observatoire de Paris, Universit\'{e} PSL, \'{E}cole normale sup\'{e}rieure, CNRS, Paris, France
}

\date{\vglue -1.5mm \today \vglue -5mm}

\abstract{\vglue -3mm 
This paper describes an improved mapmaking approach with respect to the one used for the \Planck\ High Frequency Instrument 2018 Legacy release. The algorithm \srolltwo\ better corrects the known instrumental effects that still affected mostly the polarized large-angular-scale data by distorting the signal, and/or leaving residuals observable in null tests. The main systematic effect is the nonlinear response of the onboard analog-to-digital convertors that was cleaned in the \Planck\ HFI Legacy release as an empirical time-varying linear detector chain response which is the first-order effect. The \srolltwo\ method fits the model parameters for higher-order effects and corrects the full distortion of the signal. The model parameters are fitted using the redundancies in the data by iteratively comparing the data and a model. The polarization efficiency uncertainties and associated errors have also been corrected based on the redundancies in the data and their residual levels characterized with simulations. This paper demonstrates the effectiveness of the method using end-to-end simulations, and provides a measure of the \CHANGE{systematic effect residuals} that now fall well below the detector noise level. Finally, this paper describes and characterizes the resulting \srolltwo\ frequency maps using the associated simulations that \CHANGE{are} released to the community  at \href{http://sroll20.ias.u-psud.fr}{\texttt{http://sroll20.ias.u-psud.fr}}.}


\authorrunning{Delouis et al.}

\titlerunning{\srolltwo: an solution to reduce large scale systematics in the HFI maps}

\maketitle


\section{Introduction}
\label{sec:intro}

The development of the \sroll\ global solution has been initiated within the \Planck\ High Frequency Instrument (HFI) Consortium, attempting to reduce the systematic effects that impact the large-scale polarization of the HFI frequency maps at 100, 143, 217, and 353\,GHz. The \sroll\ algorithm (hereafter \srollone) and its performance when applied to data and simulations, were first described in \citet{planck2014-a10} (hereafter \citelowell\ paper). The efficiency of \sroll\ led to significant improvements in the measurement of the cosmological reionization optical depth. The basically unchanged algorithm with marginal improvements, described in \citet{planck2016-l03} (hereafter \citedpc\ paper), was used for building the 2018 Legacy HFI frequency maps. Those maps are referred to in this paper as HFI2018 maps.

This paper presents the second generation of this algorithm, called \srolltwo, developed  \CHANGE{beyond the \Planck\ Collaboration one,} to further reduce the main residual systematic effects left in the HFI2018 maps, \NCHANGE{firstly the second-order correction of the analog-to-digital convertor nonlinearity (ADCNL);secondly the temperature-to-polarization leakage from the polarization efficiency and imperfect orientation knowledge of the bolometer;thirdly the temperature-to-polarization leakage from bandpass mismatch of foregrounds and associated foreground modeling;
fourthly the transfer function associated with very long time constants.}

This paper is organized as follows: Section~\ref{sec:method} presents the \srolltwo\ algorithm; Section~\ref{sec:result} presents the resulting \srolltwo\ maps; Section~\ref{sec:tempimprove} presents the characterization of those maps and estimation of the systematic effect residuals.

\section{The \srolltwo\ solution}
\label{sec:method}

As in \srollone, the improved \srolltwo\ is a global solution algorithm, estimating instrumental systematic effect parameters from the sky data, as described in detail in the \citelowell, and \citedpc\ papers. It works on the data in a single frequency band and produces Stokes I, Q, and U maps, using all polarization-sensitive bolometers (PSBs) at this frequency.

\srolltwo\ is better at handling systematic effects than \srollone, in particular the ADCNL effect. The nonlinearity of the analog-to-digital convertors (ADC) was measured in the ground tests, as well as in flight (during the warm phase of the \Planck\ HFI when the dilution cooler 3He tank was empty), and corrected in the first module of the Time Ordered Information (TOI) data processing. The accuracy of this correction was not good enough for the CMB channels, and left residuals which were then corrected empirically in the mapmaking process as a time varying linear detector chain gain. The 353-GHz modulated signals were more spread in analog-to-digital units (ADU) making the correction much smaller. Starting from the same TOIs, but not applying this first-order linear correction, \srolltwo\ builds a full nonlinear correction per detector and per stable pointing observation (also called ``ring'') in which, for each ADU of the signal, a nonlinear correction of the ADC transfer function model is fitted iteratively to the difference between the data and the data model at each iteration. As the \CHANGE{HFI bolometers have} never shown in tests gain variations other than the one associated with the temperature drift of the 0.1K-plate (which is corrected for), we attribute to the ADC any residual apparent gain variation, and we adopt a detector response stable in time during the mission. This is a major change which should coherently correct several systematic effects associated with the nonlinearity of the ADC residuals.  

The \srolltwo\ ADCNL correction is computed as a function of the ADU value of the demodulated signal based on a set of spline functions (Sect.~\ref{sec:ADCEQUATION}) which models semi-empirically the ADCNL residuals in each pointing period. This is done at the ring level (and not at the timeline level), reducing the computational requirements in the mapmaking process to an acceptable value. The number of spline functions needed is much smaller than the number of ADUs to correct because the electronic noise and modulation smooth the signal over a large number of ADUs. The ADCNL spline functions thus replace the correction as an empirical time variation of the detector response, and they also account for the nonlinear distortion of strong signals. This decomposition in spline functions varies in time\footnote{In most of the cases, the amplitude of the difference in ADCNL residuals between successive rings is less than  $10^{-2}\,\mu$K; in cases of solar flares, this difference can reach $\simeq 1\,\mu$K for a few rings.} to empirically adapt to the slowly changing shape of the electronic modulation of the signal, which can be considered stable during one ring.

This ADCNL determination method converges if the initial estimated frequency map used is close enough to the exact solution. This is why the ADCNL computation is always done after a computation of corrections of all other systematic effects, including the calibrated response, to produce an estimated frequency map based on the last minimization of the difference between the data rings (called HPRs for {\tt HEALPix} \citep{gorski2005} rings) and the model, providing the best correction available. The ADCNL residual semi-empirical correction is fitted as a function of the demodulated signal (not going through a full model of the ADC and the modulation electronic signal). The ADCNL correction is degenerate with the calibrated response variation. Therefore, it is not possible to fit the ADCNL correction simultaneously with the calibration which is the result of the frequency map solution from the previous step. If our assumption that the apparent gain variations are only coming from the ADCNL and not from the bolometer itself is right, the calibration response should converge to a stable one when the ADCNL correction converges. We show in  Sect.~\ref{sec:ADCEQUATION} that this is happening.

\begin{figure}[ht!]
\includegraphics[width=\columnwidth]{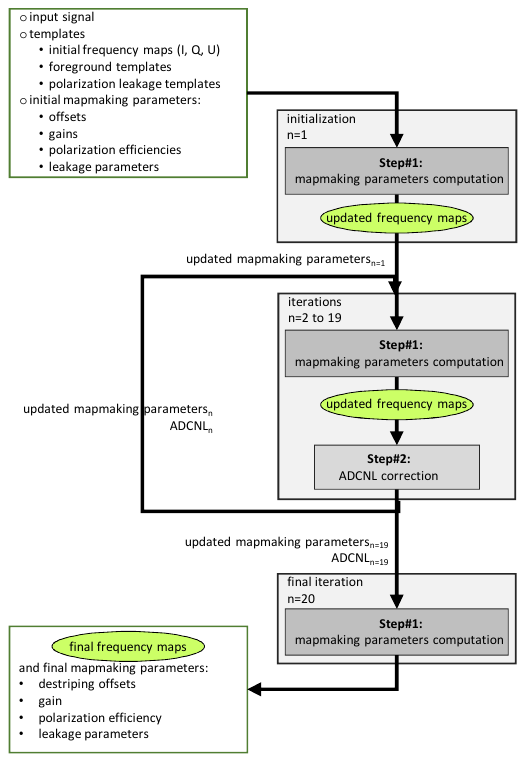}
\caption{Schematic view of \srolltwo\ iterations for building the Stokes $I$, $Q$, and $U$ maps for one frequency band. The top box contains the initial improvements of the maps. \stepone\ algorithm uses data HPRs to consistently compute the maps, the detector response, and the systematic effects correction parameters other than ADCNL. \steptwo\ computes the ADCNL correction that is propagated to the subsequent iteration.  In the middle box, \stepone\ produces maps used as input for \steptwo\ that computes the ADCNL correction. This whole process is then iterated 19 times. The bottom box is again a \stepone\ to build the final maps.}
\label{fig:srollitt} 
\end{figure}

Figure~\ref{fig:srollitt} shows the \srolltwo\ algorithm scheme composed of two steps: the first step (called \stepone) is similar to the \srollone\ algorithm, but with improvements that increase the convergence speed. It builds polarized frequency maps by optimizing mapmaking parameters for absolute bolometer calibration, polarization efficiencies, and orientations, minimizing the discrepancies between bolometer signals within a frequency band, and minimizing the difference with the data.

In order to obtain a model for the ADCNL computation, \stepone\ must be first run to produce frequency maps as input for the loop. Simulations have verified that one \stepone\ iteration before computing the ADCNL correction is enough to get convergence.
 
The second step (called \steptwo) is new: it starts from the polarized map generated in \stepone\ and refines, for each bolometer, the ADCNL parameters which are the spline function weights used to model the ADC correction.

This is contained in the central box of Fig.~\ref{fig:srollitt}. In about 20 iterations, the ADCNL converges, and the final maps are obtained by returning to \stepone.

This section details the \srolltwo\ algorithm: the first subsection explains the data model hypotheses, the second subsection details the \stepone\ that computes the calibrated detector chain response, but also the systematic correction parameters at the exception of the ADCNL ones, the third subsection describes the \steptwo\ that fits the ADCNL model parameters. The last subsections address map projection, future improvements, and simulations.

\subsection{Improved scheme}

Similarly to the \srollone\ data model (see equation~1 of \citedpc), the \srolltwo\ data model for bolometer signal is given by Eq.~\ref{eq:datamodel} where indices are:
\begin{itemize}
\item $b$ for the bolometer, up to $n_{\mathrm{bolo}}$;
\item $r$ for the stable pointing period (ring), up to $n_{\mathrm{ring}}$;
\item $p$ for the sky pixel;
\item $h$ for bins of spin frequency harmonics (up to $n_{\mathrm{harm}}$), labeled as $bin_{h=1}$ for the first harmonic, $bin_{h=2}$ for harmonics 2 and 3, $bin_{h=3}$ for harmonics 4 to 7, and $bin_{h=4}$ for harmonics 8 to 15;
\item $f$ for the polarized foreground, up to $n_{\mathrm{comp}}$ (dust and free-free).
\end{itemize}
In this paper, ``response'' refers to the detector chain response of a given bolometer to a CMB signal (in ADU/$\mu$K), calibrated on the orbital dipole signal.
\begin{eqnarray}
\label{eq:datamodel}
g_{b}M_{b,r,p} &=& A_{b,r,p} \cdot S_{p} + \sum_{h=1}^{n_{\mathrm{harm}}}\gamma_{b,h} V_{b,r,p,h} + 
\sum_{f=1}^{n_{\mathrm{comp}}}\delta L_{b,f} C_{p,f}\nonumber \\
 & & +\ D^{\mathrm{tot}}_{r,p} + F_{b,r,p}^{\mathrm{dip}}+ F_{b,r,p}^{\mathrm{gal}} + O_{b,r} + g_{b}N_{b,r,p} + g_{b}\aleph_{b,r},
\end{eqnarray}
where:
\begin{itemize}
\item $g_{b}$ is the calibrated response of a bolometer chain;
\item $M_{b,r,p}$ is the measured bolometer total signal;
\item $A_{b,r,p} = \left[1,\rho_b\cos(2\phi_{b,r,p}),\rho_b\sin(2\phi_{b,r,p})\right]$ is the pointing vector giving the observed pixel for a given bolometer in a given ring;
\item $S_{p} = \left[I_p, Q_p, U_p\right] $ is the sky signal in pixel $p$ after subtraction of the orbital dipole assumed to be known with an amplitude invariant in time. $I_p$, $Q_p$, and $U_p$ represent the common sky maps seen by all bolometers (excluding the Solar dipole)\footnote{$(\cdot)$ refers to the scalar product.};
\item $\rho_b$ is the ground-measured polarization efficiency;
\item $\phi_{b,r,p}$ is the ground-measured detector polarization angle for bolometer $b$ with respect to the north-south axis;
\item $V_{b,r,p,h}$ is the spatial template of the empirical time-transfer function computed from timeline convolution with long time constant (e.g., $>1\,$s);
\item $\gamma_{b,h}$ is the empirical time-transfer function complex amplitude;
\item $C_{p,f}$ are the input foreground component spatial templates which are stable for a given \srolltwo\ run;
\item $\delta L_{b,f}$ is the bolometer bandpass foreground color-correction coefficient relative to the bolometers average bandpass, for foreground $f$, that is, for each foreground component, we set $\sum_{b=1}^{n_{\mathrm{bolo}}} \delta L_{b,f} = 0$. The component brightness map from the frequency bandpass averaged over the bolometers is part of the frequency sky signal $S_{p}$;
\item $D^{\mathrm{tot}}_{r,p}$ is the total CMB dipole signal (sum of solar and orbital dipoles), with $D^{\mathrm{sol}}_{p}$ being the template for the solar dipole with a constant direction and amplitude and $D^{\mathrm{orb}}_{r,p}$ being the template of the orbital dipole;
\item $F_{b,r,p}^{\mathrm{dip}}$ and $F_{b,r,p}^{\mathrm{gal}}$ are respectively the total dipole and the Galactic signal integrated over the far sidelobes (FSL);
\item $O_{b,r}$ is the offset per pointing period $r$ used to model the $1/f$ noise, and we set $\sum_{b=1}^{n_{\mathrm{bolo}}} \sum_{r=1}^{n_{\mathrm{ring}}}O_{b,i}=0$, since the \Planck\ data provide no information on the monopole;
\item $N_{b,r,p}$ is the pixel white noise (electronic and photon noises), with variance ${\sigma_{b,r,p}}^2$;
\item $\aleph_{b,r}$ is the signal distortion induced by the ADCNL residual effect following the TOI correction for the bolometer $b$, at ring $r$.
\end{itemize}
The $\aleph_{b,r}$ term, new with respect to \srollone\ data model, is added to describe the ADCNL residual systematic effect as a function of the signal level. This model and the method to fit its parameters from the data redundancy is the key of the correction of the ADCNL systematic effect, including both the time variability and the nonlinearity of strong signals.

\subsection{\stepone: computing the bolometer response and associated leakage coefficients}

The \stepone\ process is similar to the \srollone\ algorithm but dropping the empirical time varying gain which was a first-order correction of the ADCNL systematic effect. Thus, \srolltwo\ models the detector response as a constant value. In \srolltwo, the ADCNL systematic effect is cleaned in the \steptwo\ process (see Sect.~\ref{sec:ADCNLGEN}). 

The \stepone\ of \srolltwo\ is an upgraded version of the \srollone\ algorithm that improves the response convergence, and decreases the number of required iterations. Solving for a single response per bolometer for the whole mission necessarily involves solving a nonlinear least-squares equation. Like \srollone, \srolltwo\ uses an iterative scheme to solve for the response $g_{b}$. We therefore introduce $g_{b,n}$  where $n$ stands for the iteration number. At iteration $n$, we set:
\begin{eqnarray}
\label{eq:ge}
g_{b, n} = g_{b} + \delta g_{b, n},
\end{eqnarray}
where $\delta g_{b, n}$ is the difference between the responses $g_{b,n}$ and the response $g_{b}$ associated with \srolltwo\ inputs (e.g., the foreground templates) and the instrumental model.

The goal is to iteratively fit $\delta g_{b, n}$. In order to reduce the complexity of Eq.~\ref{eq:datamodel}, we first remove the low-amplitude signal contribution of the FSLs; this does not induce any visible remaining signature signal in specific null tests. We also remove the best estimate of $\aleph_{b,r,n}$, initially set to $\aleph_{b,r,n=0}=0$, and then iteratively computed in \steptwo\ (see Sect.~\ref{sec:ADCNLGEN}). The reduced bolometer total signal $M'_{b,r,p}$ becomes:
\begin{eqnarray}
g_{b}~M'_{b,r,p} = g_{b}~M_{b,r,p} -\ D^{\mathrm{tot}}_{r,p} - F_{b,r,p}^{\mathrm{dip}}- F_{b,r,p}^{\mathrm{gal}} -\aleph_{b,r}.
\label{eq:fsl}
\end{eqnarray}

We define the systematic effects $\epsilon_{b,r,p}$, including detector time constant amplitudes as:
\begin{eqnarray}
\epsilon_{b,r,p} & = &\sum_{h=1}^{n_{\mathrm{harm}}}\gamma_{b,h} V_{b,r,p,h} + \sum_{f=1}^{n_{\mathrm{comp}}} \delta L_{b,f} C_{p,f} + F_{b,r,p}^{\mathrm{dip}}+ F_{b,r,p}^{\mathrm{gal}} + \aleph_{b,r},
\label{eq:epsilon}
\end{eqnarray}
where ($\delta L_{b,f} \, C_{p,f}$) is the correction to the coefficient intensity to polarization leakage due to the bandpass mismatch between the polarized detectors with respect to ($L_f \, C_{p,f}$) which is the averaged foreground ($f$) signal over all bolometers within a frequency. ($L_f \, C_{p,f}$) is part of the signal $S_{p}$. Those systematic effects are most often much smaller than the signal. To avoid extra leakage from frequency bandpass mismatch and/or beam anisotropy related to strong signal brightness gradients, a mask is used to ensure that $\epsilon_{b,r,p}$ is small enough on the sky used to compute the systematic effect parameters. This is particularly important at 353\,GHz where the Galactic plane has strong gradients. The masks are the same as those used in \citedpc, and they retain 86.2\%, 85.6\%, 84.6\%, 86.1\% of the sky for the 100, 143, 217, and 353-GHz maps, respectively. Then, for all the used pixels, we have:
\begin{eqnarray}
\epsilon_{b,r,p} &\ll & A_{b,r,p} \cdot S_{p}+D^{\mathrm{tot}}_{r,p} \,.
\label{eq:smaller}
\end{eqnarray}

Using Eqs.~\ref{eq:datamodel}, \ref{eq:ge}, and \ref{eq:smaller}, Eq.~\ref{eq:fsl} now becomes:
\begin{eqnarray}
g_{b,n}~M'_{b,r,p} &=& \left( 1 + \frac{\delta g_{b, n}}{g_{b}} \right)~g_{b}~M'_{b,r,p} \nonumber \\
& = & A_{b,r,p} \cdot S_{p} +\sum_{h=1}^{n_{\mathrm{harm}}}\gamma_{b,h,n} V_{b,r,p,h} + \sum_{f=1}^{n_{\mathrm{comp}}}\left(\delta L_{b,f,n} C_{p,f} \right) \nonumber \\
& & +\ O_{b,r,n} + \frac{\delta g_{b, n}}{g_{b}}~ \left(A_{b,r,p}\cdot S_{p}+D^{\mathrm{tot}}_{r,p}+ \epsilon_{b,r,p}\right) \nonumber \\
& & + g_{b,n}~N_{b,r,p}.
\label{eq:SimplifiedModelbis}
\end{eqnarray}

The term $\frac{\delta g_{b, n}}{g_{b}}$ should converge toward zero after several iterations. This is only possible if $(A_{b,r,p}\cdot S_{p}+D^{\mathrm{tot}}_{r,p}+ \epsilon_{b,r,p})$ is known. Equation~\ref{eq:smaller} gives the value of $\epsilon_{b,r,p}$, and it can be verified to be negligible. Then, in Eq.~\ref{eq:SimplifiedModelbis}, ($\delta g_{b, n}/g_{b}$) can be solved only knowing $S_p$ and $D^{\mathrm{tot}}_{r,p}$. The quantity ($A_{b,r,p}\cdot S_{p}+D^{\mathrm{tot}}_{r,p}$) is modeled and refined iteratively.

The \stepone\ algorithm is based on the redundancy of different observations of the signal in pixel p, $S_{p}$, made by the same detector at different times, and different detectors in the same pointing period, to determine the response of each bolometer. The key quantity for extracting the response is the observation ring vector residual model $R_{b,r,p}$ for detector $b$ and pointing period $r$:
\begin{eqnarray}
\label{eq:srollchi2}
R_{b,r,pn,} & = & g_{b,n}~M'_{b,r,p} - T_{b,r,p} M'_{p} \nonumber \\
& = & \frac{\delta g_{b, n}}{g_{b}}~\left(A_{b,r,p} \cdot S_{p}+D^{\mathrm{tot}}_{r,p}\right) - T_{b,r,p}\frac{\delta g_{n}}{g_{p}} \left(A_{p} \cdot S_{p}+D^{\mathrm{tot}}_{p}\right) \nonumber \\
& & + O_{b,r,n} - T_{b,r,p} O_{p,n} + \sum_{h=1}^{n_{\mathrm{harm}}}\left( \gamma_{b,h,n} V_{b,r,p,h} - T_{b,r,p} \gamma_{h,n}V_{p,h}\right) \nonumber \\
& & + \sum_{h=1}^{n_{\mathrm{comp}}}\left(\delta L_{b,f,n} C_{p,f} - T_{b,r,p} \delta L_{f,n} C_{p,f}\right) \nonumber \\
& & + \left(g_{b}+\delta g_{b, n}\right)\left(N_{b,r,p} - T_{b,r,p} N_{p}\right),
\end{eqnarray}

where:
\begin{itemize}
\item $T_{b,r,p}$ defined as the projection matrix $\left( A_{b,r,p} \left( A_{p}^\top A_{p}\right)^{-1}A_{p}^\top \right)$ from frequency map to rings (we define $A_{p}$ as the pointing matrix built from all $A_{b,r,p}$ for the pixel $p$);
\item $\frac{\delta g_{n}}{g_{p}} \left(A_{p}\cdot S_{p}+D^{\mathrm{orb}}_{p}\right)$ is the matrix built from all $\frac{\delta g_{b, n}}{g_{b}}~\left(A_{b,r,p}\cdot S_{p}+D^{\mathrm{orb}}_{r,p}\right)$ weighted by the corresponding $1/{\sigma_{b}}^2$ within one pixel $p$;
\item $M'_{p}$, $\gamma_{h,n}V_{p,h}$, $ \delta L_{f,n}C_{p,f}$, $O_{p,n}$, and $N_{p}$ are the matrices respectively built from observations of pixel $p$ by all $g_{b,n}~M'_{b,r,p}$, $\gamma_{b,h,n}V_{b,r,p,h}$, $L_{b,f,n}C_{p,f}$, $O_{b,r,n}$, and $N_{b,r,p}$, each being weighted by the corresponding $1/ \sigma_{b}^2$ (e.g., $O_{p} \equiv \sum_{b}\sum_{r} O_{b,r,n,p\in r}/\sigma_b^2 $).
\end{itemize}
It is important to note that retrieval of the \srolltwo\ bandpass mismatch parameter $L_{f}$ is based on measurements from different detectors in one frequency band giving coherent results. In practice, \srolltwo\ only cleans the difference in coefficients between each bolometer and their average. The average of the corrections is  \CHANGE{therefore} set to zero in order to make the system invertible.

\begin{figure}[ht!]
\includegraphics[width=\columnwidth]{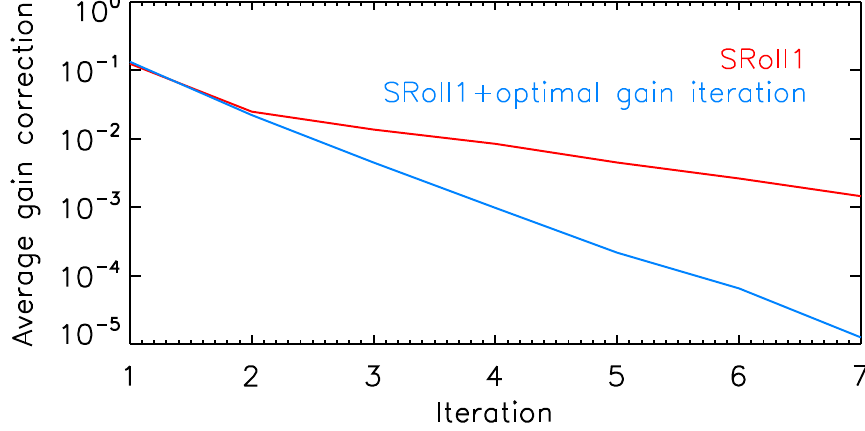}
\caption{Average on all bolometers of the response correction at each iteration of the 353-GHz map when using only the total dipole (in red) and using the sky model to fit the response (in blue).}
\label{fig:splinedistri} 
\end{figure}

The model used to fit the parameters we want to measure (e.g., the gain correction $\delta g_{b, n}$), is to minimize the difference for each pixel of the signal for a given bolometer and ring with the signal from the average of all bolometers and rings. The quantity to minimize is defined as\footnote{Quantities appearing with a tilde ({\textasciitilde}) refer to a given parameter measure affected by noise and instrumental systematic effects.}:
\begin{eqnarray}
\label{eq:srollchi3}
\tilde{R}_{b,r,p,n} & = & \frac{\delta g_{b, n}}{g_{b}}{\tilde{S}}^{\mathrm{tot}}_{r,p} - \frac{\delta g_{n}}{g_{p}} T_{b,r,p} \, {\tilde{S}}^{\mathrm{tot}}_{p} + O_{b,r,n} - T_{b,r,p} O_{p,n} \nonumber \\
& & + \sum_{h=1}^{n_{\mathrm{harm}}} \left(\gamma_{b,h,n}V_{b,r,p,h} - T_{b,r,p} \gamma_{h,n}V_{p,h}\right) \nonumber \\
& & + \sum_{h=1}^{n_{\mathrm{comp}}}\left(\delta L_{b,f,n} C_{p,f} -T_{b,r,p} L_{f,n} C_{p,f}\right) \,,
\end{eqnarray}
with ${\tilde{S}}^{\mathrm{tot}}_{r,p}$ being a model for $(A_{p} \cdot S_{p}+D^{\mathrm{tot}}_{p})$. The response value converges if 
\begin{equation}
\eta =1-\frac{{\tilde{S}}^{\mathrm{tot}}_{r,p}}{A_{p}\cdot S_{p}+D^{\mathrm{tot}}_{p}} < 1 .
\end{equation}
In \srollone, ${\tilde{S}}^{\mathrm{}}_{r,p}$ was taken as the total dipole only. Although $\eta$ was indeed small for 100\,GHz, this was not true at 353\,GHz where the Galactic signal was not taken into account, leading to a slow convergence. In \srolltwo, the ${\tilde{S}}^{\mathrm{tot}}_{r,p}$ is replaced by the ring signal extracted from $1\deg$ smoothed HFI2018 frequency maps. The consistency of this signal model with the real sky brightness is much better than 1\% at all frequencies, thus $\eta \ll 0.01$, and the convergence speed is spectacularly improved as illustrated in Fig.~\ref{fig:splinedistri}.
%

After each \stepone, a solution for the ADCNL model is obtained by a \steptwo\ algorithm. The calibrated response is thus slightly modified by the change of the ADCNL, and, at the subsequent iteration, the response converges toward another value inside the \stepone.

At each \stepone, we minimize the following $\chi^{2}$:
\begin{eqnarray}
\label{eq:finalchi2}
\chi^{2} = \sum_{b,r,p} { \frac{\left(R_{b,r,p,n} - \tilde{R}_{b,r,p,n} \right)^2}{\left(g_{b,n}{\sigma_{b}}\right)^2 + T_{b,r,p}{\left(g_{p,n}\sigma_{p}\right)}^2}}.
\end{eqnarray}
In Eq.~\ref{eq:finalchi2}, we introduce the $g_{b,n}$ terms in the weight at the denominator to compensate the $g_{b,n}~N_{b,r,p}$ term present in the numerator. Without this response term in the weights, the $\chi^{2}$ minimization would be biased towards a smaller value of $g_{b,n}$.
 
The  $\chi^{2}$ minimization through a positive and symmetric matrix inversion is performed with a parallel conjugate gradient. This provides an estimate of $\tilde{\delta g_{b, n}}$, and also $\tilde{O}_{b,r}$, $\tilde{\gamma}_{b,h}$, and $\tilde{\delta L}_{b,f}$. 

Subsequently, the cleaned and calibrated signal  $\mathcal{M}_{b,r,p}$ is built using
\begin{eqnarray}
\label{eq:projmap2}
\mathcal{M}_{b,r,p} & = & g_{b,n}~M'_{b,r,p} - \frac{\tilde{\delta g_{b, n}}}{g_{b,n}} \tilde{T}_{b,r,p}\, {\tilde{S}}^{\mathrm{tot}}_{p} - \tilde{O}_{b,r,n} \nonumber \\
& & - \sum_{h=1}^{n_{\mathrm{harm}}}\tilde{\gamma}_{b,h,n} V_{b,r,p,h} - \sum_{h=1}^{n_{\mathrm{comp}}}\tilde{\delta L}_{b,f,n} C_{p,f} \,,
\end{eqnarray}
and a new map $\tilde{S}_{p}$ can be computed from the clean and calibrated timeline:
\begin{eqnarray}
\label{eq:projmap}
\tilde{S}_{p} & = & \left( \left( A_{p}^\top A_{p}\right)^{-1}A_{p}^\top \right) \mathcal{M}_{p} \,.
\end{eqnarray}

\subsection{\steptwo: correcting for the ADCNL systematic effect}
\label{sec:ADCNLGEN}

The ADCNL systematic effect is due to the imperfect knowledge of the analog-to-digital signal conversion. The essential \srolltwo\ improvement is to correct for this effect by fitting a model of the ADCNL in the data as done for the other instrumental systematic effects. This subsection describes the model, and how we extract its parameters from the data.

\subsubsection{ADCNL model}
\label{sec:ADCSDESCR}

The bolometer analog signal is sampled at 7.2\,kHz, and modulated at 90\,Hz. The analog-to-digital conversion is performed at 7.2\,kHz, and a 180\,Hz digital signal is computed as the sum over 40 samples (called ``fast samples") for each half modulation \citep{lamarre2010}. One TOI sample is described as: 
\begin{eqnarray}
\label{eq:adcnlphys}
g_{b,n}~{M'}_{b,r,p} & = & \sum^{40}_{m=1}{\left({S}_{b,r,p}+{E}_{m} + {N}_{m} + \beth\left({S}_{b,r,p}+{E}_{m} + {N}_{m}\right)\right)} \nonumber \\
 & = & 40\,{S}_{b,r,p} + \sum^{40}_{m=1}{\left({E}_{m} + {N}_{m}\right)} + \Delta_{b,r,p} \,,
\end{eqnarray}
where:
\begin{itemize}
\item $m$ is the fast sample index within one half-modulation;
\item ${S}_{b,r,p}$ is the signal and photon noise;
\item ${E}_{m}$ is the positive/negative modulation electronic signal at 90\,Hz, initiated as a square but distorted by the stray entrance capacitance after demodulation;
\item ${N}_{m}$ is the electronic noise per sample, assumed to be a white noise with $\sigma \approx 4$ digital units;
\item $\beth\left({S}_{b,r,p}+{E}_{m} + {N}_{m}\right)$ is the bias introduced by the ADCNL for a given level of ADC input electronic signal (${S}_{b,r,p}+{E}_{m} + {N}_{m}$) at the fast sampling rate;
\item $\Delta_{b,r,p}=\sum^{40}_{m=1}{ \beth\left({S}_{b,r,p}+{E}_{m} + {N}_{m}\right)}$ is the ADCNL bias at the TOI sampling frequency.
\end{itemize}
In order to remove the ADCNL bias from the measured value ${M'}_{b,r,p}$, we have to compute $\Delta_{b,r,p}$. To avoid biasing the ADCNL correction by the noise, we do not compute the $\Delta_{b,r,p}$ function for each TOI sample, but  average over a large enough number of samples (typically 50 samples) to make the noise negligible. 

For a period $t$, while the transients of the modulation ${E}_{m}$ are not changing significantly, the average bias is:
\begin{eqnarray}
\label{eq:adcnlphys2}
<\Delta_{b,r,p,t}> & \approx & \left( \tilde{E} \otimes N_{m} \otimes \beth\right) \left({S}_{b,r,p}+<E_{m}>\right) \,,
\end{eqnarray}
where
\begin{itemize}
\item $\tilde{E}$ is the distribution of the $E_{m}$ function over the 40 samples around the mean of the $E_{m}$;
\item $\otimes$ is a convolution;
\item $N_{m}$ is the Gaussian distribution of the electronic noise;
\item (${S}_{b,r,p}+<E_{m}>$) at each sample $p$ for the ring $r$ is computed using the mean of the known digital 180-Hz value sample at the pixel $p$.
\end{itemize}
In Eq.~\ref{eq:adcnlphys2}, the convolution of the signal by the noise and the electronic modulation leads to $<\Delta_{b,r,p,t} >$ being smoothed over approximately several hundred of ADUs.

As demonstrated by simulations, the ADCNL correction for the demodulated signal can be fitted on a limited number of spline bases in ADU level (called 1D spline). When all other systematic effects are removed, the residuals between the sky signal and the data model are attributed to the ADCNL residuals averaged over all observations with this bolometer for which the demodulated signal is in the same range of ADUs.  \CHANGE{The modulation signal varies slowly and is stable on one ring}.

\begin{figure}[ht!]
\includegraphics[width=\columnwidth]{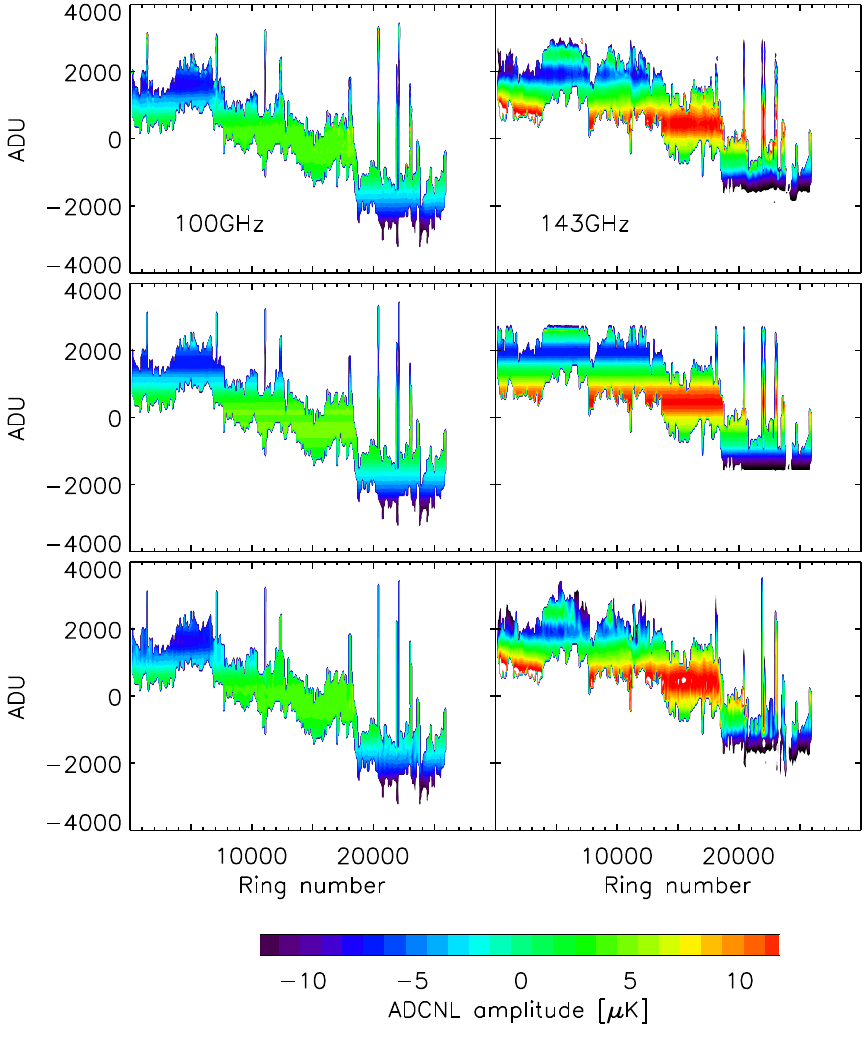}
\caption{Simulated ADCNL residuals (color coded) for the ADU level on the ADC (ordinate axis) along the time, expressed as ring number. Two bolometers are shown: left column for the 100-4b and right column for the 143-1a. First row: input of the simulation, second and third rows: their fitted 1D or 2D spline corrections without noise. Typical calibration is $\simeq 6\,\mu$K/ADU.}
\label{fig:adcnlvsadu} 
\end{figure}

When needed, an additional time-dependant description (called 2D spline) with a limited number of stable correction ranges of rings is introduced. Figure~\ref{fig:adcnlvsadu} illustrates that, for the bolometer 100-4b, the 1D spline catches all patterns of the ADCNL residuals. On the contrary, the 2D spline catches small time-dependant effects for a few bolometers, like the bolometer 143-1a, especially at the beginning of the mission. This is due to the variability of the modulation signal.

We note that the function ($\tilde{E} \otimes N_{m} \otimes \beth$) has a linear component that is degenerate with the linear part of the response. This is the reason for not computing this function in \stepone. 

The electronic modulation changes during the mission. In particular, the 4-K lines  \CHANGE{vary along the mission}. Electrical coupling between the \HeJT\ cooler drive and the bolometer readout appears as very narrow lines (called ``4-K lines'') in the power spectrum of the TOI at harmonics of 10\,Hz. These are removed from the TOI \citep[see, e.g.,][]{planck2014-a08}. Those lines modify the electronic modulation, and thus slightly modify the effective ADCNL residuals  \CHANGE{for a given ring}. A given pixel is observed by one bolometer several times within a ring, but with different 4-K line phases when the spacecraft spin frequency is not synchronized with the 4-K line frequency. In order to avoid the effect that would be induced when they are synchronized, the rings for which a 4-K line is a harmonic of the spacecraft spin frequency were not used to build the HFI maps to avoid corruption \citep{planck2013-p03}. Thus, the mean ADCNL residual for a given ring is only affected by the mean of the 4-K line variation which, not being synchronous with the sky, tends to cancel out. Therefore, the 4-K line variation is a marginal effect at the ring level and is described as a long time variation between pointing periods.

Finally, the 4-K lines only induce a small variance increase: for the most intense lines and the most sensitive to 4-K lines, bolometer 143-3a displays twice the variance of the electronic noise, otherwise around 25\,\% of the electronic noise. For the 143-3a, the amplitude is multiplied by four during the mission, leading to a smoother ADC residual response at the end of the mission. Therefore, the only issue is to sample well enough the ADCNL function of ADU and time to catch all  \CHANGE{the nonlinearity variability}. In the following section, we show a test on simulation of one of the worst bolometers to determine how many spline parameters are needed to describe the $\beth$ function.

\subsubsection{Fitting the ADCNL model parameters}
\label{sec:ADCEQUATION}

An estimated residual signal model $\delta\mathcal{M}_{b,r,p}$ is computed using the parameters and the cleaned and calibrated signal from the previous \stepone:
\begin{eqnarray}
\label{eq:modeldef}
\delta\mathcal{M}_{b,r,p} &=& \mathcal{M}_{b,r,p} - A_{b,r,p}\tilde{S}_{p} - \delta g_{b, n} \tilde{T}_{b,r,p}\, {\tilde{S}}^{\mathrm{tot}}_{p} \nonumber \\
 & & - \sum_{h=1}^{n_{\mathrm{harm}}}\gamma_{b,h,n} V_{b,r,p,h} -
\sum_{f=1}^{n_{\mathrm{comp}}}L_{b,f,n} C_{b,r,p,f}.
\end{eqnarray}
We identify this residual as being due to the ADCNL correction which is the systematic effect not treated in the first step due to its degeneracy with the gain, and we define:
\begin{eqnarray}
\label{eq:modeldefbis}
\delta\mathcal{M}_{b,r,p}  &=& {\mathcal{A}}_{b,r,p}\,{a}_{b} + \aleph_{b,r}\left({\mathcal{A}}_{b,r,p}\right) + A_{b,r,p}\, \delta{S_{p}}\nonumber\\
 & &+ O_{b,r} + N_{b,r,p} ,
\end{eqnarray}
where:
\begin{itemize}
\item $\mathcal{A}_{b,r,p}$ is the value in ADU of the signal $M_{b,r,p}$.
\item $ {a}_{b}$ is the bolometer linear response of the ADU that is an adjustable addition to the response. The computed detector response in the first \stepone\ iteration is affected by the full ADCNL residual which is not corrected yet. Even in the further iterations, the gain is still converging to its final value. To allow the \steptwo\ to retrieve the ADCNL function not affected by small detector gain residual, it is necessary to include a linear part of the ADCNL which should converge to zero. To avoid any divergence of the algorithm, this $a_{b}$ is computed again at each iteration.
\item $\aleph_{b,r}\left(\mathcal{A}_{b,r,p}\right)$ is the model of the ADCNL for the bolometer $b$ at the sample ring $r$ for the corresponding $\mathcal{A}_{b,r,p}$.
\item $ \delta{S_{p}}$ is the residual signal that is due to the imperfect estimation of all parameters because the ADCNL correction, which is taken to be null in the first iteration, has not yet reached the best estimation at this stage of the algorithm.
\end{itemize}
Because of the noise and electronic modulation smoothing, the effective ADCNL mismatch can be expressed as a smooth function $\beth$ of ADU for each bolometer,  \CHANGE{and is not a pure ADCNL correction}(see Sect.~\ref{sec:ADCSDESCR}).  Thus the iteration, in  \CHANGE{the} \steptwo, on the weights of the spline description of $\aleph_{b,r}$ defined in Eq.~\ref{eq:adcdef}, converges when $\delta{S_{p}}$ becomes small enough. The overall gain iteration converges when the signal variations between two iterations become sufficiently small. Equation~\ref{eq:modeldefbis} shows that $\aleph_{b,r}$ is fully determined when $\delta{S_{p}}=0$.

We describe $\aleph_{b,r}$ as a spline function:
\begin{eqnarray}
\label{eq:adcdef}
\aleph_{b,r}\left(\mathcal{A}_{b,r,p}\right) &=& \sum_{t=1}^{n_{\mathrm{time}}}\sum_{s=1}^{n_{\mathrm{spline}}} w_{t,b,s,n} \, \alpha_{b,s}(\mathcal{A}_{b,r,p})\, \beta_{b,t} \,,
\end{eqnarray}
where
\begin{itemize}
\item $t$ is the index of the spline weight varying with time;
\item $s$ is the spline index of the bolometer $b$;
\item $w_{t,b,s,n}$ is the spline weight at the itteration $n$;
\item $\alpha_{b,s}(\mathcal{A}_{b,r,p})$ are the spline functions of ADU level for the bolometer $b$ to correct the whole signal  \CHANGE{around} level $M_{b,r,p}$, \CHANGE{therefore around the corresponding} ADU level;
\item $\beta_{b,t}$ are the spline functions of time.
\end{itemize}
The iteration, in \steptwo, on the weights of the spline description of $\aleph_{b,r}$ converges when $\delta{S_{p}}$ becomes small enough. The overall gain iteration converges when the signal variations between two iterations become sufficiently small.

For a bolometer, the model for ADCNL residual can be represented by different numbers of spline functions for different frequencies. This is due to the difference in digital range for the signals: for 100, 217, and 353\,GHz, the ADCNL is described as 32 third-degree splines in ADU, where at 143\,GHz time variation is needed, and ADCNL are described as 8 third degree splines in ADU and 128 splines in time.

From Eq.~\ref{eq:modeldef}, \ref{eq:modeldefbis}, and \ref{eq:adcdef}, and using again the redundancies in the data, we minimize the following $\chi^{2}$ including a implicit binning in ADU due to the smoothness of the splines:
\begin{eqnarray}
\chi^{2} = \sum_{b,r,p} \frac{1}{\sigma_{b}^2} \left(\delta\mathcal{M}_{b,r,p} - O_{b,r,n}- \sum_{t, s} w_{t,b,s,n} \, \alpha_{b,s}(\mathcal{A}_{b,r,p}) \beta_{b,t} \right)^{2} ,
\label{eq:chi2adc}
\end{eqnarray}
and we obtain a new estimation of $\tilde{O_{b,i,n}}$, and an estimation of the totality of $w_{t,b,s,n}$. $\tilde{O_{b,i}}$ changes because the ADCNL correction based on the ADU level that follows the very slow focal plane thermal evolution is degenerate with the very low frequency noise. After this \steptwo, a new iteration, with a new gain and a new ADCNL model, can start.

\begin{figure}[ht!]
\includegraphics[width=\columnwidth]{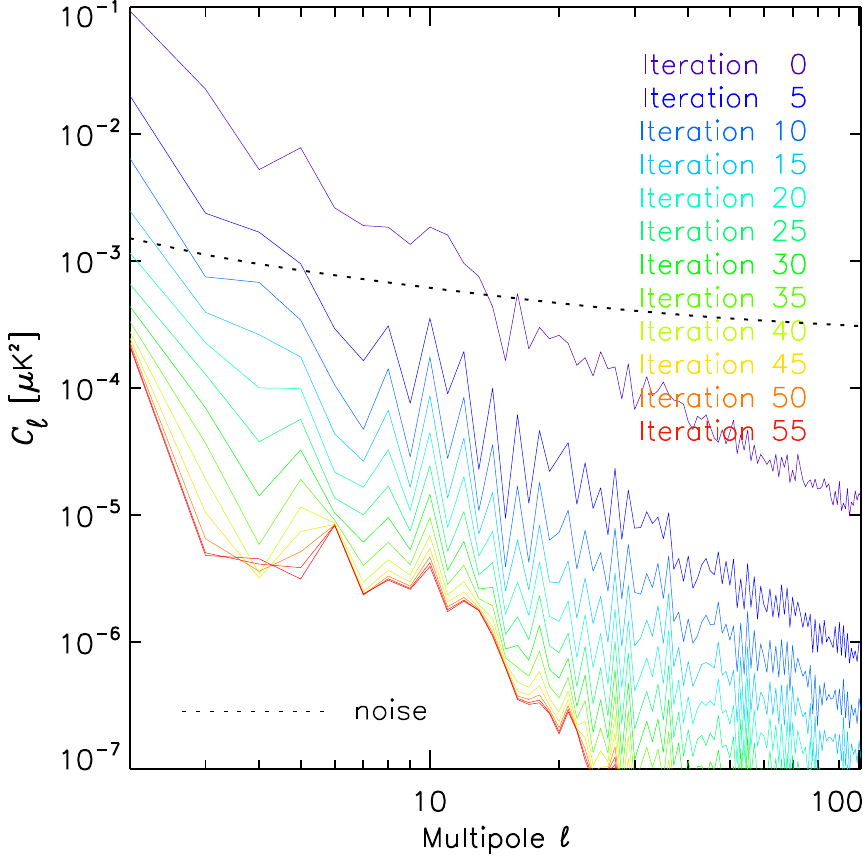}
\caption{$EE$ power spectra for difference between two successive iterations of residual maps from a noiseless simulation including only the ADCNL effect and \srolltwo\ mapmaking for increasing number of iterations at 143\,GHz. 20 iterations are enough to reach a good enough convergence.}
\label{fig:adcconverge} 
\end{figure}

Figure~\ref{fig:adcconverge} shows the residual between the simulated HPR data and the model given the number of iterations for a simulation without noise.
It demonstrates that ADCNL effect can be cleaned very well, as the convergence, measured by the residual spectrum at $ \ell =3,$ falls below the noise for 10 gain iteration. We therefore limit the number of iterations to 20. Increasing the number of iterations does not improve the results. These ADCNL residuals describe the level of uncertainty reached if one ignores all other systematics at low multipoles which are shown to be smaller in Sect.~\ref{sec:allsyste}.

\subsection{Map projection}

At the last iteration, only the \stepone\ is performed, and the final map is given by Eq.~\ref{eq:projmap}. The resulting \srolltwo\ frequency maps are presented Fig.~\ref{fig:maps_all} and discussed in the following section.

\subsection{Future improvements}
\label{sec:otherimprov}

Using  a nonlinear correction of the ADCNL as a substitute for the gain variation in time, the improvements of the polarization leakage and the transfer function are exhibited in Sect.~\ref{sec:tempimprove} where we show that these systematic effects are under control.

It is also important to note that CO template accuracy is critical as CO has a strong impact on 100-GHz map computation, and a lesser impact at 217 and 353\,GHz. At all these frequencies, CO maps can be built from the different single detector response to CO lines, either from ground measurements or using large-enough CO  \CHANGE{maps} from radioastronomy observations as was done in \citedpc. Such CO templates  \CHANGE{for the two main  isotopologues} used in this analysis are described in \citedpc\ characterization, but are not used in the corresponding HFI2018 maps. The changes that could be brought to CO templates are discussed in Sect.~\ref{sec:tempstudy}. At this level of sensitivity, no other molecular line is significant.

Further improvements will be made to reduce the large-scale detector noise due to the undetected cosmic ray glitch tails that now dominate at low multipoles for some frequencies.

\subsection{Simulations}
\label{sec:s2sims}

\srolltwo\ simulations, hereafter \sims, built for the characterization of \srolltwo\ data, follow the same schematic as the one described in \citedpc\ for the FFP10 simulations. The differences are improved inputs: \NCHANGE{firstly updated thermal dust template as the ``fiducial'' FFP10 one (unlike the 300 FFP10 iterations ones); secondly solar dipole as $(d,l,b) = (3362.71\,\mu{\rm K}, \text{264\pdeg21} , \text{48\pdeg253})$; thirdly inclusion of a kinetic Doppler quadrupole whose amplitudes, in $\mu$K, are 0.84 at 100\,GHz, 1.00 at 143\,GHz, 1.35 at 217\,GHz, and 2.1 at 353\,GHz.}

All the other simulation pipeline parameters (scanning strategy, optical beams, spectral transmissions, polarization efficiencies, conversion factors, thermal baseline, noise realizations and auto-correlation, ADC nonlinearities, etc.) are unchanged. Besides the inputs, the most significant improvement with respect to the FFP10 simulations is the use of the \srolltwo\ algorithm at the mapmaking step.

As the sky simulations do not perfectly mimic data, a final post processing is needed to add a small correlated white noise between detectors. Having no correlated noise between bolometers in the timeline simulation allows for a much greater level of computational parallelization.

\begin{figure*}[ht!]
\includegraphics[width=\textwidth]{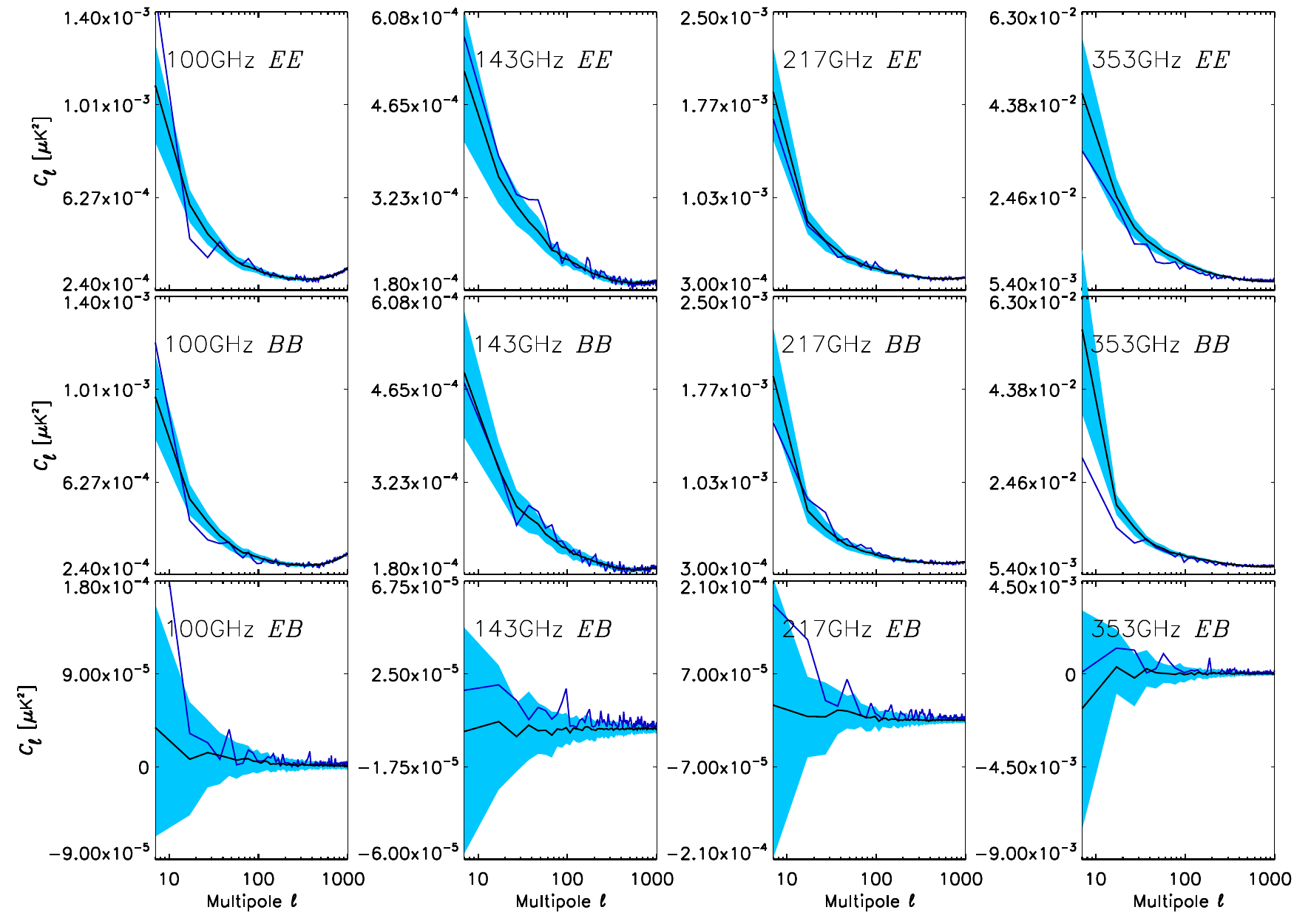}
\caption{Noise and systematic residuals in $EE$, $BB$, and $EB$ spectra, for difference maps of the half-mission null test binned by $\Delta \ell =10$. Spectra are rescaled to the full mission, and also rescaled to full sky from the unmasked sky fraction $f_{sky}=0.43$ used. Data spectra are represented by thick blue lines, and the average of simulations by thin black lines. The blue band shows one $\pm \sigma$ equivalent of the statistical distribution of the simulations. The linear $y$-axis scales are adapted to show the full data range on each panel.}
\label{fig:simspec}
\end{figure*}

Figure~\ref{fig:simspec} illustrates the quality of those simulations compared to the data for the half-mission null test.
Overall, simulations show good consistency with the data. We see a low frequency noise rise, much reduced with respect to figure 17 of \citedpc\ below  $\ell=50$, as expected. At 143\,GHz, at low multipoles, there is a small discrepancy related to the mask used to compute the spectra. This lack of low multipole noise in the simulations is mainly in the temperature spectra while the polarization noise and systematic effect residuals are well modeled. At 353\,GHz the simulation spectra show a higher level, demonstrating an overestimation of systematic effects at $\ell < 10$. The largest 353-GHz systematic effect is the dust signal distortion which cannot be modeled properly since the sky model is not representative of the real data at very large scale.

\section{Results}
\label{sec:result}

\subsection{\srolltwo\ frequency maps}

\begin{figure*}[ht!]
\includegraphics[width=\textwidth]{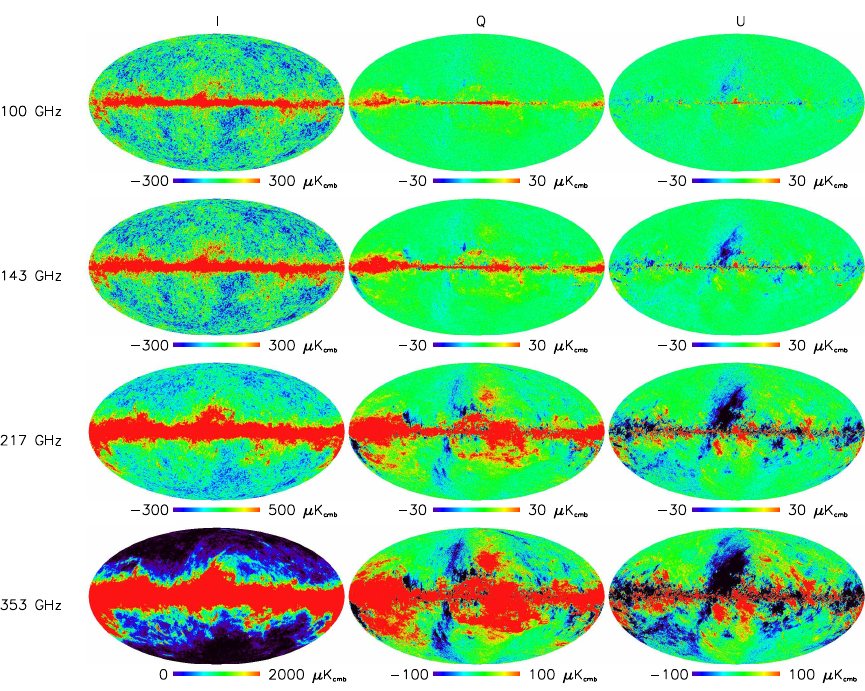}
\caption{Full sky, full mission \srolltwo\ maps at 100 to 353\,GHz (in rows), for Stokes $I$, $Q$, and $U$ (in columns).}
\label{fig:maps_all} 
\end{figure*}

Figure~\ref{fig:maps_all} presents the \srolltwo\ ``full mission'' maps at 100, 143, 217, and 353\,GHz maps, the total dipole being removed.

Along with those full mission maps, we build half data maps by splitting the data in two sets for which both give a complete set of frequency maps. We can then build null tests with differences of the two sets of maps in a pair to obtain residuals in which the uncorrelated noise and the signal distortion due to systematic effects not identical in the pair will appear, providing information about the noise and systematic effect residuals.

\NCHANGE{The half-mission (hm) null test is sensitive to the slow temporal variation of instrumental effects (e.g., the ADCNL effect). The detector-set (detset) null test separates the four PSB detectors, at 100, 143, 217, and 353\,GHz, into two detector sets. The difference of detset maps can be used as a very useful test of systematics arising from the detector chain specificities not fully accounted for (e.g., calibration or bandpass differences). The odd-even ring (oe) null test, using the difference of two pointing periods that just follow each other, one odd and the other even in the numbering of pointing periods. The difference should be the closest to maps built with the detector noise only. The odd and even six-month surveys are scanning nearly the same ring in opposite directions six months apart. This is very useful to test how well time constants (including very long ones) or asymmetrical FSL systematic effects are corrected. The caveat with this null test is that a map built with only odd (or even) surveys will not be able to properly deal with the polarized systematic effects due to the reduced redundancy in polarization angles. We thus use odd-even survey differences with the systematic effects estimated on the full mission and then build the half maps taking only the HPRs from odd (or even) surveys. }

\subsection{Comparison of HFI2018 and \srolltwo\ frequency maps}

Here we discuss the temperature maps. The noise in the \srolltwo\ 143-\, and 217-GHz temperature maps is higher than the one in the HFI2018 maps. This is expected as \srolltwo\ only uses 8 PSBs where \srollone\ was using both PSBs and SWBs, namely 11 bolometers at 143\,GHz and 12 at 217\,GHz.
 
The solar dipole directions and amplitude were measured with high accuracy in \citedpc. The bias in amplitude for 100 simulations of \srollone\ was $1.5\times10^{-4}$ for the CMB channel, thus about $1\,\sigma$ measured on the dispersion on the 100 simulations, and corrected. This marginally significant correction was applied to maintain coherence with the \Planck\ 2015 determination. In the present analysis, the solar dipole was not extracted again.
The solar dipole amplitude residuals show differences, for each frequency, between HFI2018 and \srolltwo\ maps estimated using a sky fraction of $f_{sky}$=0.43: \NCHANGE{  $-0.87\,\mu$K ($-2.4\times10^{-4}$) at 100\,GHz; $-0.62\,\mu$K ($-1.7\times10^{-4}$) at 143\,GHz; $+0.79\,\mu$K ($2.1\times10^{-4}$) at 217\,GHz.}

The negative average bias on the three CMB frequencies is $-2\times10^{-4}$, and reduces to  $-0.5\times10^{-4}$ when \srollone\ bias is removed. In summary, \srolltwo\ does not show any sign of bias with respect to the \citedpc\ best value without debiasing \srollone.

Dipole biases due to \srollone\ miscalibration residuals were estimated to be $\pm 1\,\mu$K at CMB frequencies (see \citedpc). For the HFI2018 and \srolltwo\ simulations, the input absolute calibration  was taken to be the same, and thus no calibration solar dipole residual should be present. The solar dipole amplitude differences are at a comparable level as the uncertainties may be driven by other systematic effects than the calibration errors found for the \citedpc. Systematic effects like ADCNL modify the dipole amplitude. The foreground removal, assuming a constant spectral energy distribution (SED) on the whole sky, should be performed using the gradient of SED found in \citedpc\ which could induce a dipole error $\pm 1\,\mu$K. Thus, the amplitude of the residual solar dipoles, at CMB frequencies, between the HFI2018 and \srolltwo\ maps should not be interpreted as a calibration mismatch but as being due to a sum of comparable errors in foreground templates and other weak instrumental systematic effects. This is illustrated in Fig.~\ref{fig:detpolcorr}. The higher difference at 353\,GHz by an order of magnitude ($-13\,\mu$K or $-3.6\times 10^{-3}$) than at CMB frequencies, is at least partly explained by the uncertainties on the polar efficiencies and the improvement of the time transfer function model. 

The $1\,\mu$K accuracy on the solar dipole amplitude is thus a limit of \srollone\ and \srolltwo\ algorithms. The only way to improve previous dipole estimation is to introduce a full component separation in the mapmaking to avoid bias introduced by degeneracy between systematics and foreground dipoles and taking into account the foreground SED variations at large scales.

\begin{figure}[ht!]
\includegraphics[width=0.24\textwidth]{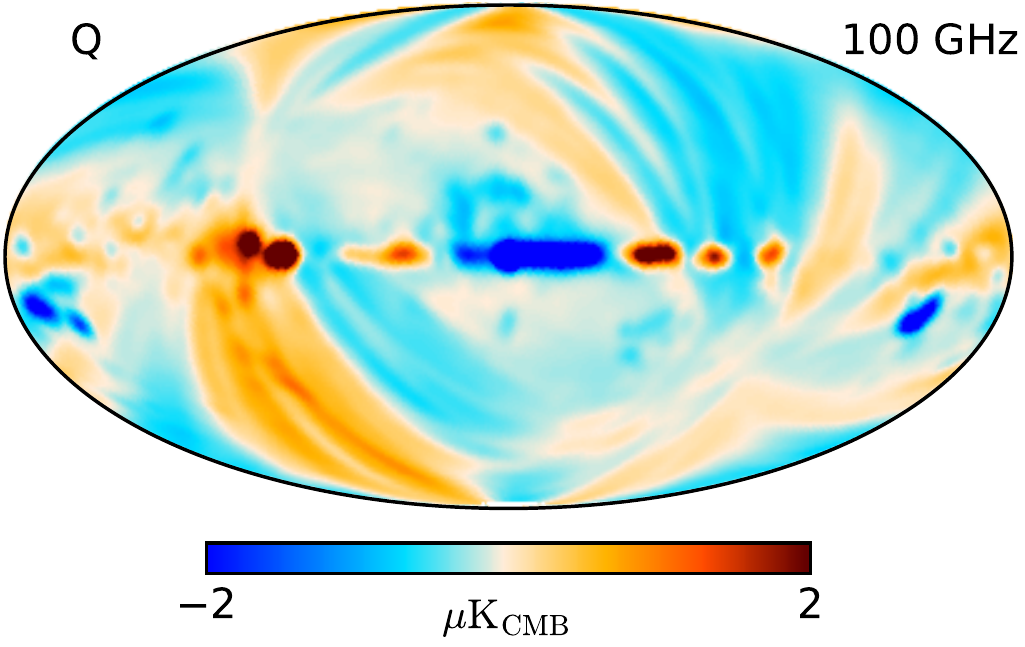}
\includegraphics[width=0.24\textwidth]{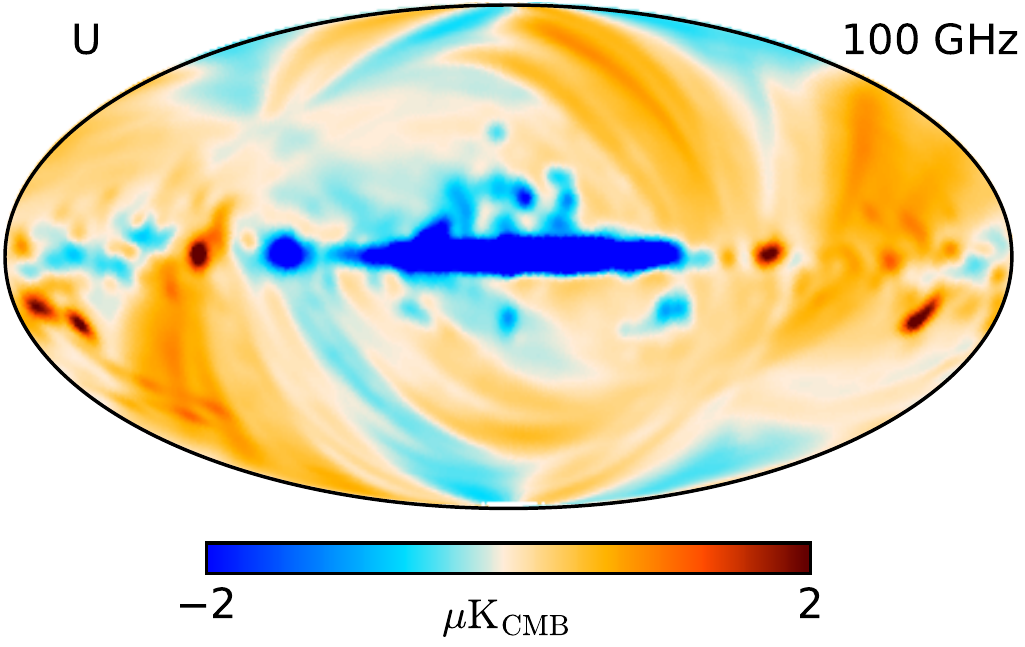}\\
\includegraphics[width=0.24\textwidth]{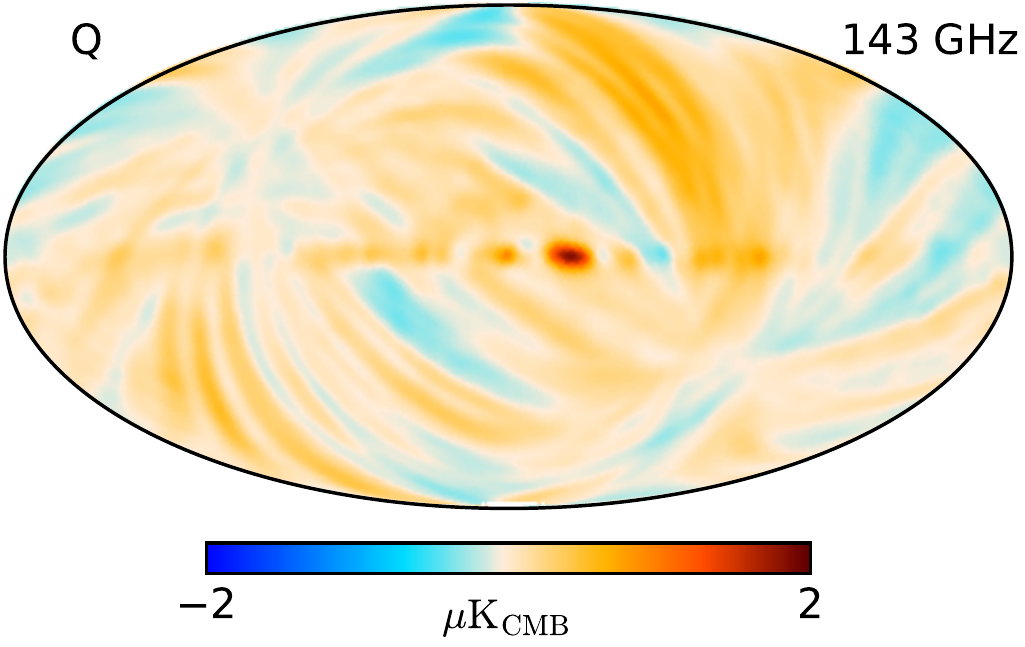}
\includegraphics[width=0.24\textwidth]{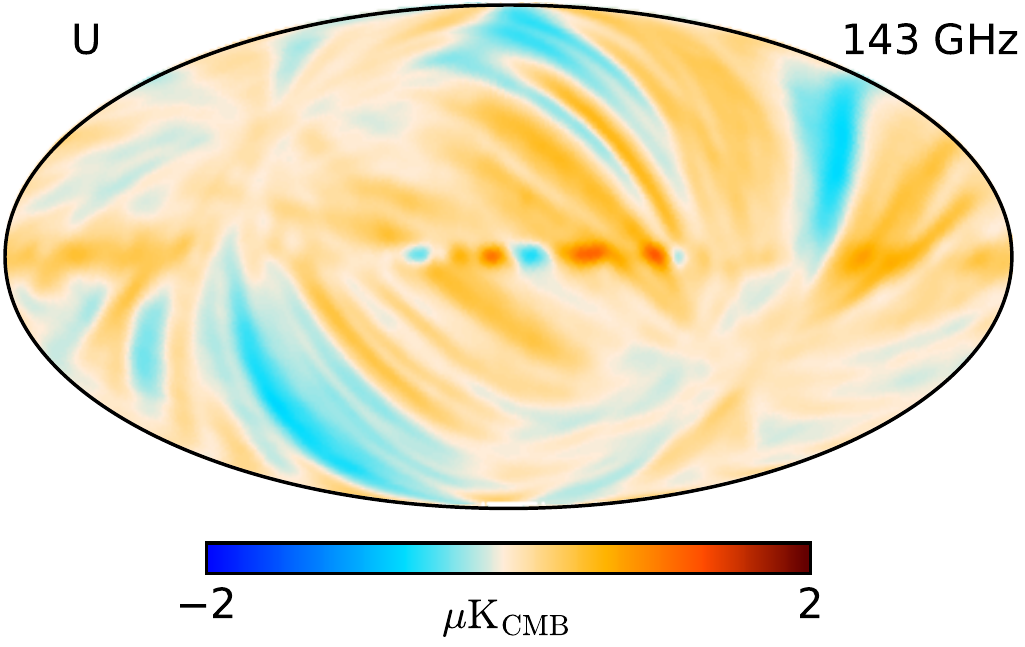}\\
\includegraphics[width=0.24\textwidth]{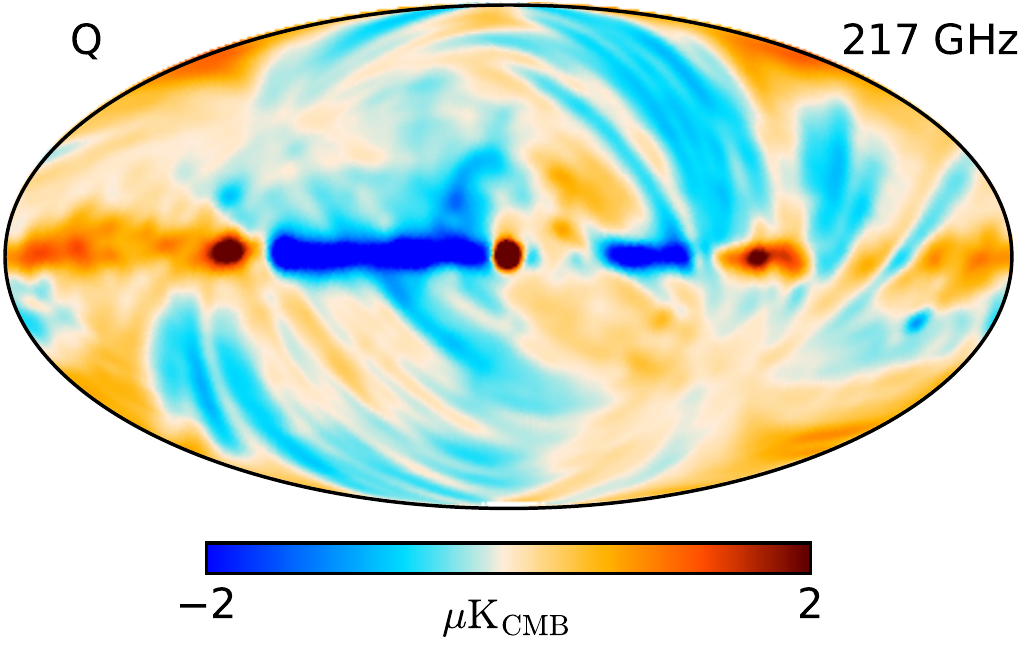}
\includegraphics[width=0.24\textwidth]{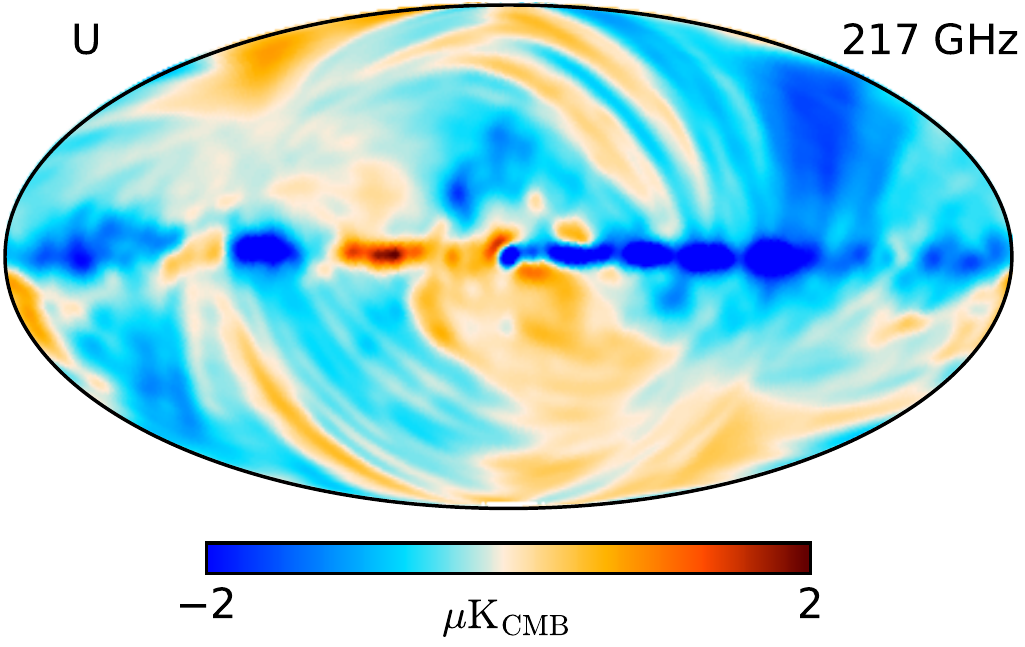}\\
\includegraphics[width=0.24\textwidth]{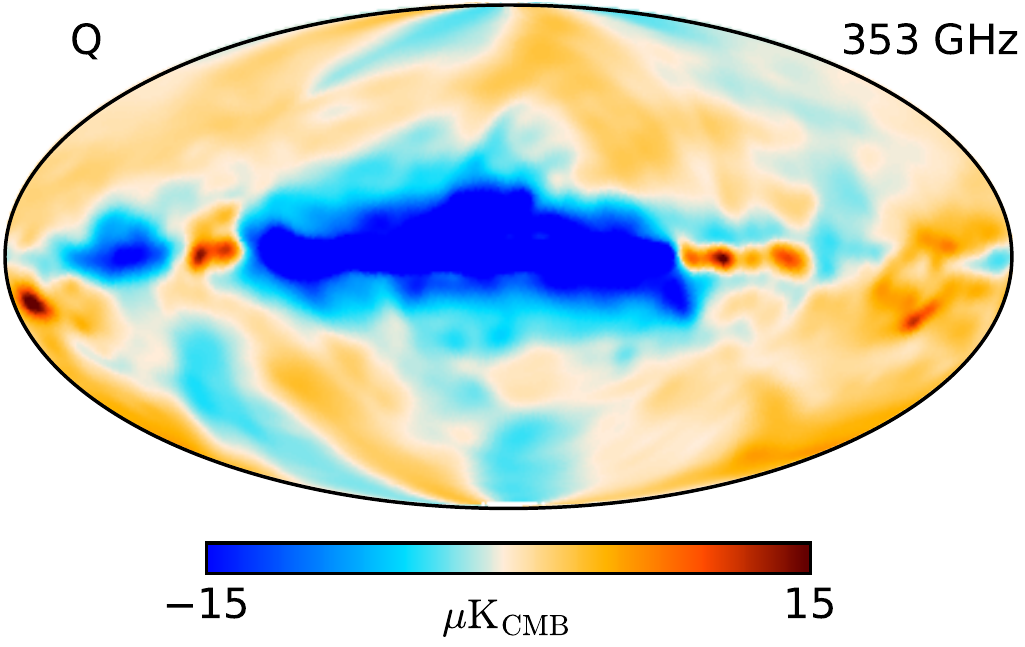}
\includegraphics[width=0.24\textwidth]{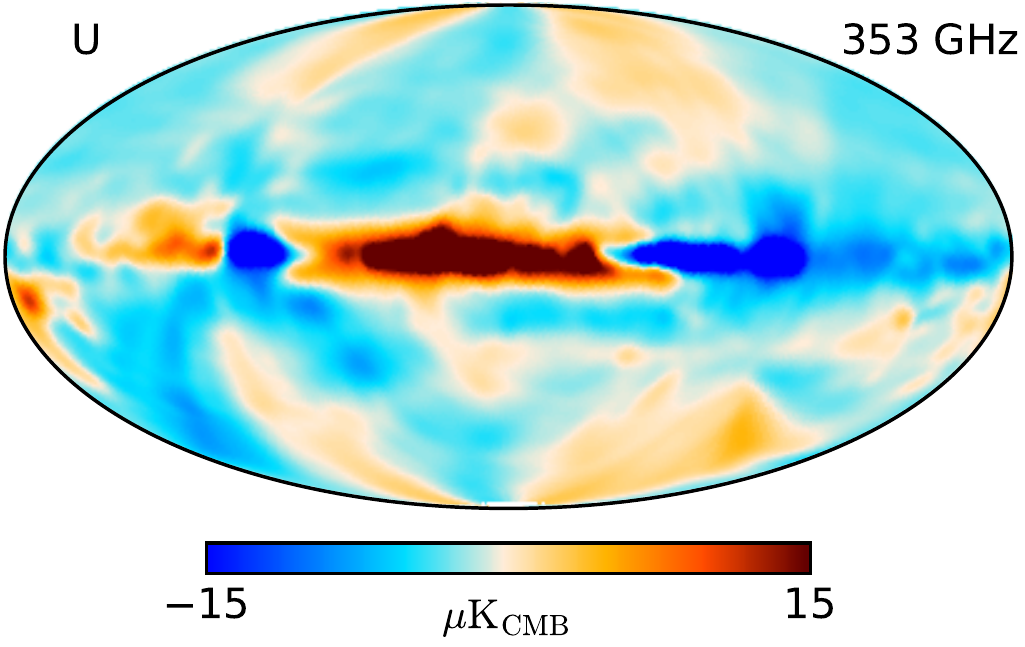}
\caption{HFI2018 minus \srolltwo\ difference maps at 100 to 353\,GHz in Stokes $Q$ (left column) and $U$ (right column), smoothed by a $5\deg$ beam.} 
\label{fig:cmpImaps} 
\end{figure}

Figure~\ref{fig:cmpImaps} shows the difference between the HFI2018 and the \srolltwo\ frequency maps in polarization.
These difference maps are expected to show mostly the patterns associated with the systematic effects better handled in the \srolltwo\ processing.

At 100\,GHz, in comparison with 143\,GHz where there is no CO line contribution, the improvement in the CO component can be seen in the Taurus and Orion molecular clouds below the Galactic disk, and in the Ophiuchus cloud above the Galactic disk in the central region. This is very likely due to the improved CO full sky map templates, including the $^{13}$CO, not included before. These are built using the bandpass mismatch coefficients extracted from a specific run using, as input templates, the CO maps of the Taurus molecular cloud from \citet{Goldsmith2008}. The use of this small but very well measured CO region to determine better bandpass mismatch is described in Sect.~3.1.3 of \citedpc.

For the dust component, residuals should appear in the same regions as the CO, if the dust was not properly corrected. These are not apparent at 217\,GHz, and appear marginally at 353\,GHz. This is consistent with the \srolltwo\ input dust template being the same as the \srollone\ one.

At 100, 143, and 217\,GHz, the striping scanning patterns (banana-like patterns) are associated with ADCNL systematic effect residuals. At 353\,GHz, those stripes are not visible as the ADCNL effect residuals are much smaller than other instrumental systematic effects due to a wider spread of the signal range on the ADC.

At all frequencies, improvements show up in polarization, thanks to \srolltwo\ specific polarized parameters adjustments.

\begin{figure}[ht!]
\includegraphics[width=0.24\textwidth]{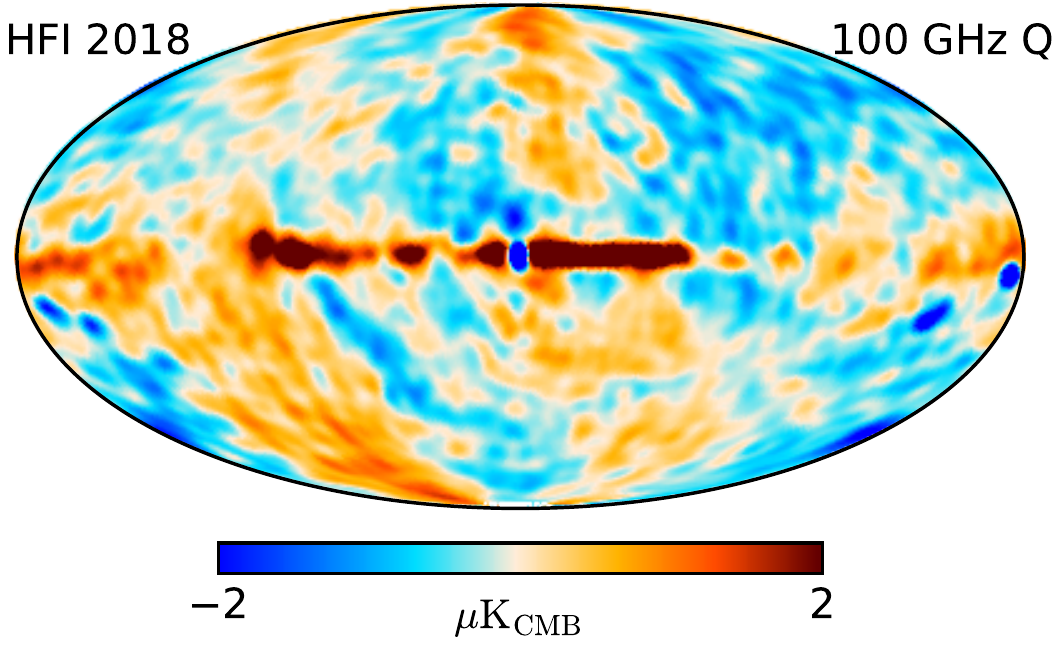}
\includegraphics[width=0.24\textwidth]{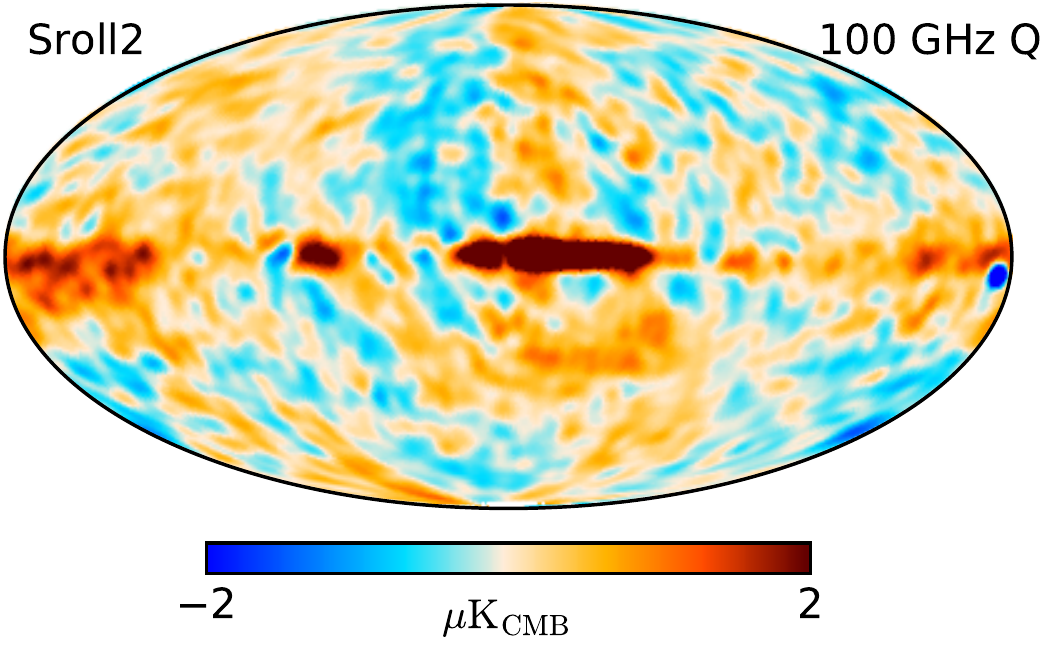}\\
\includegraphics[width=0.24\textwidth]{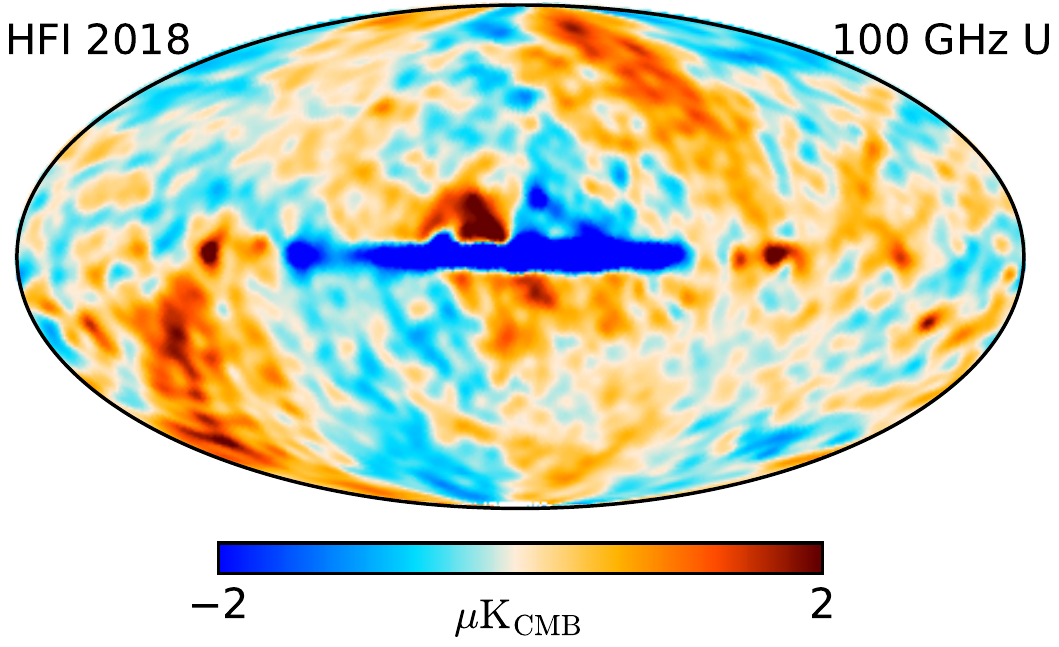}
\includegraphics[width=0.24\textwidth]{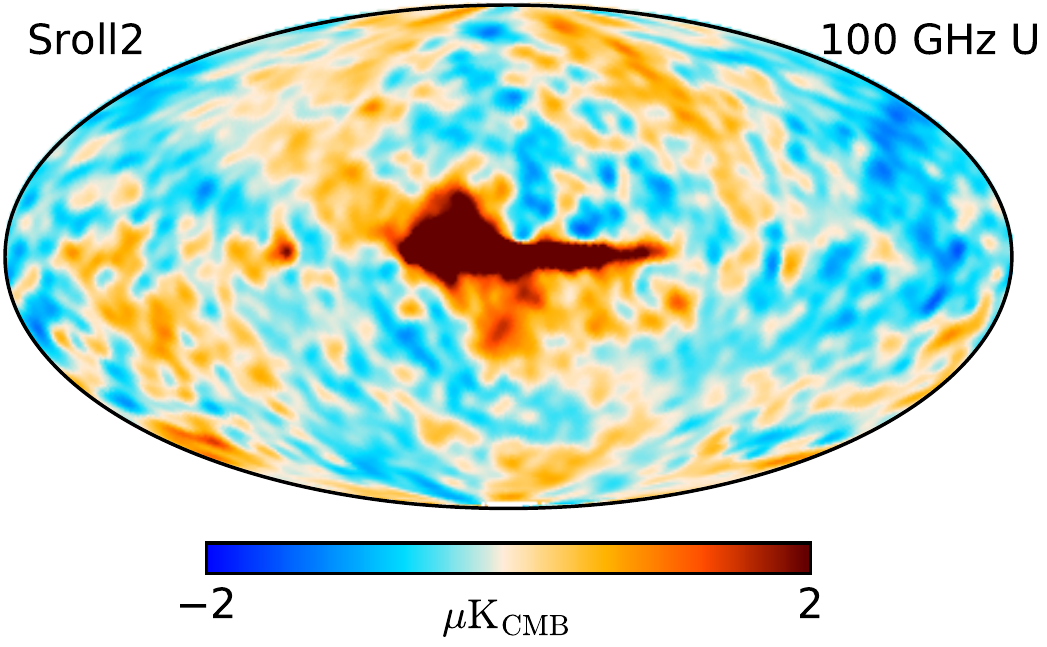}\\
\includegraphics[width=0.24\textwidth]{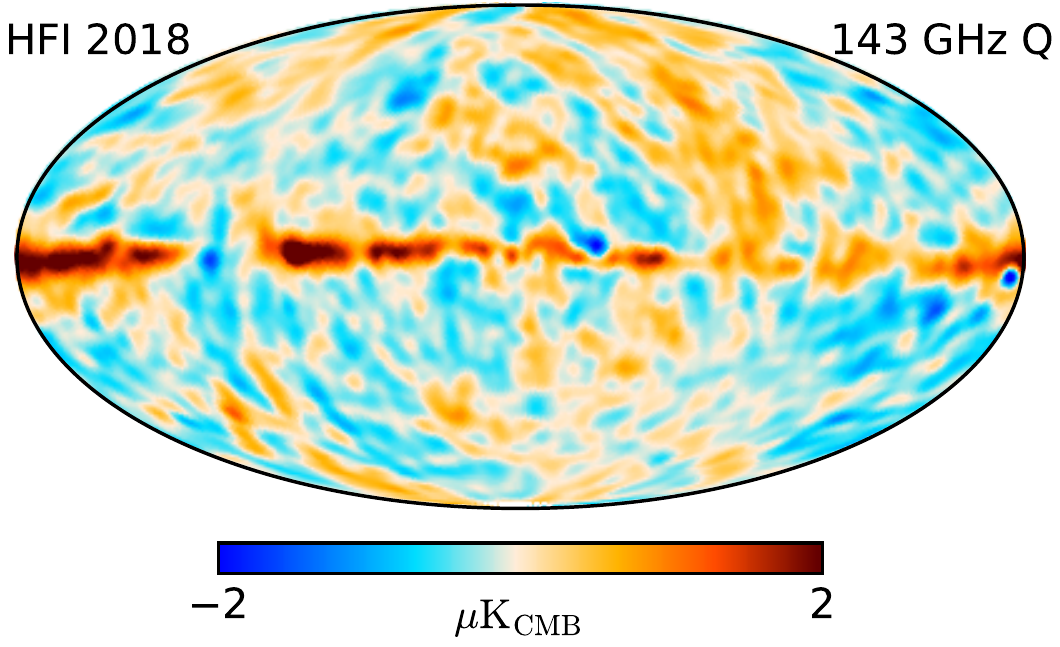}
\includegraphics[width=0.24\textwidth]{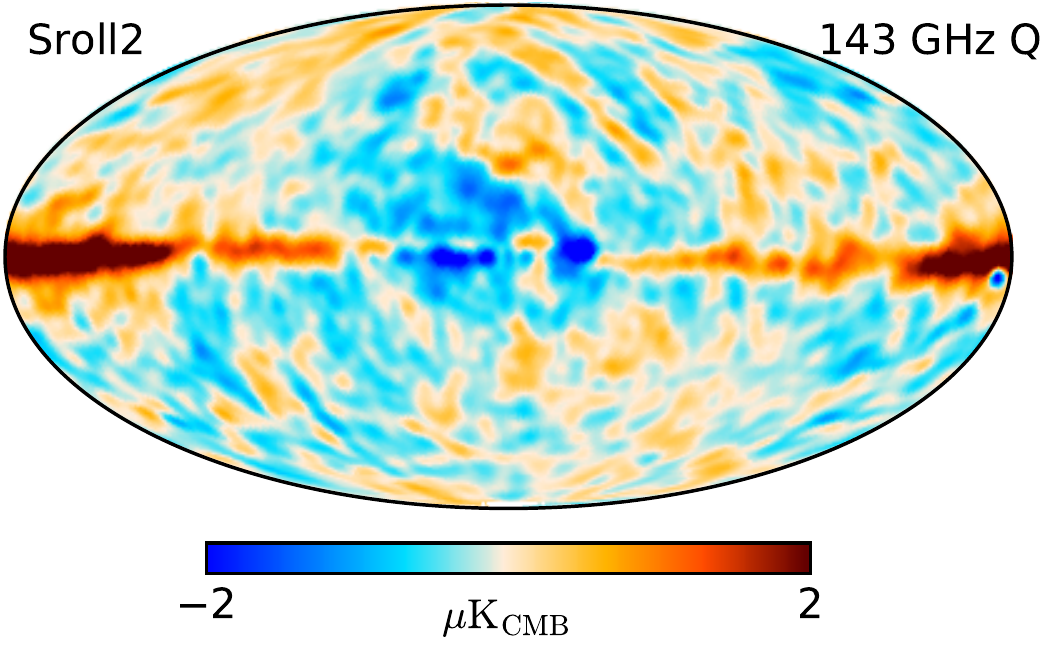}\\
\includegraphics[width=0.24\textwidth]{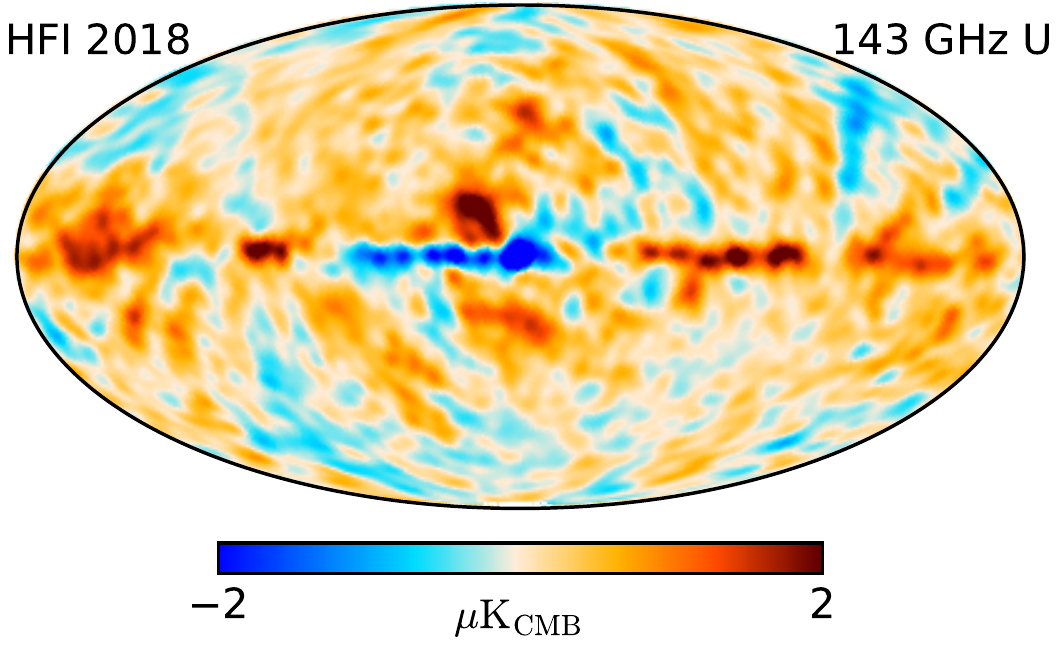}
\includegraphics[width=0.24\textwidth]{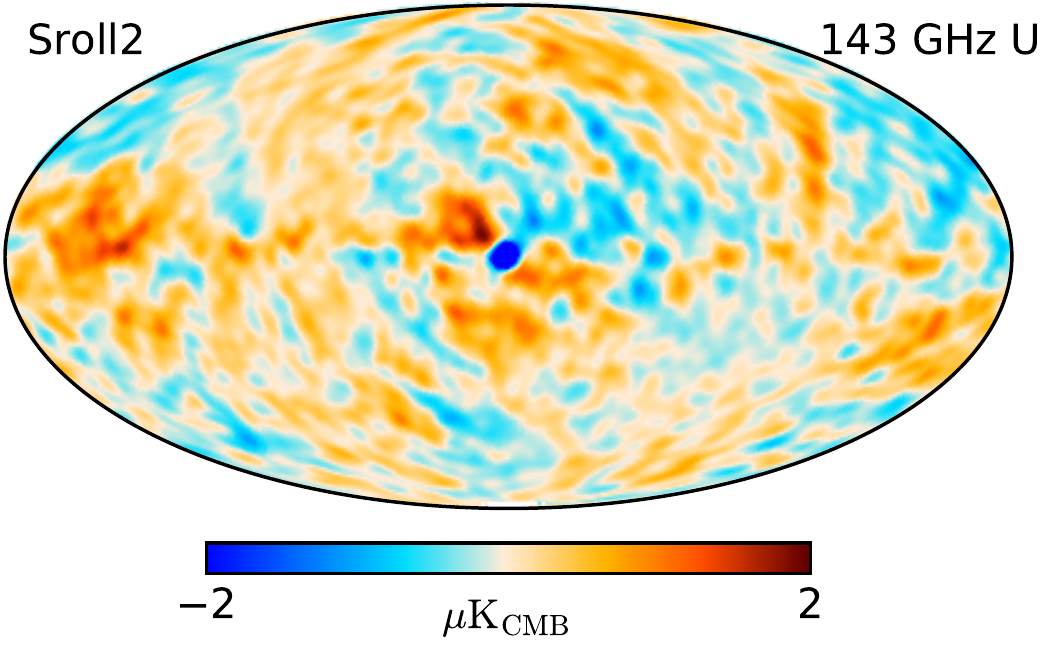}\\
\includegraphics[width=0.24\textwidth]{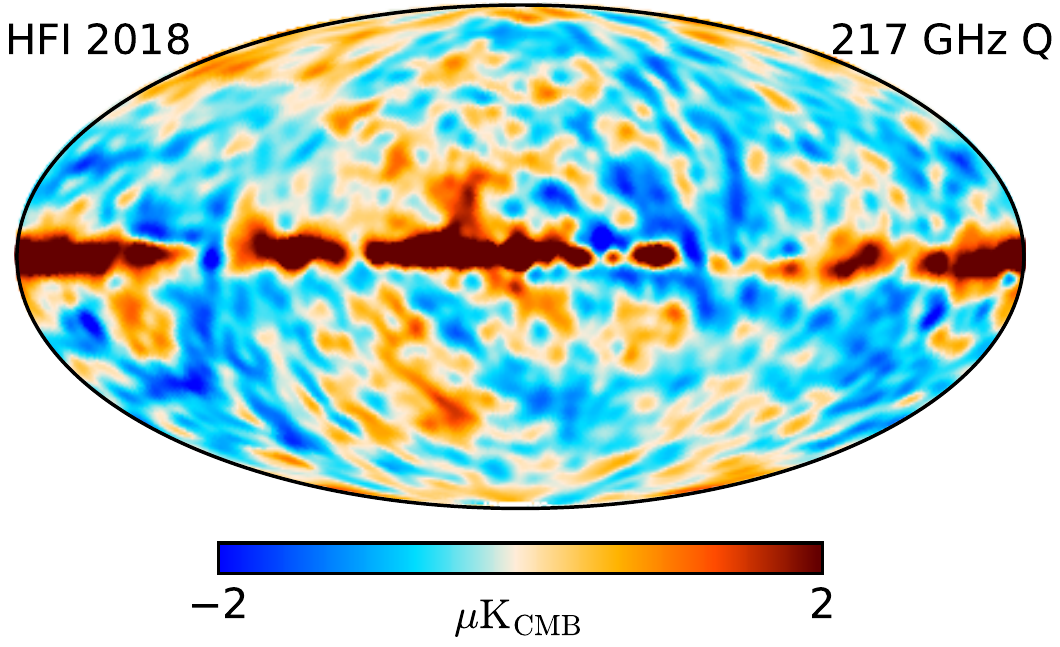}
\includegraphics[width=0.24\textwidth]{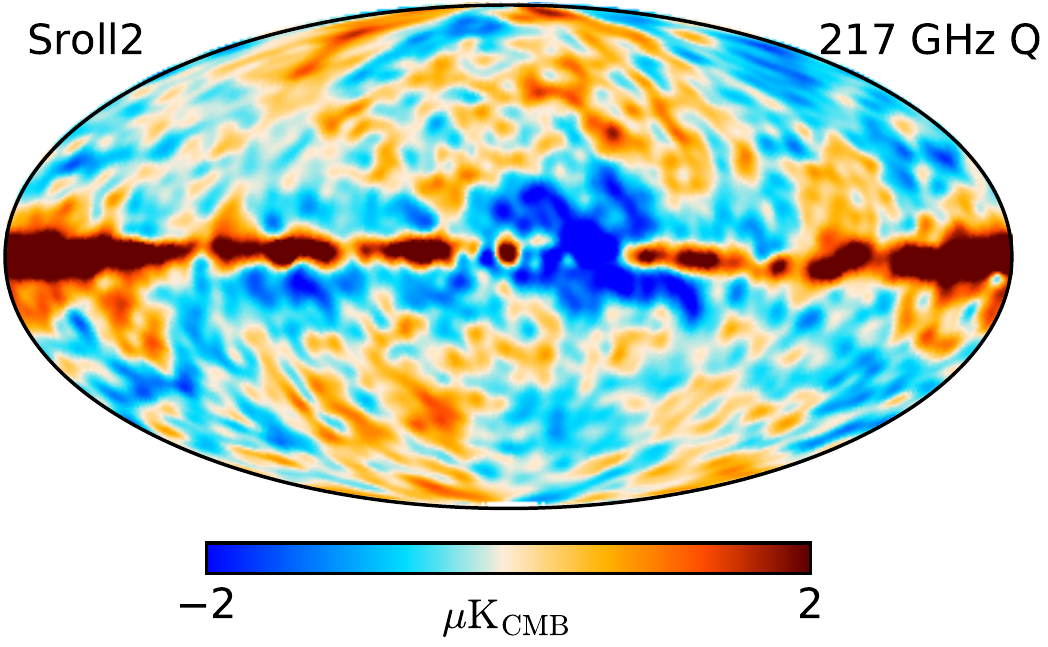}\\
\includegraphics[width=0.24\textwidth]{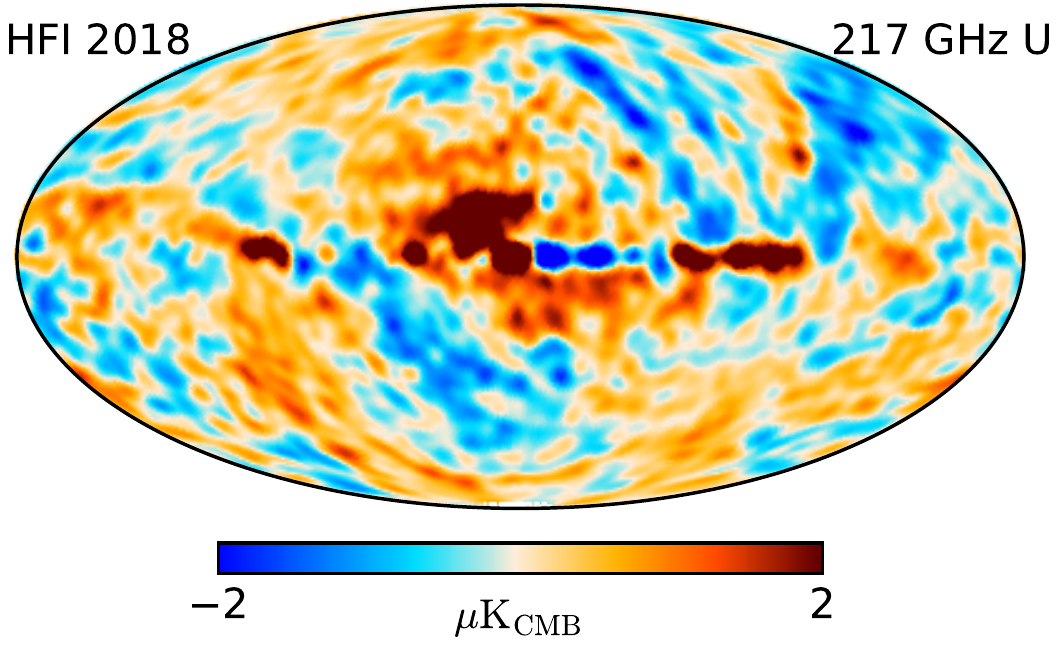}
\includegraphics[width=0.24\textwidth]{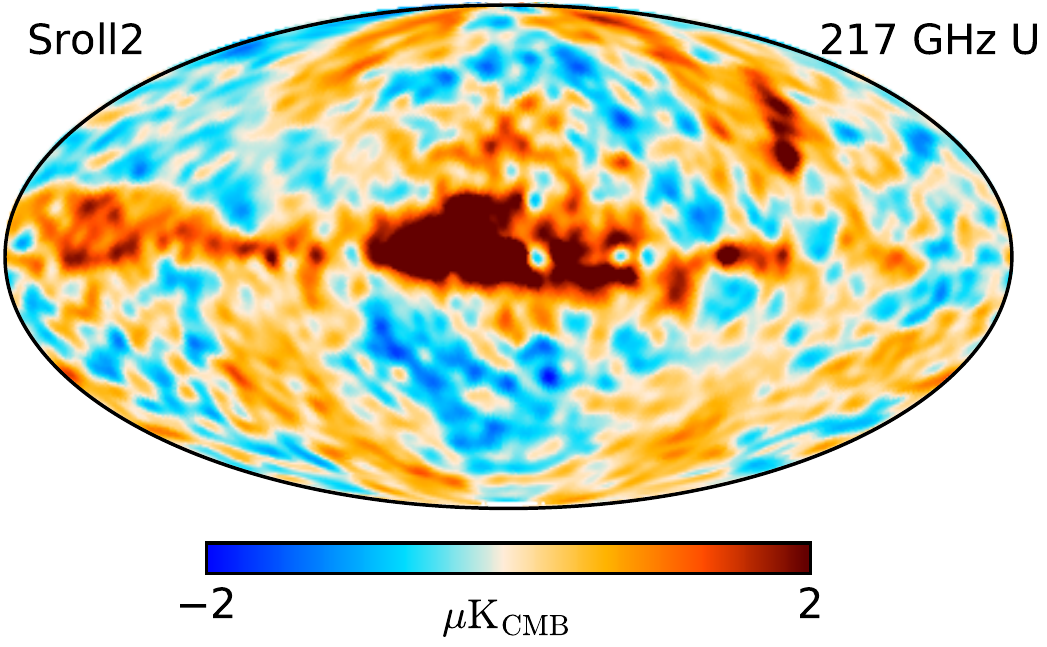}
\caption{Comparison between 100 to 217-GHz HFI2018 and \srolltwo\ $Q$ and $U$ maps, cleaned by the 353-GHz maps smoothed by a $5\deg$ beam. 
}
\label{fig:cmpQUmaps}
\end{figure}

Figure~\ref{fig:cmpQUmaps} shows the $5\deg$ smoothed polarized map\CHANGE{. For foreground cleaning at 100\,GHz, no synchrotron cleaning is performed.} After cleaning the dust signal using a single all-sky SED emission law approximation to propagate the 353-GHz polarized information to other frequencies obtained by the correlation factors between CMB frequencies and the 353-GHz after removing the 100-GHz to make it fully dominated by dust.
The maps shown are expected to be consistent with noise, signal and foreground or instrumental residuals. We note that the polarized CMB brightness is very small at large scale (less than $0.5\,\mu$K). HFI2018 maps had striping patterns along the scanning direction that become much smaller in the \srolltwo\ maps demonstrating the effectiveness of the algorithm in reducing large-scale instrumental systematic effect residuals.

Even at 217\,GHz, the  \CHANGE{differences} between the HFI2018 and \srolltwo\ maps  \CHANGE{show} less large structures for the $U$ signal. Nevertheless, the 217-GHz map dust cleaning using a single coefficient for the whole sky to be applied to the 353-GHz map cannot remove the dust well enough at large scales. In the \Planck\ Legacy release, to get consistency on the solar dipole measured at all frequencies from 100 to 545\,GHz, it was necessary to introduce, for the dust removal using the 857\,GHz map, an SED variation, on very large scales (dipoles and quadrupoles). It was noted that a similar correction is also indicated by the frequency map intensity ratios (figures 20 and 21 of \citedpc). The proper cleaning of the dust foreground at 217\,GHz needs to use more parameters at large scales to project the 353\,GHz signal to take into account the dust temperature distribution variation over the sky.

\begin{figure}[ht!]
\includegraphics[width=0.24\textwidth]{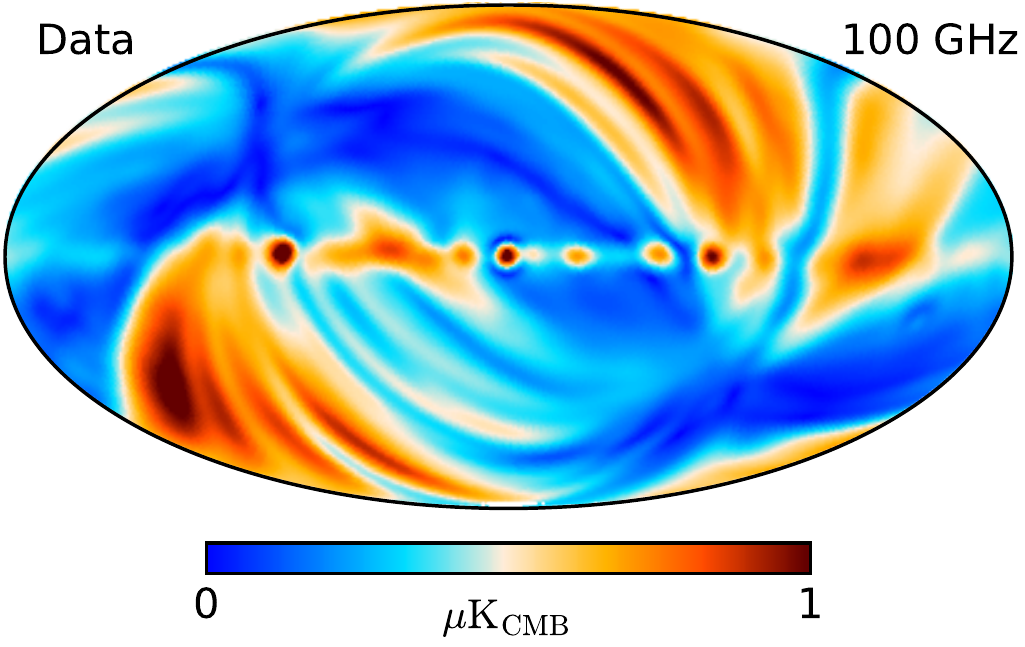}
\includegraphics[width=0.24\textwidth]{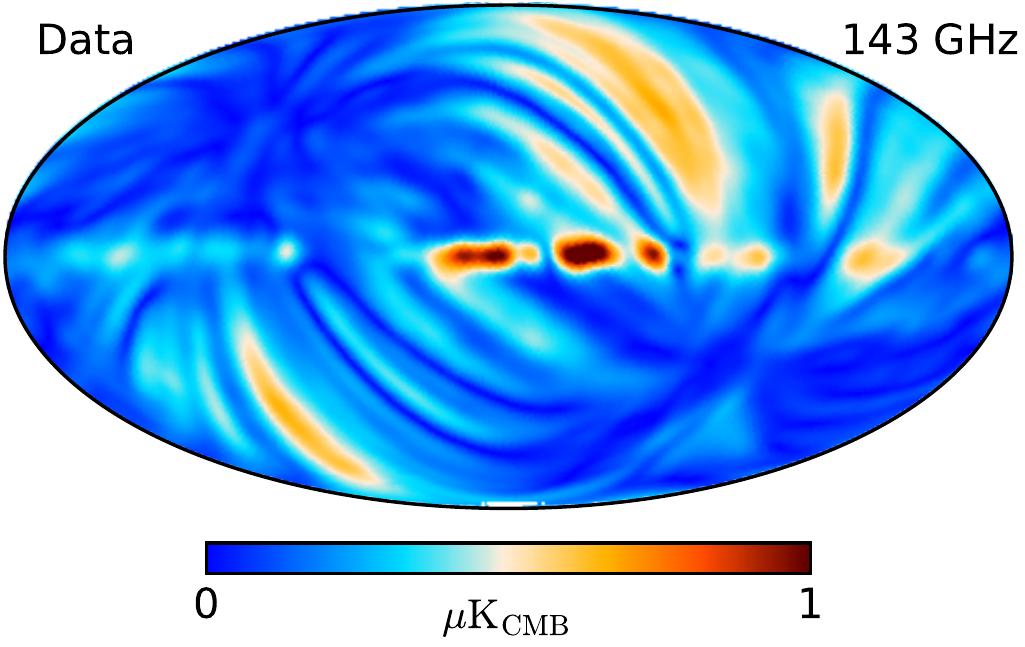}\\
\includegraphics[width=0.24\textwidth]{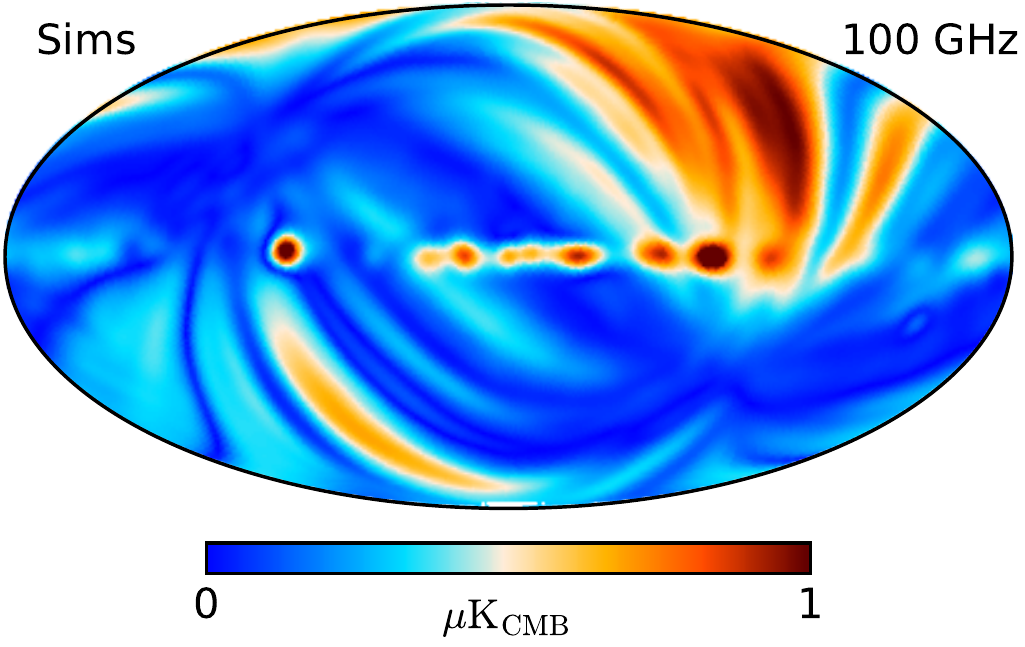}
\includegraphics[width=0.24\textwidth]{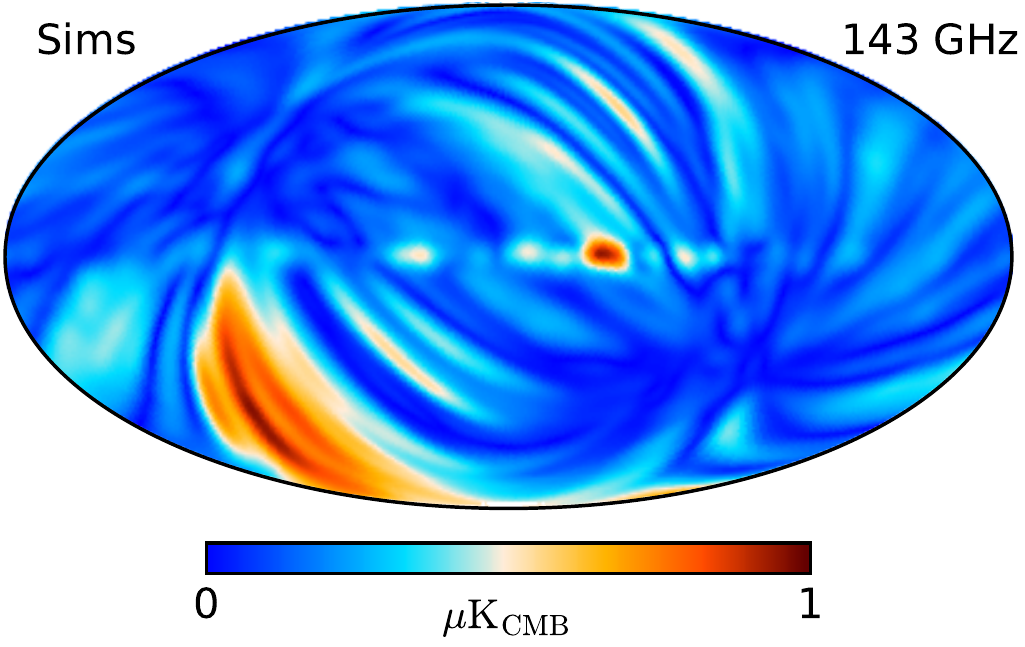}
\caption{Polarization intensity map difference between \srollone\ and \srolltwo\ algorithms at 100 and 143\,GHz (in columns), applied to the data (first row) and to one realization of the end-to-end simulations (second row).}
\label{fig:compsims1S2} 
\end{figure}

Figure~\ref{fig:compsims1S2} shows the spatial pattern consistency between the HFI2018 and \srolltwo\ maps difference at 100 and 143\,GHz and the same for a simulation of the instrumental systematic effects better corrected in \srolltwo.
The single realization chosen shows similar pattern and amplitude when compared to the data demonstrating the effectiveness of the \srolltwo\ simulations in capturing the main systematic effects.

\subsection{Power spectra of null test maps}
\label{sec:allscalesdata}

\begin{figure}[ht!]
\includegraphics[width=\columnwidth]{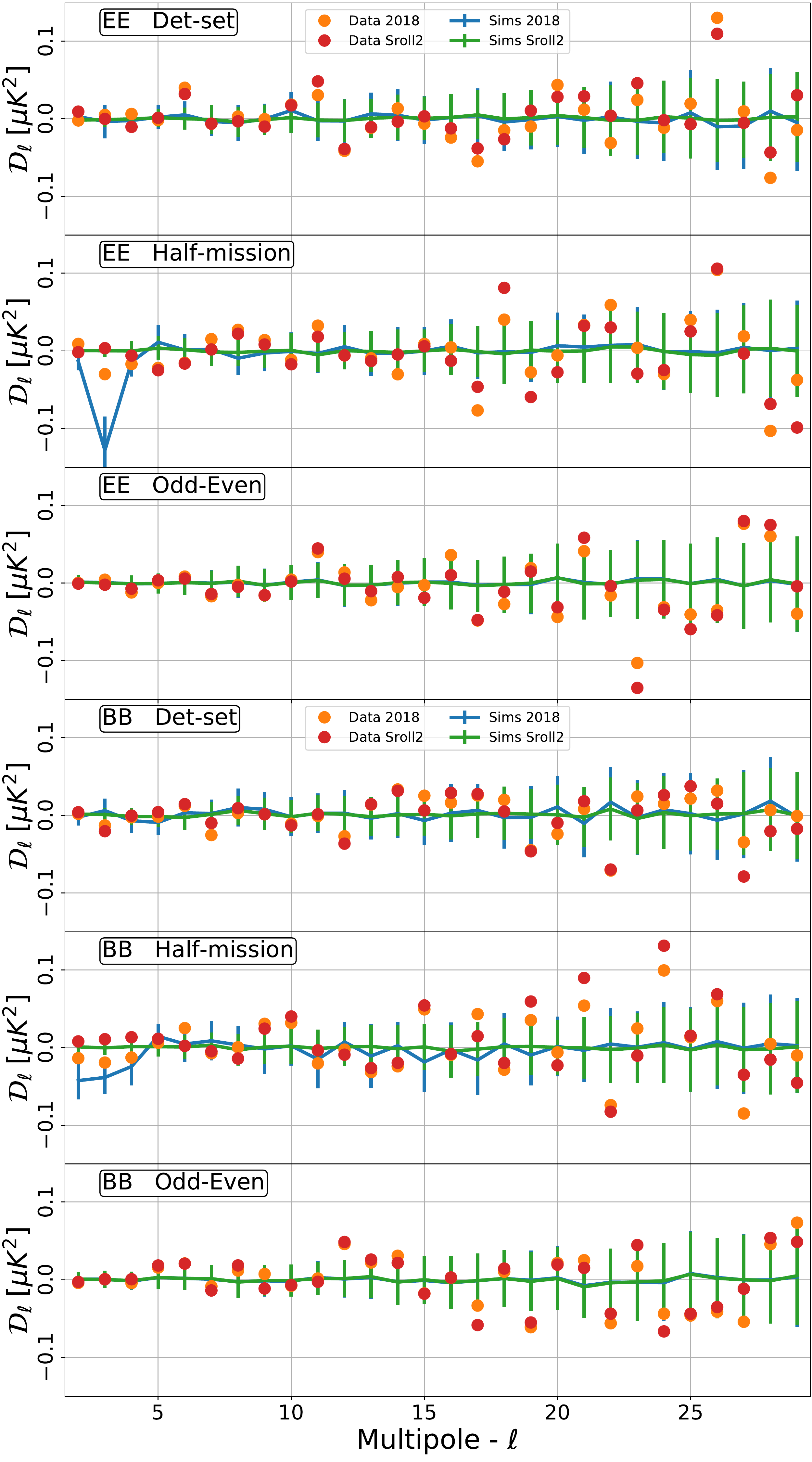}
\caption{$EE$ (upper 3 plots) and $BB$ (lower 3 plots) power spectra of difference of null test maps for: half mission, detsets and odd-even rings, for HFI2018 (orange and blue) and \srolltwo\ (red and green). Data are overplotted on top of the simulations (average and dispersion).}
\label{fig:JKspectra}
\end{figure}

Figure~\ref{fig:JKspectra} shows two sets of 100$\times$143 power spectra at large scales, for half-mission, detsets, and odd-even rings null tests, $EE$ and $BB$, for HFI2018 and for \srolltwo. Data points are overplotted, and simulation average and error bars are shown. Three hundred simulations are available for the HFI2018, and 500 are built for \srolltwo\ processing.
The very large scales $\ell <5$ show significant deviations with respect to zero both in the data and in the simulations for HFI2018; these are highly improved by \srolltwo. HFI2018 simulations show higher variance than \srolltwo\ at very large scales due to ADCNL correction only to first order. \srolltwo\ data do not show the obvious biases which were seen and predicted by the simulations for the 2018 release version of the data.

The two data distributions are consistent with the simulations for both data sets as shown by the PTEs for two multipole ranges $\ell=2$--10, and $\ell = 2$--29 in Table~\ref{tab:PTE}.
\begin{table}[htbp!]
\newdimen\tblskip \tblskip=5pt
\caption{PTEs for HM, DS, and OE null tests, for $\ell_{\rm max}=10$ and 29 (in parenthesis). The spectra are obtained by crossing difference maps.}
\label{tab:PTE}
\vskip -8mm
\footnotesize
\setbox\tablebox=\vbox{
\newdimen\digitwidth
\setbox0=\hbox{\rm 0}
\digitwidth=\wd0
\catcode`*=\active
\def*{\kern\digitwidth}
\newdimen\signwidth
\setbox0=\hbox{+}
\signwidth=\wd0
\catcode`!=\active
\def!{\kern\signwidth}
\newdimen\pointwidth
\setbox0=\hbox{.}
\pointwidth=\wd0
\catcode`?=\active
\def?{\kern\pointwidth}
\halign{\hbox to 1.5cm{#\leaderfil}\tabskip 2em&
\hfil#\hfil\tabskip 1.5em&
\hfil#\hfil&
\hfil#\hfil\tabskip 0em\cr
\noalign{\doubleline}
\omit\hfil null test\hfil&half-mission&detset&odd-even ring\cr
\noalign{\vskip 3pt\hrule\vskip 5pt}
$EE$ & $51.1\,(36.7)$ & $41.7\,(55.0)$ & $99.4\,(65.7) $\cr
$BB$ & $44.7\,(16.3)$ & $36.0\,(58.8)$ & $80.0\,(76.3)$\cr
\noalign{\vskip 3pt\hrule\vskip 5pt}}}
\endPlancktable
\end{table}
Measurements of cosmological parameters (e.g., $\tau$) on the reionization peak using \srolltwo\ maps will have a negligible bias with respect to the error budget.

Furthermore, the half-mission map difference power spectrum, which is related to the level of ADCNL residuals and noise, is now consistent with the detector noise dominating the odd-even rings map difference. This test demonstrates that the level of ADCNL residuals is now lower than the $1/f$ detector noise.

\begin{figure}[ht!]
\includegraphics[width=\columnwidth]{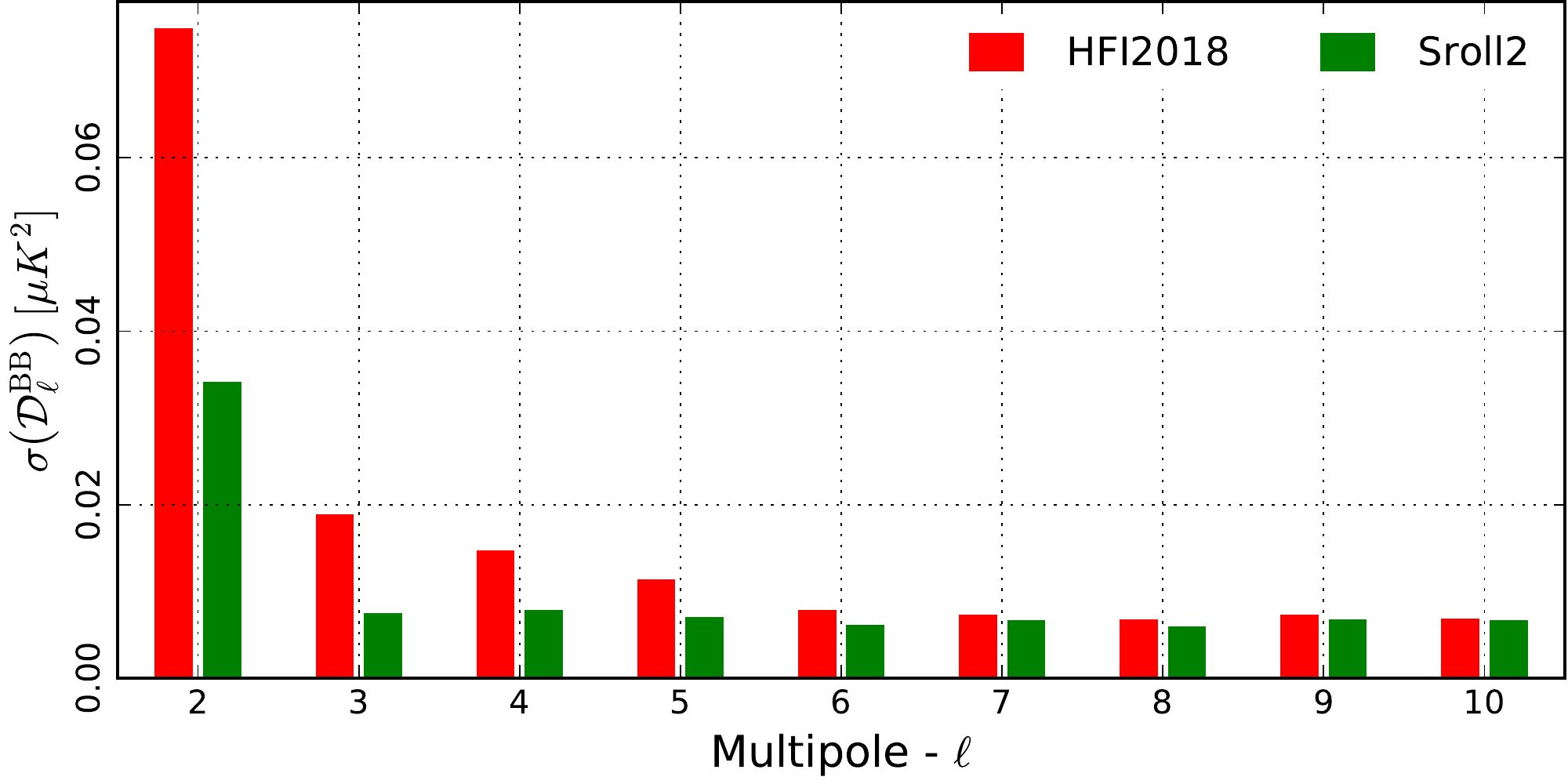}
\caption{Comparison between HFI2018 and \srolltwo\ $100\times143$ B-modes variances, computed analyzing the respective E2E simulations.}
\label{fig:errorBB}
\end{figure}

The very low multipole variances are substantially reduced by more than a factor of two for $\ell < 5$ between HFI2018 and \srolltwo. B-modes error bars, assuming $r=0$, are computed for the 100$\times$143-GHz power spectra, and shown in Fig.~\ref{fig:errorBB}. 

\begin{figure*}[ht!]
\includegraphics[width=\textwidth]{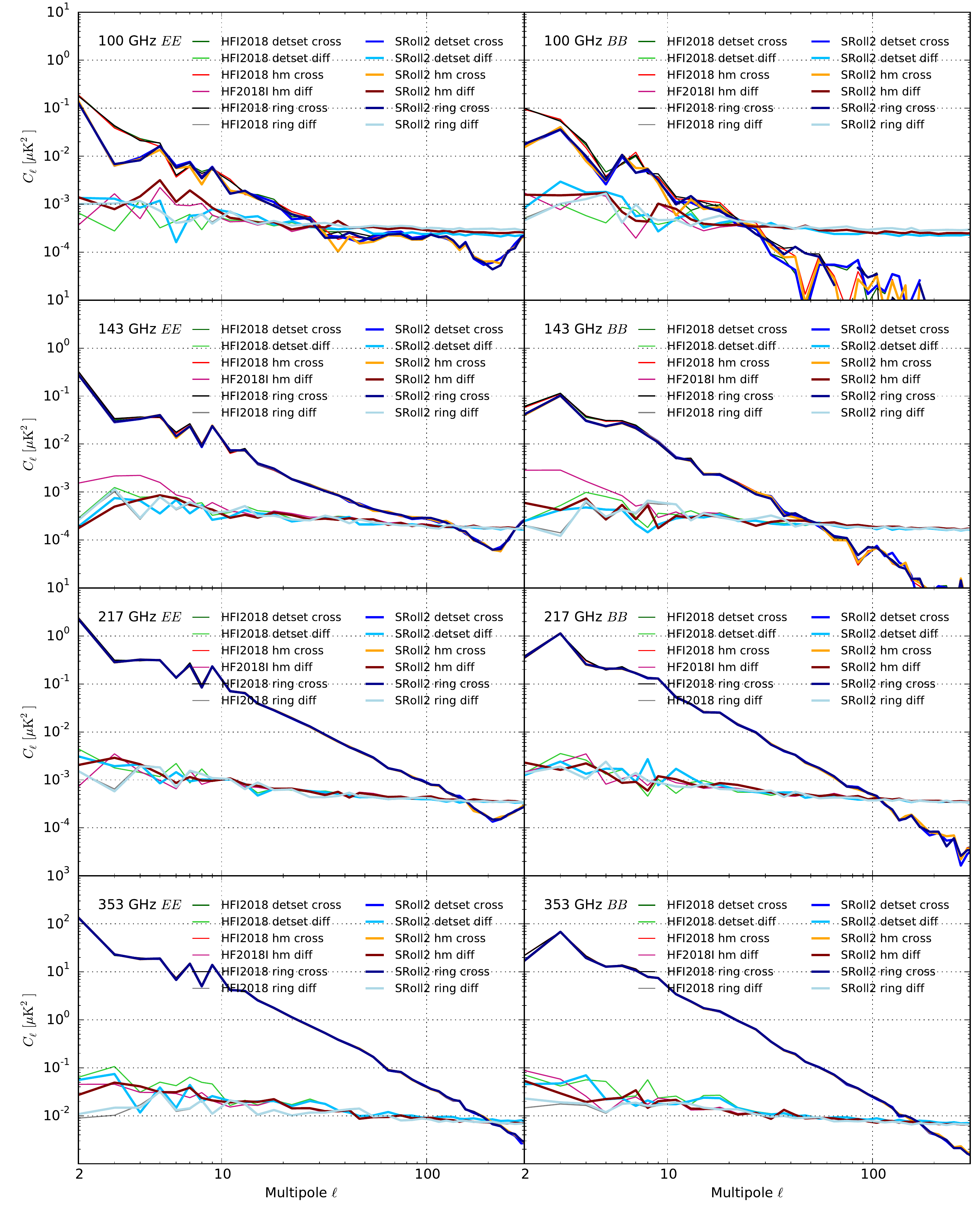}
\caption{$EE$ (left column) and $BB$ (right column) power spectra of HFI2018 and \srolltwo, based on half split data sets (detests, half-mission, and odd-even rings) maps at 100, 143, 217, and 353\,GHz. Auto-spectra of the difference maps show the noise plus systematic effect residuals, rescaled to full mission equivalent and corrected for sky fraction used here (43\,\%). The cross-spectra between the same maps are displayed and show the signal. The binning is: $\delta\ell=1$ for $2\leq\ell<30$; $\delta\ell=5$ for $30\leq\ell<50$; $\delta\ell=10$ for $50\leq\ell<160$; $\delta\ell=20$ for $160\leq\ell<1000$; and $\delta\ell=100$ for $\ell>1000$.}
\label{fig:JKspectraDX11vsRC4}
\end{figure*}

Figure~\ref{fig:JKspectraDX11vsRC4} uses a suite of maps, for both HFI2018 and \srolltwo\ data, built from half split-data sets, namely detsets, half missions, and odd-even rings.
 The figure shows $EE$ and $BB$ cross-spectra of the two halves of the split maps showing signal, and the power spectra of the difference of those maps showing noise plus residuals of systematic effects. This is another sensitive and quantitative estimate of the level of improvements in the \srolltwo\ processing over the \srollone\ one.
 
White noises measured at $\ell=$100--300 are compatible with HFI2018 data. The low-frequency noise plus systematic residuals in the range $\ell=$6--15 is now compatible with the detector $1/f$ noise. This is an indication that systematic effects have all been brought to a significantly lower level. This is discussed in Sect.~\ref{sec:tempimprove}.

The odd-even ring difference uses the same systematic effect-removal parameters based on the whole mission and should give the measurement closest to the one limited by the detector noise. The half mission difference gives a very good estimate of the ADCNL distortion of the signal varying on long time scales for which the induced residuals were clearly seen at 143\,GHz in the HFI2018 maps. At 143\,GHz, such residuals have almost disappeared in the \srolltwo\ data. This again advocates \CHANGE{for} future improvements concentrating  on reducing the large-scale detector noise (see Sect.~\ref{sec:otherimprov}).

At other frequencies, where the half mission differences were more similar to the other null test in the HFI2018 data, the reduction of the residuals is of course smaller.

\subsection{Caveats when using the \srolltwo\ maps}

Although the \srolltwo\ algorithm concentrates on improving the large-scale mapmaking, several systematic effects at small scales have also been reduced by the indirect effect of the improvements discussed above through the degeneracies between the two sets of systematic effects.

Several instrumental effects, already identified in \citedpc, which could improve the maps have not been implemented in \srolltwo. \NCHANGE{Firstly the Beam anisotropy: one should be aware that the beam leakage may correlate temperature and polarized signals at small scales.  Secondly the $1/f$ detector noise at $\ell \le 100$ shown to be dominated by undetected glitches: no attempt has yet been made to clean this Gaussian noise, and this dominates over the white noise. Thirdly the 4-K line cleaning has not been improved and known features are still present in the maps. Finally the sub-pixel effect of the CO template that affects a few pixels in regions of very strong gradients in the Galactic center and molecular cloud cores is still present. As the 143-GHz map is of course not affected, the 100- and 217-GHz polarized signal in such regions should not be used.}

We note that, for the \srolltwo\ frequency maps, the foregrounds bandpass mismatch coefficients $L_{f}$ have been cleaned. For this, all detectors have been brought to a common response for each foreground which is the bandpass average of all detectors in the band. Foreground maps cannot be computed from the difference between any single bolometer map\footnote{For polarized bolometers, we build intensity HPRs and corresponding maps by subtracting the current polarization model from the total HPRs.} built from the cleaned HPRs. 

We also note that the \srolltwo\ data processing includes nonlinear steps that make the data much more complex than a signal with a simple additive residual corrections on top of a Gaussian noise. As for all previous \Planck\ releases, it is mandatory to validate any use of these data by testing against the results of simulations.

Furthermore, the simulation sky model has a limited representativeness due to the adopted simple foreground emission laws and is also incomplete (e.g., the $^{13}$CO emission is not simulated).

\clearpage
\clearpage
\section{Characterization}
\label{sec:tempimprove}

\subsection{Foreground residuals}
\label{sec:tempstudy}

To model foreground induced systematic effects, the \srollone\ and \srolltwo\ algorithms both rely on sky templates to fit their correction parameters from the data redundancies, and then for the implementation of the corrections. The quality of the frequency maps produced depends on the representativity of these input foreground templates with respect to the sky ones. The main foreground templates for the HFI data ($^{12}$CO, $^{13}$CO, and dust) are used to correct for bandpass mismatch leakage in polarization. In \srolltwo, the all-sky templates are the dust thermal and the free-free emissions, based on the Commander \Planck\ 2015 component separation products. The $^{12}$CO and $^{13}$CO line emission all-sky templates are built with the leakage coefficients obtained from the Taurus molecular cloud maps from \citep{Goldsmith2008} as described in \citedpc, and used in the characterization of the HFI2018 maps. 

Here, we develop a test of the impact of the input foreground templates which are not modified within one \srolltwo\ run \CHANGE{on the final maps}. This is done by implementing an iterative approach to \CHANGE{improve} the accuracy of the foreground templates outside \srolltwo. This test is based on a limited characterization which itself is based on an addition to the foreground templates, and a simple model with two foregrounds: the 100-GHz map, assumed to be dominated by the CO bandpass leakage, and the 353-GHz map, dominated by the dust bandpass leakage.

In the first iteration of this toy model, we degrade the input CO and dust foreground templates by adding a colored noise map with the same power spectrum as the foreground but with random phases. This makes the initial CO foreground leakage template significantly different from the CO input maps shown in \citedpc\ to be very good by testing them within the Taurus region mapped by \citet{Goldsmith2008}. A similar degradation is done for the input $I$, $Q$, $U$ dust maps obtained by component separation methods and used as the input foreground component maps in this simulation. Subsequently, we perform a full \srolltwo\ run to build a set of frequency maps. We then build new foreground templates. The CO templates are built using the bolometer efficiency coefficients improved in the first \srolltwo\ run, while the dust map is simply the difference between the 353\,GHz and the 100\,GHz maps, after being cleaned from the new CO templates. We then use these new foreground templates in the following iteration of \srolltwo\ run. Subsequently, we test that the residuals of the foreground leakage converge to the solution obtained with the nondegraded input templates.

The implementation is done by integrating the two CO template maps in an extended definition of the pointing vector $A$:
\begin{eqnarray}
\label{eq:polparam}
A^{CO}_{b,r,p} = \left[1,\rho_b\cos(2\phi_{b,r,p}),\rho_b\sin(2\phi_{b,r,p}),L_{b,{}^{12}{CO}},L_{b,{}^{13}{CO}}\right].
\end{eqnarray}

\begin{figure}[ht!]
\includegraphics[width=\columnwidth]{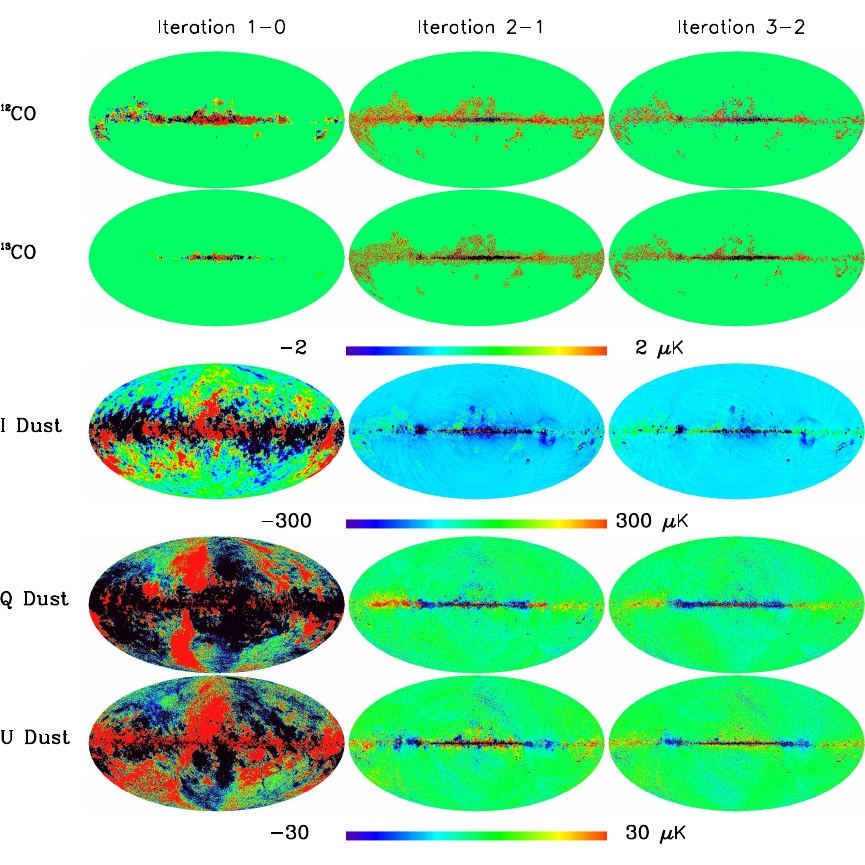}
\caption{\CHANGE{Differences between two runs of \srolltwo, one using the best available CO and dust templates, and another using a degraded version of these initial ones (as described in the text) designated as iteration 1. We display the foreground templates evolution: $^{12}$CO, $^{13}$CO (two top rows) and $I$, $U$, $Q$ dust template (three lower rows). The first column is the difference between the best templates (iteration 0) and the run with the degraded templates (iteration1). \srolltwo\ is then run with foreground templates built from the sky maps and leakage coefficients from the iteration 1 displayed in the middle column. The results from a third iteration are displayed in the right column.}}
\label{fig:templateiter}
\end{figure}

The first run starts with the degraded templates. Then, in the following runs, the values of the leakage coefficients $L_{b,{}^{12}{CO}}$ and $L_{b,{}^{13}{CO}}$ are given by the previous iteration of the \srolltwo\ fit. Using Eqs.~\ref{eq:projmap2}, \ref{eq:projmap}, and \ref{eq:polparam}, we build the frequency maps and the two new CO leakage template maps. At each step, the 100-GHz single bolometer HPRs are used to compute both the $^{12}$CO and the $^{13}$CO templates. The 100- and 353-GHz frequency maps are then used to build the new dust template. This procedure is \CHANGE{then iterated.} We show in Fig.~\ref{fig:templateiter} that \CHANGE{four} runs, one starting with the best estimates of foreground template and \CHANGE{three iterations starting with the degraded templates, lead to a set of foreground template maps showing decreasing differences between iterations}.
This demonstrates that, despite the strongly degraded quality of the initial template, the procedure provides a method able to quickly build much better component templates for both CO and dust. 

\begin{figure}[ht!]
\includegraphics[width=\columnwidth]{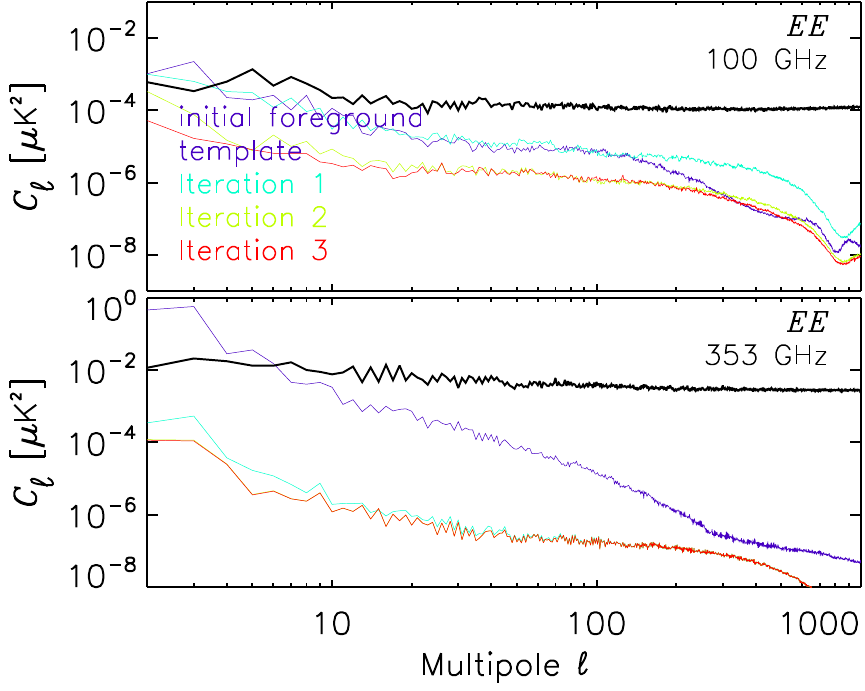}
\caption{Power spectra of the initial templates and then differences between two successive iterations: first iteration in cyan, second in green, and third in red. The differences becomes quickly much smaller than the noise (black line) that is estimated from half mission difference computed with $f_{sky}=0.43$ and realigned at the full mission level.}
\label{fig:dustiter}
\end{figure}

To be more quantitative on the convergence, Fig.~\ref{fig:dustiter} shows the $EE$ power spectra of the difference of the 100- and 353-GHz polarized frequency maps built using the nominal input templates and the maps obtained with the degraded ones.
The two panels show in purple the starting "bad" input templates. At 100\,GHz, the first iteration (in blue), shows, for $\ell<100,$ a similar level as the foreground power spectrum purposely added to distort the associated template. The second and third iterations show much smaller differences showing quantitatively the fast convergence of the templates down to differences much smaller than the noise. The convergence toward zero of the dust foreground error at 353\,GHz is almost reached as soon as the first iteration.

We thus demonstrate here on a simple model with two frequencies and two foregrounds, that, after a few iterations, the foreground templates produced by \srolltwo\ after adding very different initial templates (random phases) converge to the initial one. This does not demonstrate that the component templates found by this method are the optimal component maps that could be built, but that they are sufficiently good to clean the polarized data from foreground bandpass mismatch leakage to a level much lower than the noise.
This provides a path to integrate component separation to the mapmaking, by combining all detectors from different frequencies in the same mapmaking run with the caveat that each foreground component should be negligible in at least one frequency band.

\subsection{Transfer function, polarization angles, and efficiencies}

\subsubsection{The \srolltwo\ transfer function}

\begin{figure}[ht!]
\includegraphics[width=\columnwidth]{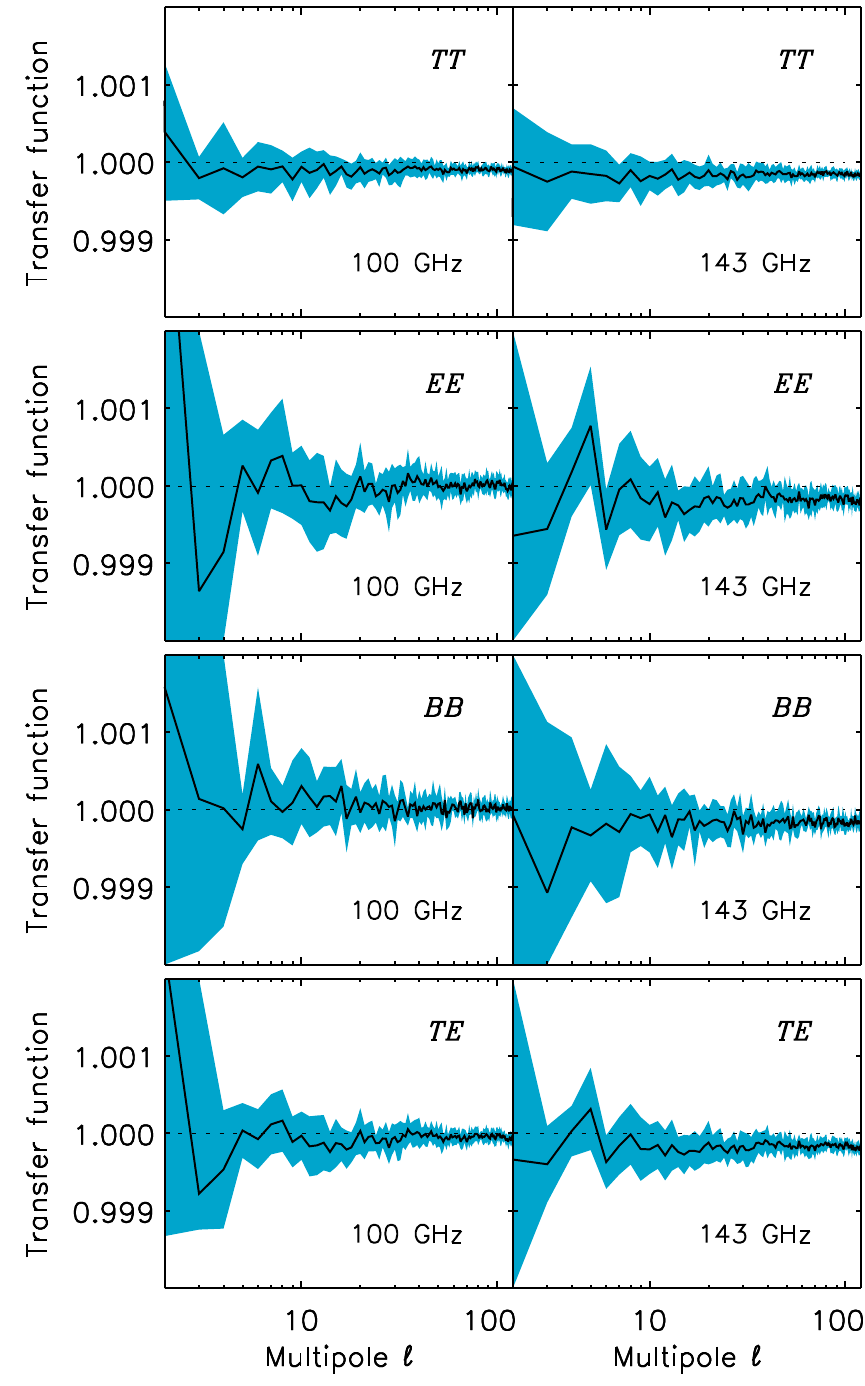}
\caption{Transfer function of the \srolltwo\ 100- and 143-GHz maps. Thick line shows the average of ten simulations while the envelop shows the peak-to-peak dispersion.} 
\label{eq:pltesttf}
\end{figure}

The transfer function evaluates how much of the sky signal is either absorbed by the mapmaking into a systematic effect correction, or distorted by the processing. This cannot be done by pure simulation as the models are not statistically reliable models for systematic effects and foregrounds. We thus run \srolltwo\ with the sky data to which we add a white noise map in $I$, $Q$, and $U$ of comparable rms. We subtract from these maps the ones obtained with the sky data alone, and build the power spectra from the resulting map. The ratio of these power spectra to the power spectra of the additional white noise simulated signal gives the transfer function in conditions close to those of the sky. Figure \ref{eq:pltesttf} shows these transfer functions measured on the auto-power-spectra at 100\,GHz and 143\,GHz.
This result demonstrates that the signal distortion is of the order of or smaller than $10^{-3}$ at $\ell=3$--10, and a few  $10^{-4}$ above. We also note a very small suppression of power at the level of few $10^{-4}$ in the 143-GHz power spectra. We conclude that there is no evidence for any transfer function effect to be corrected for.

\subsubsection{Thermal time constants}

\begin{figure}[ht!]
\includegraphics[width=0.9\columnwidth]{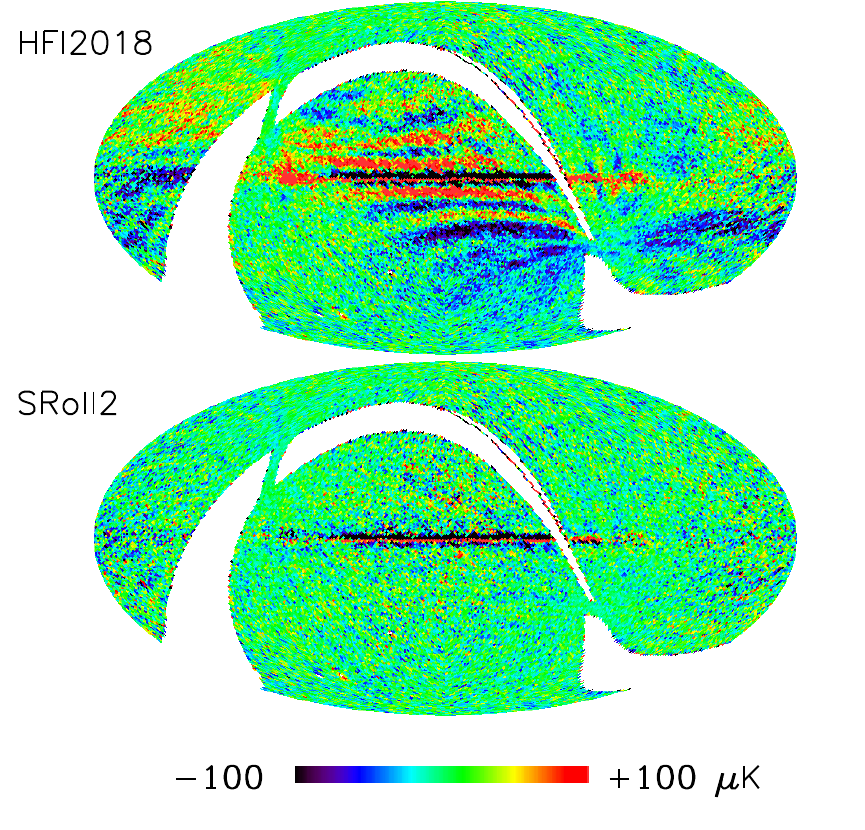}
\caption{Intensity map of the odd-even survey difference at 353\,GHz from HFI2018 (top) and \srolltwo\ (bottom) after changing the model of the real part of the transfer function.}
\label{fig:cleanzebra}
\end{figure}

The bolometer thermal chain includes the copper bolometer housing and feed horns, and its interface with the bolometer plate. These have very long time constants (the main time constant of the HFI bolometers is a few milliseconds long). When testing the instrument, the heaters of the temperature control system were used to inject power into the bolometer plate and measure the time delay of the bolometers reaction. Those show a distribution of time constants not associated in propagation in the plate of order 10 to 30\,s. The previous mapmaking algorithms, including \srollone, could not correct them well. This needed to be improved to reach the detector noise at the largest scales. The shorter time constants identified in the ground tests before flight (up to 2\,s) were already corrected in the timeline processing in the previous versions of the data processing. In the HFI2018 maps, the very long time constants were corrected through a temporal transfer function only modeled in three ranges of harmonics of the spin frequency. Within each range, the time transfer function was assumed to be constant. The best null test for evaluating time constant correction is the comparison of odd and even surveys which scan the sky in opposite directions. At 353\,GHz, this null test showed (see \citedpc) striping not aligned with the scanning pattern (the so called ``zebra effect''), and systematic calibration discrepancies between odd and even surveys. This description of the transfer function does not properly correct the very long time constant systematic effect, although Fig.~\ref{fig:cleanzebra} shows that replacing the description of the transfer function in steps with a continuous complex transfer function description associated with a single time constant (of the order of 25\,s) removes the zebra effect.

In \srolltwo, a new description of the complex time transfer function has been added. The imaginary part for dipoles is the critial one, and we take a similar approach as the \Planck\ release to correct the phase shift only for dipoles.

The real part affects the transfer function for spectra, and is adjusted to a function with a single time constant\footnote{$T_{r}\left(\nu \right )=1/(1+{\left(2 \pi \nu \tau \right )}^{2})$ where $\tau$ is the time constant and $\nu$ is the temporal frequency.}  which accounts for the more complex behavior. Due to the sky observation period of 60\,s, it is not possible to properly identify any time constant longer than 10\,s. Figure~\ref{fig:cleanzebra} shows that using this empirical model made of two independent sets of transfer functions corrects the striping.
It also corrects the odd-even calibration discrepancy between odd and even surveys. This confirms that the \srollone\ solution was not the right one, and that just adding a single very long time constant was not accounting for the complexity of the heat transfer from the bolometers to the bolometer plate.

\begin{figure}[ht!]
\includegraphics[width=\columnwidth]{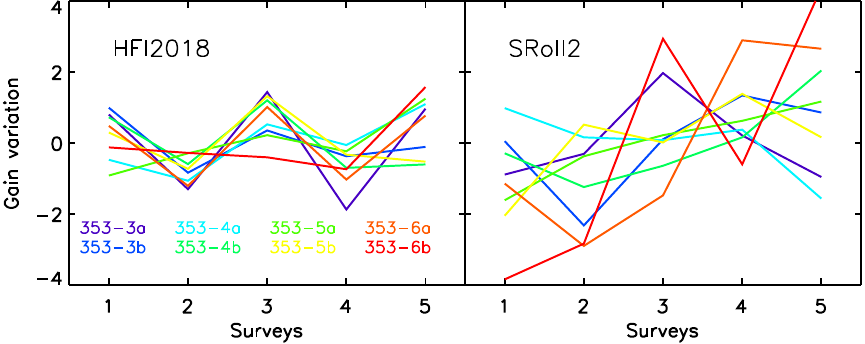}
\caption{Residual solar dipole amplitudes with respect to the average dipole, for each 353-GHz PSB, shown \CHANGE{per} survey.}
\label{fig:chapochinoi} 
\end{figure}

Figure~\ref{fig:chapochinoi} shows that the odd-even survey pattern seen in the response variation of the HFI2018 353-GHz data (shown figure~39 of \citedpc) has been cleaned in \srolltwo\ using the new time constant model.
Furthermore, the calibration dispersion between surveys is $2.6\times10^{-4}$ for HFI2018, and $5\times10^{-4}$ for \srolltwo\ data. This leads to an estimate of the uncertainty when averaging over all detectors of  $2.2\times10^{-4}$ for \srolltwo\ for which the noise is dominating. This value is in agreement with the estimate of the detector noise given in Table~7 of \citedpc. The reduced dispersion between bolometers in each survey is clearly smaller than the noise, showing that there was an overfitting, discussed below. 

\begin{figure}[ht!]
\includegraphics[width=0.9\columnwidth]{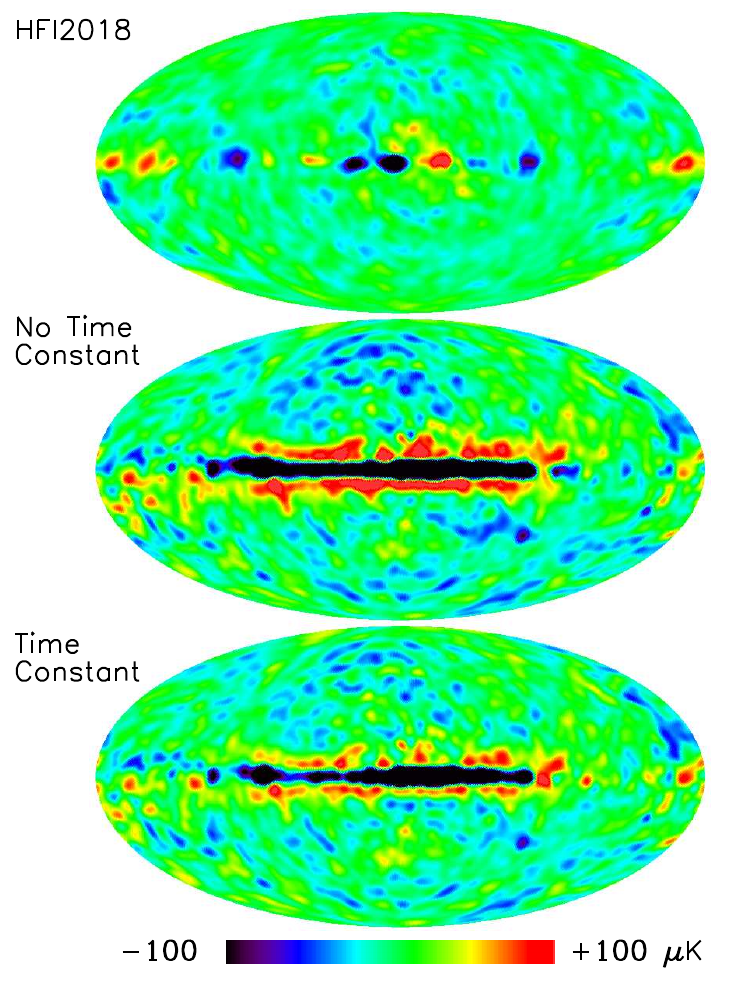}
\caption{Difference of two single-bolometer intensity maps (353-3a and 353-5b) smoothed at 5\deg. 
{\it Top}: HFI2018; {\it Middle}: \srolltwo\ without cleaning the long time constant; {\it Bottom}: \sroll2\ with cleaning the long time constant.}
\label{fig:timeconsteffect} 
\end{figure}

Figure~\ref{fig:timeconsteffect} shows the difference between intensity maps from the two single bolometers within a single PSB pair at 353\,GHz, which demonstrates the need for this procedure. 
The two lower panels show the cases with and without time constant correction. They show the reduction of the residuals brought by \srolltwo. In the top panel, the map difference for the \citedpc\ mapmaking shows a lower large-scale noise at high latitude than in the \srolltwo\ solution (bottom panel), which is also again the sign of overfitting.

The overfitting is due to the degeneracy of the low-frequency $1/f$ detector noise with the very long time constant. To avoid the overfitting, the very long time constant is fitted, within a single PSB pair, for bolometer $a$ on a template built from bolometer $b$. This removes the degeneracy between the very long time constant and the $1/f$ noise. This method is justified because the very long time constants, due to the bolometer housing interface with the plate, are common to the $a$ and $b$ bolometers.

\subsubsection{Detector polarization angles and efficiencies}

Although the polarization angle and efficiency errors are partially degenerate in their effect on the polarized maps, a global error on the overall angle of the focal plane would induce $EB$ and $TB$ correlation which are observed to be very low. This is discussed and confirmed for the \srolltwo\ maps in the following section. The absolute value of the polarization efficiency is degenerate with the polarized map calibration (\citedpc). The polarization angles are well measured and induce a smaller error than the polarization efficiency uncertainties. Furthermore, \citedpc\ demonstrated that the ground measurements of the bolometer polarization efficiency were not accurate, especially for SWB. Without an external calibrator, it is not possible to clean the induced absolute polarization mismatch. It is nevertheless possible to compute for each detector a relative polarization efficiency with respect to the average for the frequency band detector set. The following equation introduces the polarization parameter error in the Eq.~\ref{eq:datamodel} data model:
\begin{eqnarray}
\label{eq:polparam2}
\delta{\left( g_{b}M_{b,r,p}\right) } &=& \delta{\rho_b} T_{\rho} + \delta{\phi_{b,r,p}} T_{\phi},
\end{eqnarray}
where
\begin{itemize}
\item $T_{\rho} = \left[Q_p\cos(2\phi_{b,r,p}) + U_p\sin(2\phi_{b,r,p})\right]$,
\item $T_{\phi} = 2 \rho_b \left[U_p\cos(2\phi_{b,r,p}) -Q_p\sin(2\phi_{b,r,p}) \right]$.
\end{itemize}

Therefore, by fitting both $T_{\rho}$ and $T_{\phi}$ templates, \srolltwo\ measures the relative errors on polarization efficiency and angle. As mentioned above, $T_{\rho}$ is degenerate with the polarized signal calibration, and, in order to avoid removing polarization signal from the final map, the mean of \CHANGE{all $\delta{\rho_b}$ within a frequency} is set to zero. With $T_{\phi}$ being degenerate with the leakage from $E$ to $B$ modes, a similar hypothesis is to set the mean $\delta{\phi_{b,r,p}}$ to zero.

Using end-to-end simulations, we obtain the following uncertainties for the polarization efficiency per bolometer: 2\% at 100\,GHz, 0.8\% at 143\,GHz, 0.3\% at 217\,GHz, and 0.3\% at 353\,GHz. This shows that no improvement of the polarization efficiency knowledge can be obtained. Thanks to the increase of the dust emission with the frequency, the polarized signal becomes much higher than the noise, and we can thus improve its knowledge on the 2\% level quoted in \citedpc\ at 217 and 353\,GHz.

\begin{figure}[ht!]
\includegraphics[width=\columnwidth]{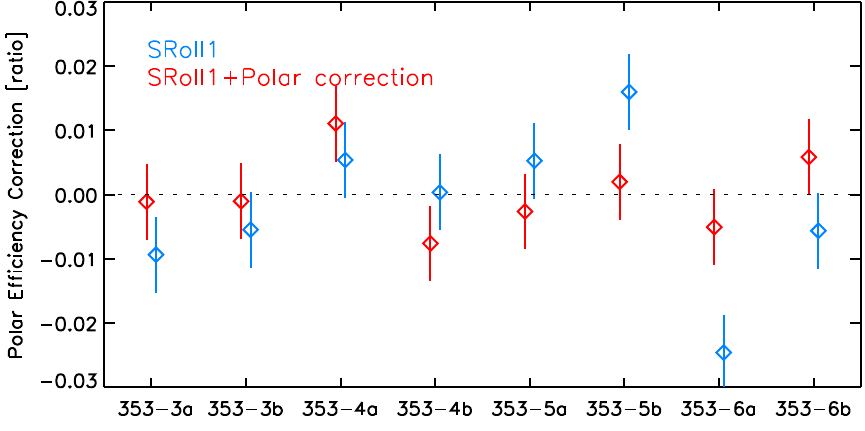}
\caption{Relative polarization efficiency with respect to ground-based measurements, extracted from single-bolometer maps for the 353\,GHz PSBs. This figure compares the relative polarization efficiency from HFI2018 (in blue) with polarization efficiency cleaning (in red). The error bars are obtained by testing the same procedure on simulations. } 
\label{fig:detpolcorr} 
\end{figure}

Figure~\ref{fig:detpolcorr} shows the efficiency of the \srolltwo\ method in cleaning residual relative polarization efficiency with respect to the ground measurements average on all 353-GHz PSB detectors seen on single-bolometer HFI maps.
The residuals after correction (in blue) are the HFI2018 ones, as shown in Figure~35 of \citedpc. The red relative residuals from \srolltwo\ are consistent with the 0.3\,\% quoted above, thus confirming the nearly complete correction of the polarization efficiencies with a reduced dispersion of $0.6\%$, compatible with the error bars.

\subsection{Global focal plane rotation}

We calibrate the effects of a possible global focal plane rotation by adding, to each simulation, the same constant polarization angle $\psi$ to all bolometers \CHANGE{at all frequencies}. This generates $E$-to-$B$ leakage, and we can then infer an absolute polarization angle correction as the one minimizing this leakage. We use the mean of the $EB$ cross-spectra for $\ell\in[60, 167]$ as an indicator of the leakage, and we verify that it behaves linearly with $\psi$. This multipole range is a best compromise driven by the fact that, at lower multipoles, dust has unknown $E$-to-$B$ leakage, and at higher multipoles, there is unknown beam leakage. 
We then produce two runs (one shifted by $\psi_{1}=-0.6\deg$, and one not shifted with respect to the ground-based measurements $\psi_{0}=0\deg$) of 100 simulations, and we use a linear interpolation to find the value of $\psi$ for which the leakage is zero.

\begin{figure}[ht!]
\includegraphics[width=\columnwidth]{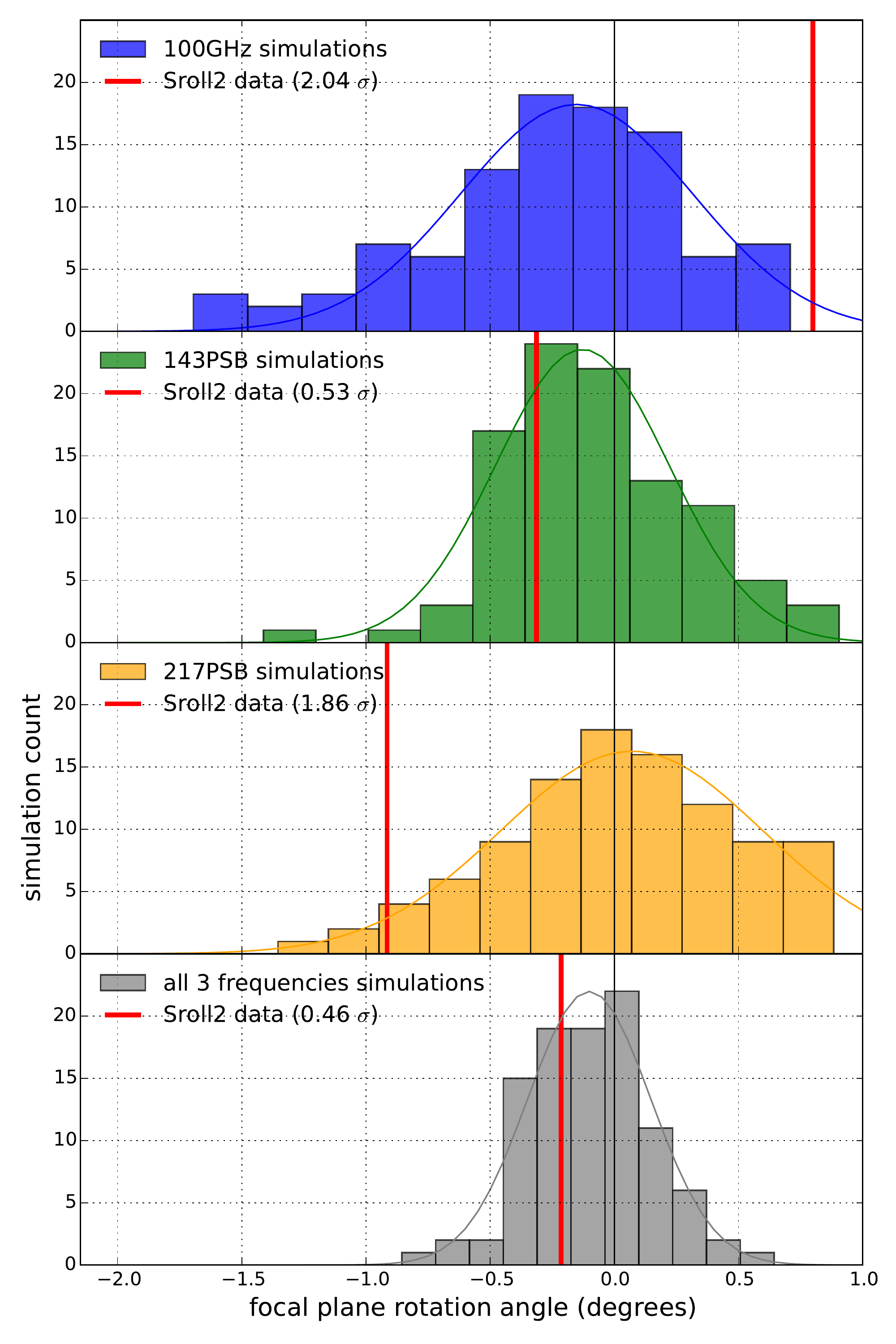}
\caption{Distribution of the polarization absolute angle shift on 100 simulations without any shift in the input. The corresponding value for the data is overplotted (red line). There is no detection of any bias of the absolute angular shift in frequency average. The data value at 100\,GHz shows $2\,\sigma$ detection, too small to be used as a correction.} 
\label{fig:fprotation} 
\end{figure}

Figure~\ref{fig:fprotation} shows the data and the dispersion of the simulations.
A similar analysis has previously been performed (see \citelowell) but without estimating the error from end-to-end simulations. It is nevertheless very important to go through the mapmaking process to take into account the other instrumental systematic effects that could generate leakage from $E$ to $B$ (e.g., beam errors).

The distribution of the simulations shows that there is no detectable focal-plane-rotation-angle bias introduced by \srolltwo. Finally, this analysis demonstrates that the detected absolute polarization angle shift is consistent with the distribution of 100 simulations without shift, and therefore the \srolltwo\ maps do not have to be corrected for any absolute angle rotation.

\subsection{ADCNL removal}
\label{sec:adcclean}

The level of the ADCNL residuals cannot be checked only on the null test (see Sect.~\ref{sec:allscalesdata}). To demonstrate the efficiency of the \srolltwo\ algorithm in removing the ADCNL systematic effect, we use the end-to-end simulations.
We compare the \srollone\ approach when the ADCNL was corrected by a linear time varying gain, with the \srolltwo\ approach in which the full ADCNL of each detector is modeled by spline functions for each ring. The spline weights are fitted using the redundancy in the input maps.

\begin{figure}[ht!]
\includegraphics[width=0.24\textwidth]{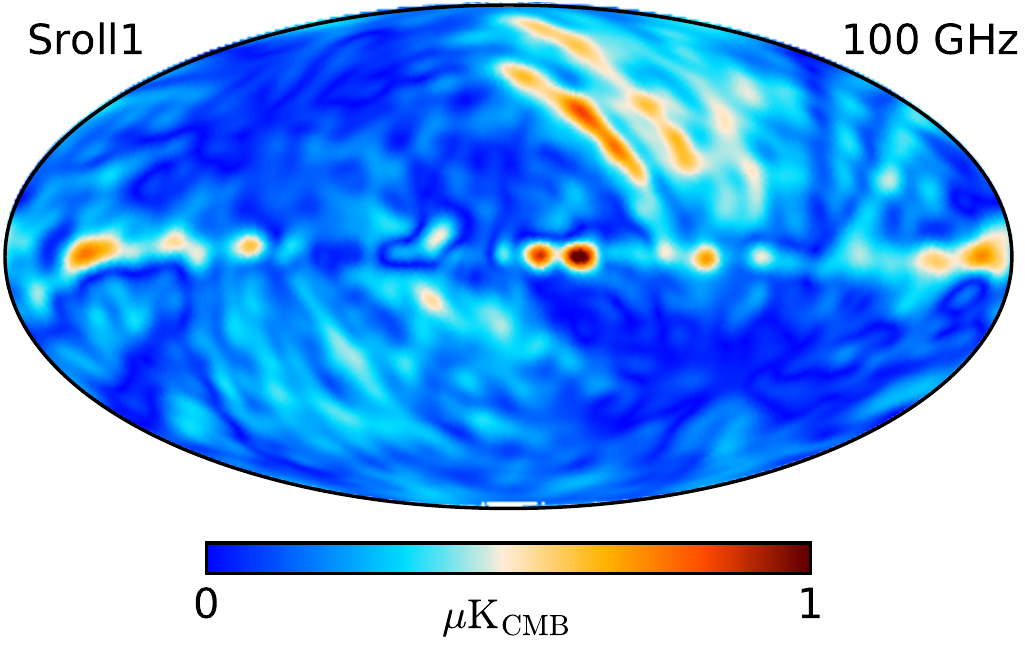}
\includegraphics[width=0.24\textwidth]{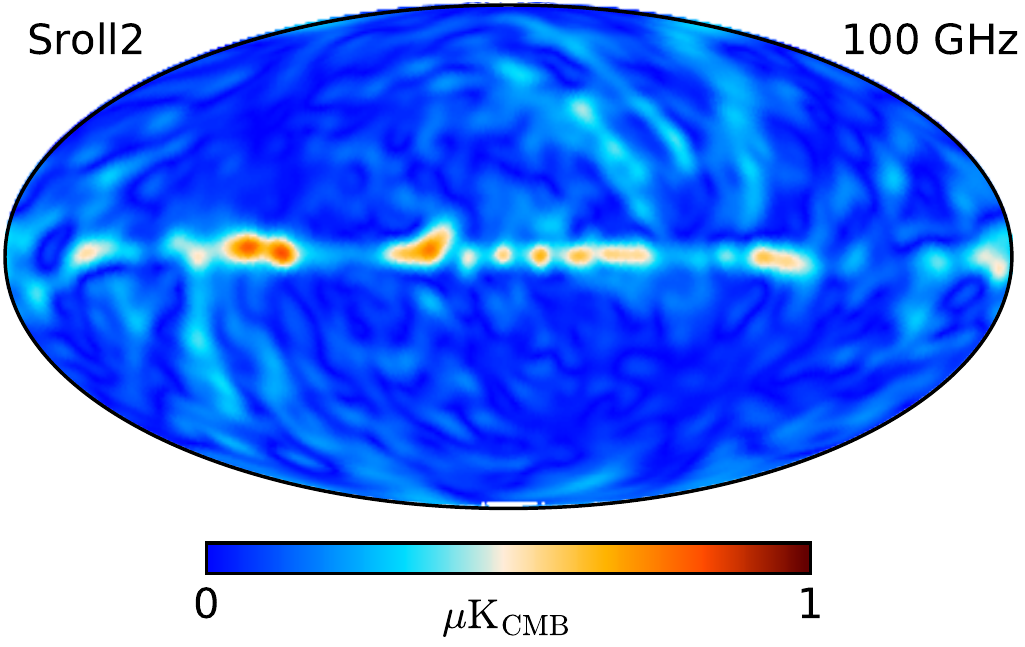}\\
\includegraphics[width=0.24\textwidth]{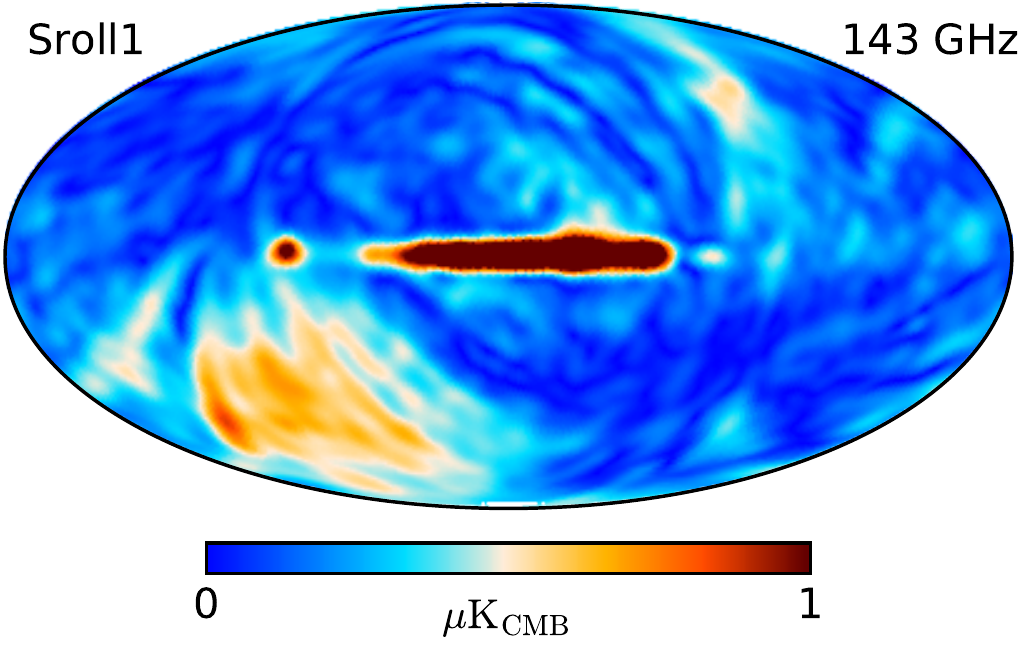}
\includegraphics[width=0.24\textwidth]{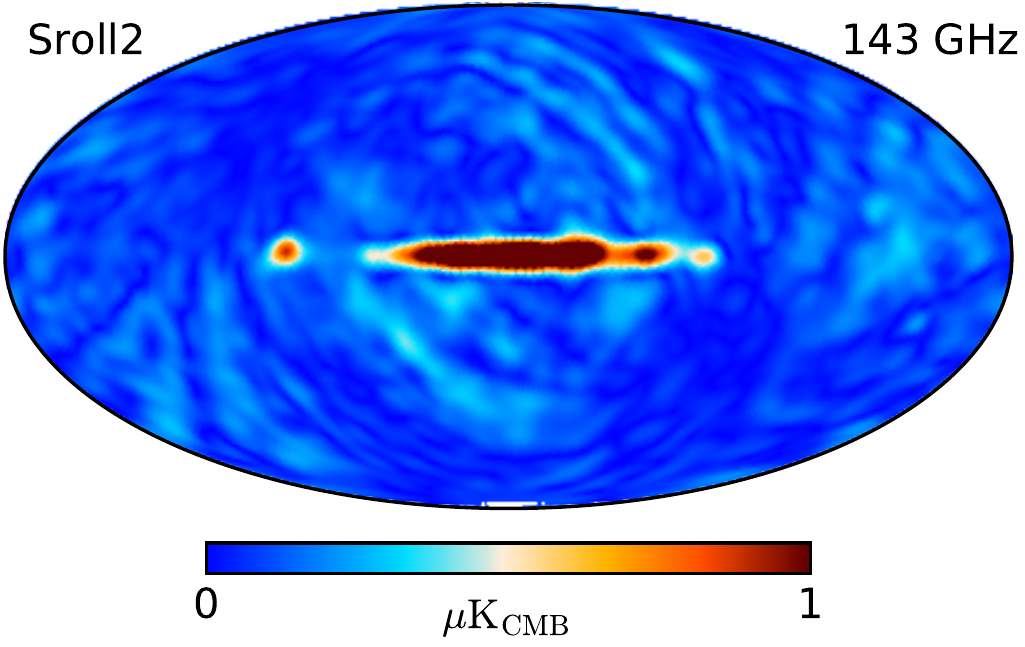}\\
\includegraphics[width=0.24\textwidth]{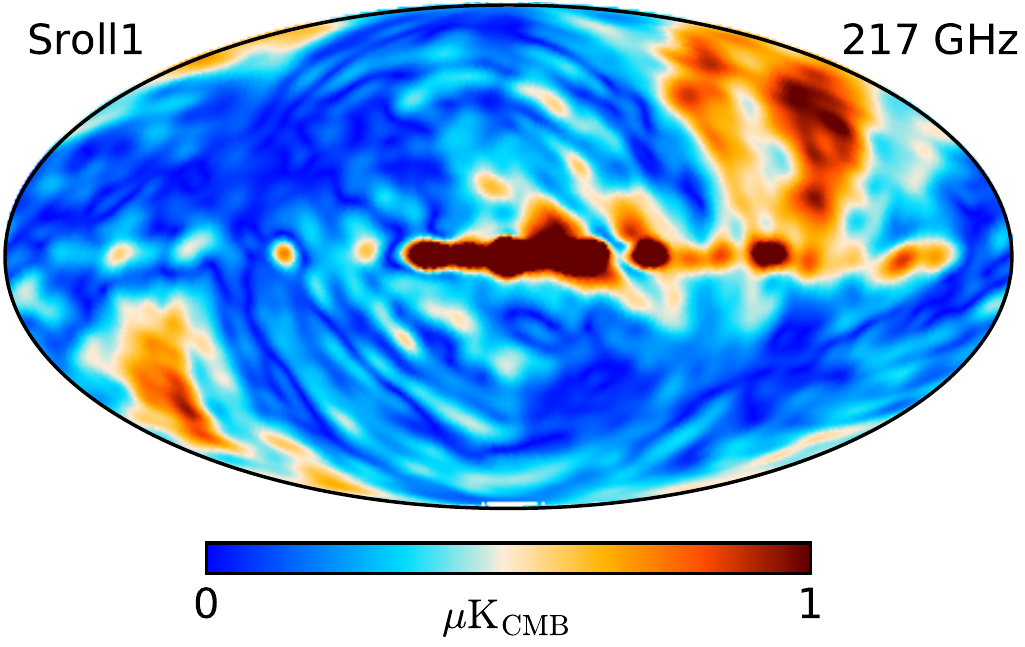}
\includegraphics[width=0.24\textwidth]{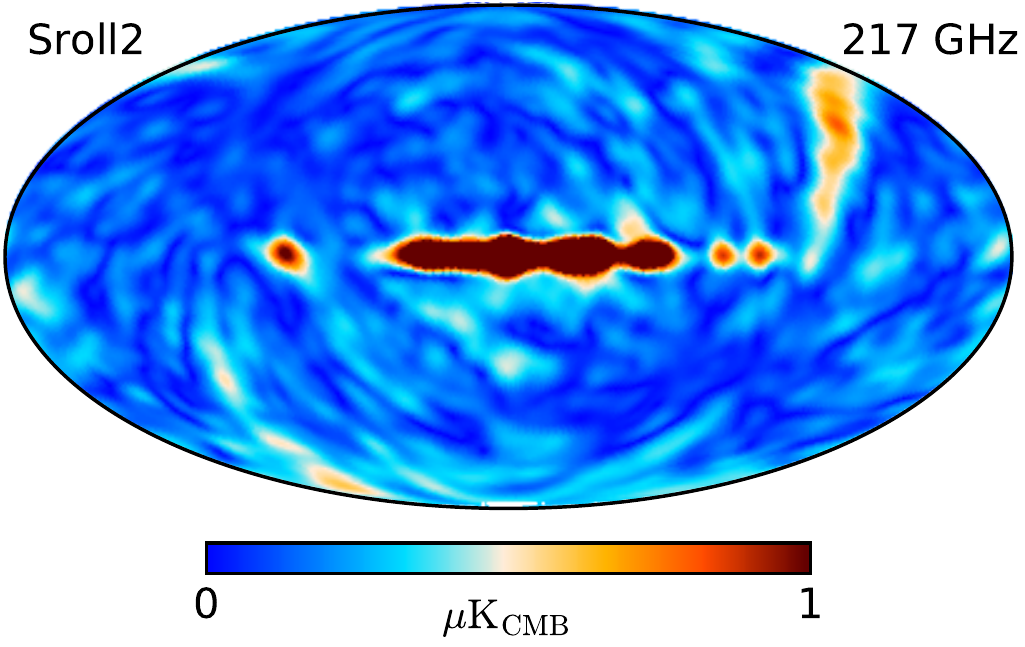}\\
\includegraphics[width=0.24\textwidth]{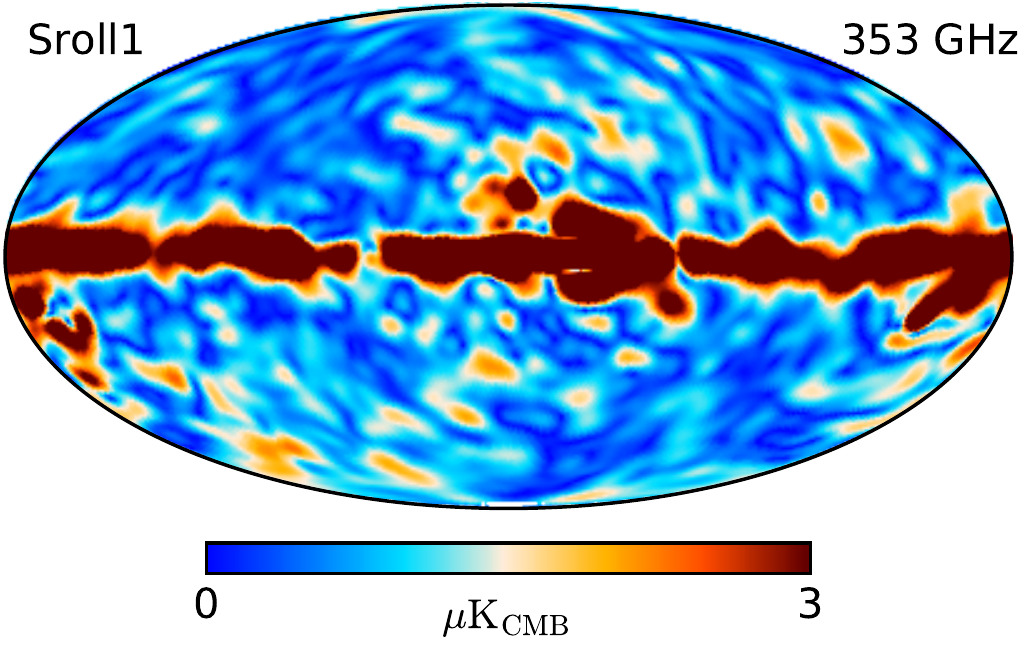}
\includegraphics[width=0.24\textwidth]{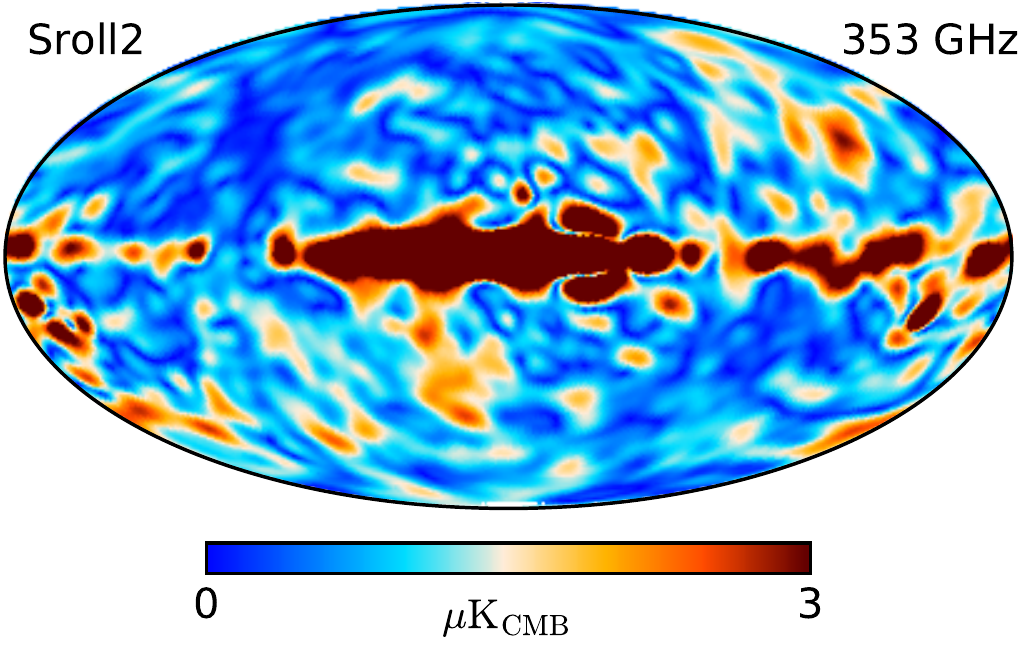}\\
\caption{\srollone\ (left column) and \srolltwo\ (right column) polarization intensity maps obtained by subtracting\CHANGE{, from the output sky map,} the input sky map from a simulation including a nonlinear ADC. The simulation input map contains dipole, sky signal, and systematic effects but electronic noise only to reveal effects lower than the full noise.}
\label{fig:sims_sroll1_and_sroll2}
\end{figure}

Figure~\ref{fig:sims_sroll1_and_sroll2} shows the residual maps obtained taking the difference between input and output of an end-to-end simulation containing only dipole, sky signal, systematic effects, and ADC electronic noise.
In the left column, \srollone\ is used to process the simulated data, while in the right one, we adopt the \srolltwo\ method. At CMB frequencies, the large-scale residuals present in the \srollone\ computation are much lower in the \srolltwo\ approach, leading to residuals below $0.5\,\mu$K outside the Galactic plane. At 353\,GHz, the ADCNL effect is not dominant, and the $1/f$ noise dominates. 

The central part of the Galactic plane is not improved as the signal gradient is too large and induces strong sub-pixel effects which dominate. The level of the residual is however very small (below 1\%) compared to the Galactic plane signal.

\begin{figure}[ht!]
\includegraphics[width=\columnwidth]{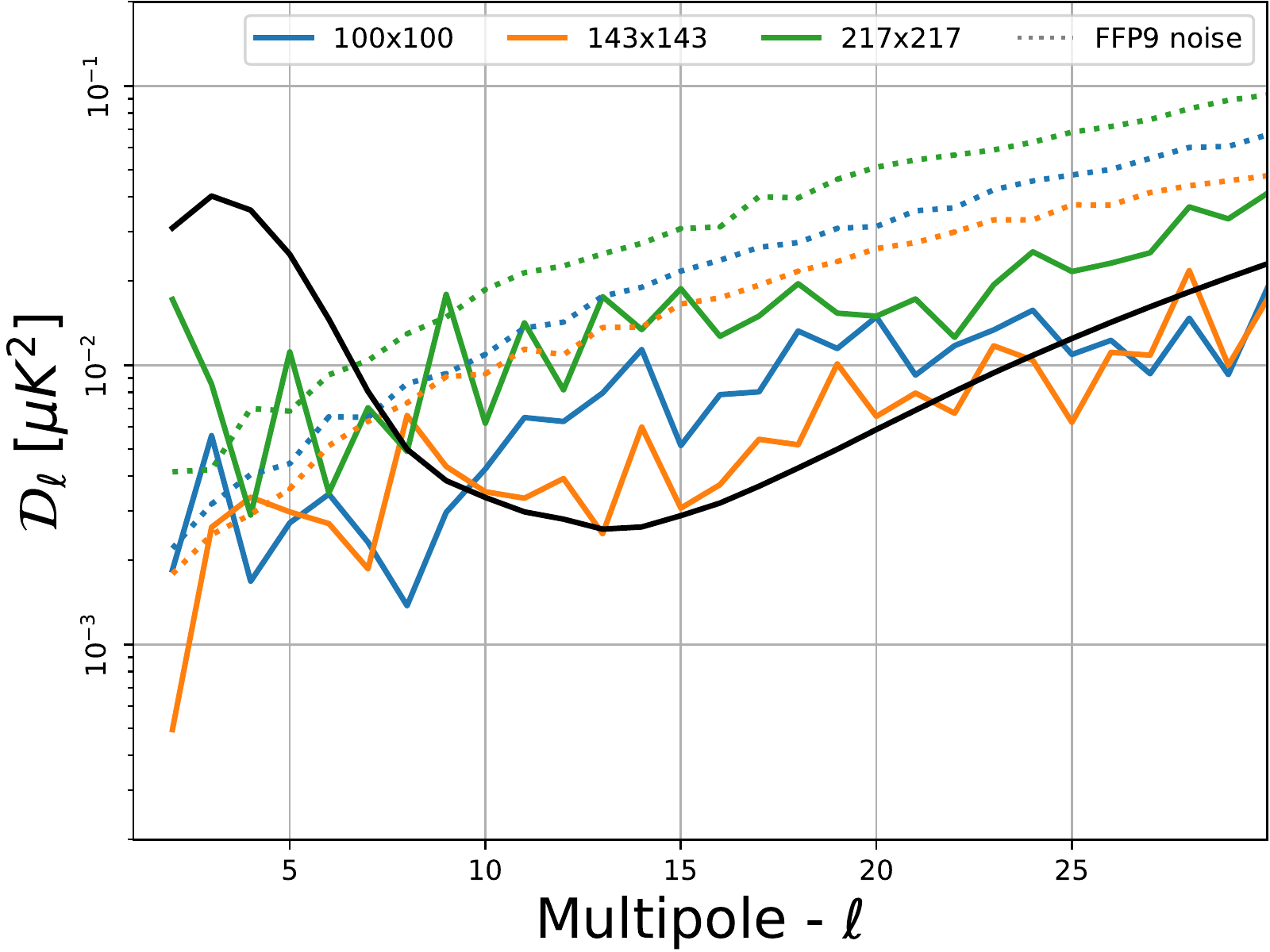}\\
\includegraphics[width=\columnwidth]{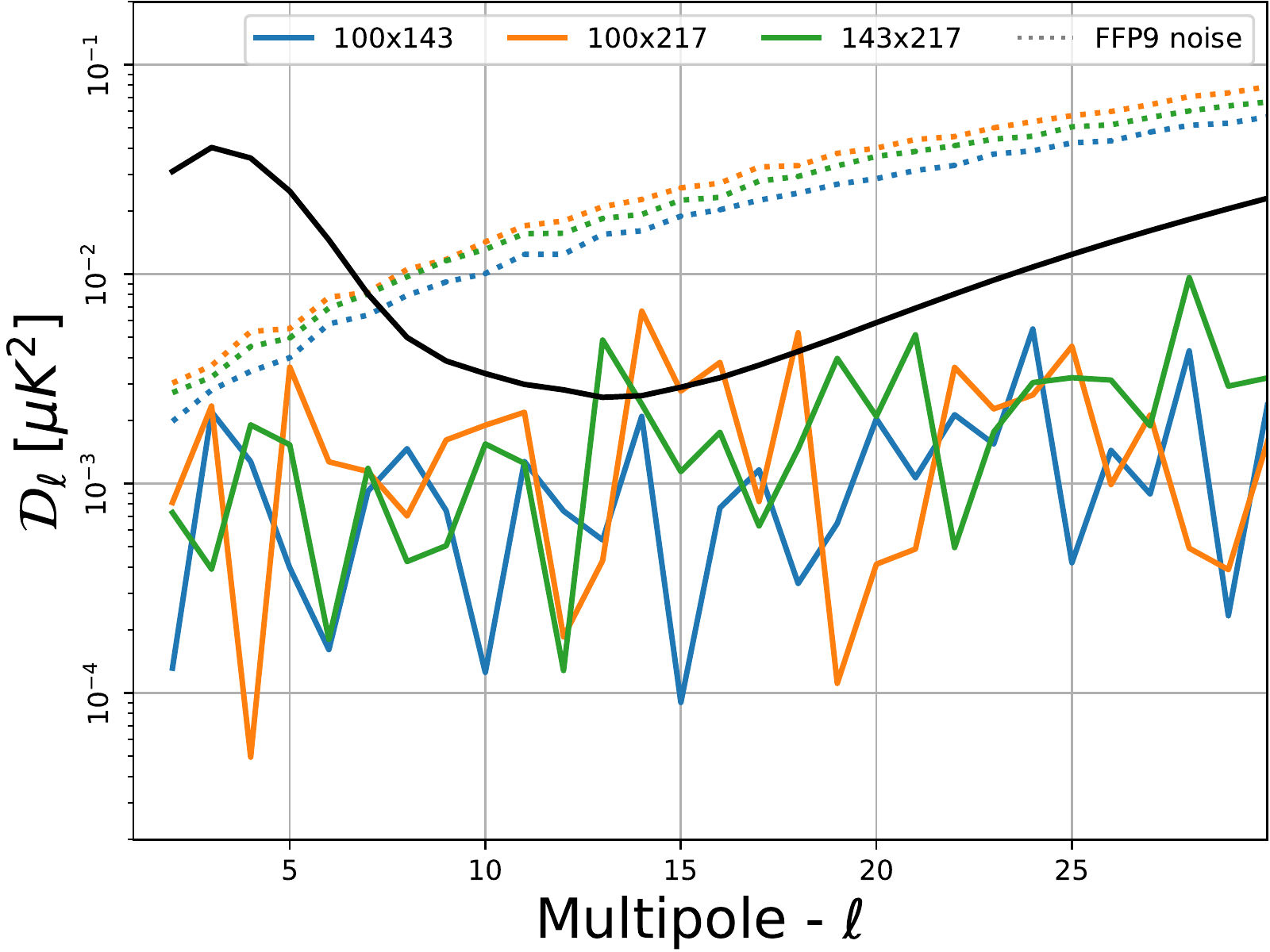}
\caption{$EE$ pseudo spectra (auto-spectra in the top panel, cross-spectra in the bottom panel) evaluated for 100\,GHz, 143\,GHz, and 217\,GHz on the simulations shown in Fig.~\ref{fig:sims_sroll1_and_sroll2} with a symmetric Galactic cut of 20\,\deg, 66\% of the sky. The black solid line corresponds to a $EE$ power spectrum with $\tau=0.055$ and $10^{10}A_s=21.14$.}
\label{fig:spectrum_sims_sroll2}
\end{figure}

The top panel of Fig.~\ref{fig:spectrum_sims_sroll2} shows the $EE$ power spectrum of the \srolltwo\ residual maps presented in the right column of Fig.~\ref{fig:sims_sroll1_and_sroll2}.
At 217\,GHz, the residual power left in the maps by \srolltwo\ is below the noise given by simplified FFP9 simulations \citep{planck2014-a14} for $\ell >10$ but dominates at $\ell<5$. At 100 and 143\,GHz, the ADCNL residuals are below the FFP9 noise down to $\ell=3$.

The bottom panel of Fig.~\ref{fig:spectrum_sims_sroll2} shows the ADCNL residuals for cross-power spectra which fall significantly below the auto spectra, showing that those residuals are weakly correlated between frequencies, and well below the fiducial cosmological signal. The much lower level of these cross spectra demonstrates that cross frequency spectra should be used for cosmology studies based on power spectra.

\begin{figure}[ht!]
\includegraphics[width=\columnwidth]{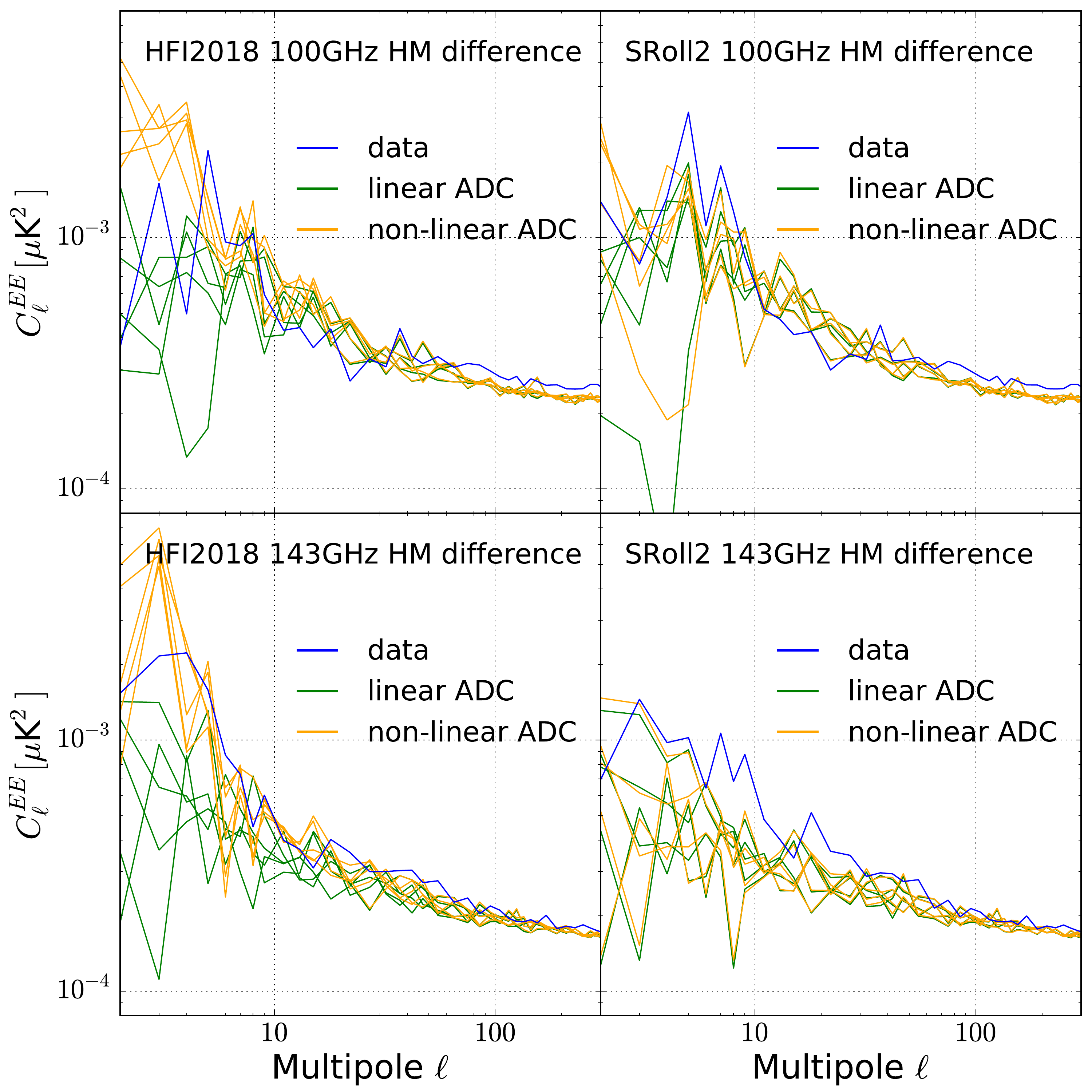}
\caption{Power spectrum of simulated half-mission null test. The plots in the left column are computed using \srollone\ while those in the right column are computed with \srolltwo. The two lower frequencies are shown in the two cases when the ADCNL is included in the input and when it is not. These plots demonstrate the efficiency of the \srolltwo\ method to remove the ADCNL residuals while the previous method left some artefacts at very large scale.}
\label{fig:adcnonoisetest}
\end{figure}

Figure~\ref{fig:adcnonoisetest} shows half-mission null test power spectra of five \srolltwo\ simulations with and without ADCNL.
This figure shows that \srolltwo\ cleans the ADCNL effect in such a way that the levels of the power spectra of the residual maps are comparable for simulations including or not ADCNL. This demonstrates the ability of \srolltwo\ to clean the ADCNL residual at the detector noise level. The half-mission null test is no longer affected by the ADCNL residuals and any detected residuals in this null test are caused by systematic effects other than the ADCNL.

In conclusion, this section shows that the ADCNL correction is very efficient and produces maps cleaned from the associated polarization leakage with residuals much lower than the noise amplitude, especially if we use cross-frequency spectra. This test  confirms that large-scale residuals are now dominated by the $1/f$ detector noise.

\subsection{All systematics}
\label{sec:allsyste}

\begin{figure*}[ht!]
\includegraphics[width=\textwidth]{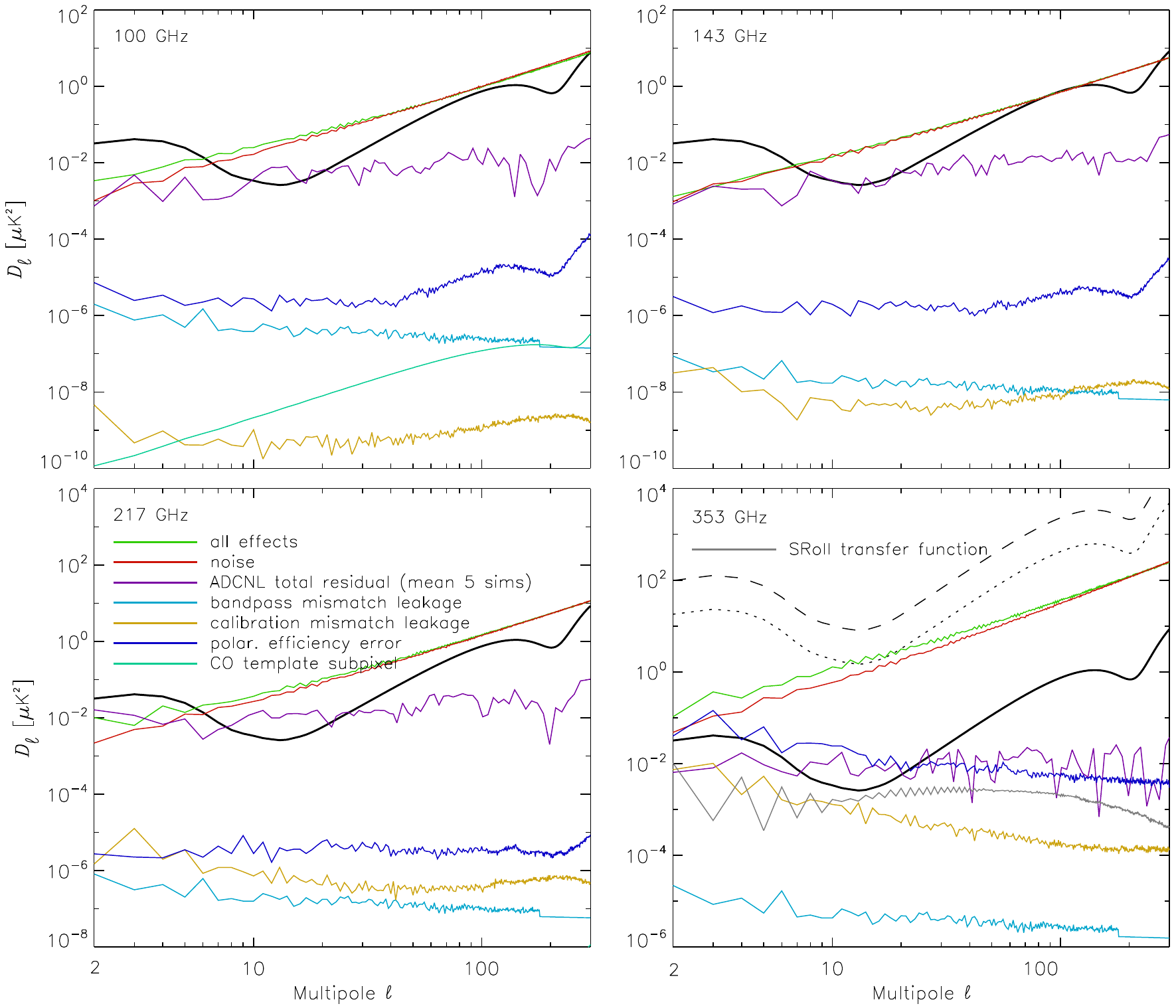}
\caption{Polarization power spectra in $D_{\ell}^{EE}$, showing the noise (red) and the main systematic residuals: ADCNL remaining after cleaning (purple), polarization efficiency (dark blue), bandpass-mismatch leakage (light blue), leakage from calibration mismatch (orange), and the sum of all these (green). \srolltwo\ residual empirical transfer function (gray), and CO template subpixel effect (turquoise) have not been included in the sum. The fiducial CMB power spectrum is shown as a black lines. At 353\,GHz, the systematic residuals are dominated by the polarization efficiency knowledge that has been fitted at the noise level for the lowest multipoles. At other frequencies, ADCNL is still the instrumental systematic effect that has the highest residuals.}
\label{fig:fig17} 
\end{figure*}

Finally, it is important to identify the main systematic effects that impact the data for each frequency. Figure~\ref{fig:fig17}, which can be compared to Fig.~52 of \citedpc, shows the level of each systematic effect for each frequency.
For all frequencies, at large scales, the $1/f$ detector noise becomes the main limitation, while this was not the case with previous versions of the \Planck\ HFI data. A classical 1/f cleaning that uses offsets on timescales shorter than one minute has not been included  in \srolltwo\  because we verified that it removes \CHANGE{some} signal due to insufficient redundancy in the \Planck\ scanning strategy. Implementing this would require end-to-end simulations to assess the trade-off of reducing the 1/f noise versus the knowledge of the induced amplitude transfer function. For the three lower frequencies, the main systematic effect below the $1/f$ noise is the ADCNL residual. At 353\,GHz the main systematic effect is the distortion of the dust emission by the polar efficiency and angle errors. Another very important improvement on the previous mapmaking is the very small level of leakage residuals, thanks to a better 353-GHz map due to the identification of the degeneracy between the gain determination and the very long time transfer functions.

This demonstrates that \srolltwo\ maps are \CHANGE{the best that can be achieved before removing} the $1/f$ detector noise. \CHANGE{This} cannot be carried out before the systematic effects at large scales can be brought down to a level where the degeneracies between $1/f$ noise and systematic effects do not remove signal. To improve the map \CHANGE{at large angular scales} the next generation of mapmaking should undertake an improved model reducing the dimensionality of the $1/f$ noise parameters to be computed and make it compatible with the information brought by the redundancy of the data.

Another \CHANGE{limitation}, that is especially acute at 353\,GHz, lies in our ability to build statistically realistic simulations of the interstellar dust. The next generation of maps should integrate the component-separation tools inside the mapmaking and include the higher frequencies of 545 and 857\,GHz (although they are not polarized) in order to provide the best dust foreground template. In that case, the frequency maps will be limited by the residual after removal of the foreground components which is expected to be dominated by the CMB signal after enough galactic masking (the central part of the galactic disk is probably not sufficiently sampled in frequency, or spatially, to be decomposed).

\section{Conclusions}

This paper presents the \srolltwo\ mapmaking method and the resulting frequency maps using the Planck HFI data. This method corrects or improves the correction of the instrumental systematic effects identified earlier and described in \citedpc\ but not fully cleaned in the Planck collaboration legacy release. The main goal of \srolltwo\ is to provide polarized maps for studies requiring the largest scale. This paper demonstrates and quantifies the efficiency of this method in correcting for systematic effects, and provides frequency maps at the detector noise level for all multipoles from 2 to 100 when masking the Galactic plane. The \srolltwo\ maps are free of systematic errors at large scale but, for higher multipoles, the maps need to be cleaned from other instrumental effects, such as the beam anisotropies or the 4-K line residuals, to reach a level lower than the noise.
 
We demonstrate that this new mapmaking approach provides polarized CMB maps with a variance at $\ell<6$ reduced by a factor larger than two with respect to \citedpc, and without instrumental bias outside the Galactic disk. Furthermore, significant improvements are made  for multipoles up to 100. This new data set can therefore be used to improve cosmological parameter measurements based on the re-ionization peak at very low multipoles ($\tau$ for E modes, and r and lensing for B-modes). \CHANGE{\srolltwo\ maps  (data and simulations) are available at \href{http://sroll20.ias.u-psud.fr}{\texttt{http://sroll20.ias.u-psud.fr}}}.

The future Planck-HFI data mapmaking process should integrate component separation and systematic effect correction inside the mapmaking to avoid the degeneracies induced by the foreground bandpass leakage,  for example.

\begin{acknowledgements}
This work is part of the Bware project, partially supported by CNES. The program was granted access to the HPC resources of CINES (\url{http://www.cines.fr}) under the allocation 2017-A0030410267 made by GENCI (\url{http://www.genci.fr}). The authors acknowledge the heritage of the \Planck\ HFI consortium regarding data, software, knowledge, and the IAP hosted HFI DPC computing facility. LP acknowledges the support of the CNES postdoctoral program.
\end{acknowledgements}

\bibliographystyle{aat}
\bibliography{Planck_bib}

\end{document}